%% file: main_arxiv.tex
\documentclass[11pt]{article}
\usepackage{amsmath, amssymb}
\usepackage{algorithm, algorithmic} 
\usepackage{amsfonts,mathrsfs,bm}
\usepackage{geometry}
\usepackage{listings}
\usepackage{caption}
\usepackage{subcaption}
\usepackage{url}
\usepackage[numbers, sort&compress]{natbib}
\usepackage{tabularx}
\usepackage{pdfpages}
\usepackage{emptypage}
\usepackage{authblk}
\usepackage[shortlabels]{enumitem}
\usepackage[english]{babel}
\usepackage[normalem]{ulem}

\usepackage[colorlinks,
linkcolor=blue,
citecolor=blue,
urlcolor=magenta]{hyperref}

\newcommand{\GP}{\mbox{GP}}

\newcommand{\covloss}{\ell^{\text{\small{cov}}}}
\newcommand{\varloss}{\ell^{\text{\small{var}}}}

\title{Self-Supervised Representation Learning for Longitudinal Polypharmacy Patterns}
\author[1, *]{Yunshan Duan}
\author[1]{Eero Korpela}
\author[1]{Yanxun Xu}
\affil[1]{{\small Department of Applied Mathematics and Statistics, Johns Hopkins University}}
\affil[*]{{\small yduan25@jh.edu}}
\date{} 

\begin{document}

\maketitle

\begin{abstract}
Polypharmacy, commonly defined as the concurrent use of multiple medications, is increasingly prevalent in aging populations and is associated with adverse health outcomes. Motivated by longitudinal studies of aging people with HIV (PWH), we study how medication use evolves over time and characterizes multimorbidity patterns. Longitudinal medication data present substantial methodological challenges, including high dimensionality, sparsity, irregular observation times, and structured pharmacologic relationships among medications. Existing approaches are typically task-specific and lack a unified framework for learning general-purpose representations of medication trajectories. We propose LOPEL (LOngitudinal Polypharmacy Embedding Learning), a two-stage self-supervised framework for learning low-dimensional representations of longitudinal medication data. In the first stage, visit-level embeddings are learned using a Gaussian process model that incorporates pharmacologic similarity through the Anatomical Therapeutic Chemical hierarchy. In the second stage, subject-level embeddings are constructed by modeling trajectories over time and defining similarity through a Wasserstein-based representation capturing temporal dynamics and uncertainty. Simulation studies demonstrate that LOPEL accurately recovers latent structure under realistic conditions with high-dimensional sparsity and irregular sampling. In an application to aging cohorts of PWH, LOPEL identifies clinically meaningful subgroups that differ in the timing, composition, and progression of medication use, highlighting heterogeneity relevant for risk stratification and clinical management.
\end{abstract}

\section{Introduction}

Polypharmacy, commonly defined as the concurrent use of five or more medications, is increasingly prevalent among aging populations and individuals with multiple chronic conditions.
While appropriate polypharmacy reflect evidence-based management of complex conditions, the accumulation of medications over time is strongly associated with adverse drug reactions, treatment-related morbidity, avoidable hospitalizations, and increased healthcare burden \citep{chang2020polypharmacy, delara2022prevalence}. 
From a clinical perspective, these risks are driven not only by the number of medications prescribed at a single time point, but also by how medication regimens evolve: which drug classes are introduced or discontinued, how specific drug combinations are maintained or modified over time, and how these longitudinal patterns relate to downstream outcomes such as cognitive decline, frailty, or mortality \citep{greene2014polypharmacy}. 
Characterizing longitudinal medication trajectories is therefore important to a wide range of clinical and public health tasks, including identifying high-risk subgroups, investigating harmful drug-drug interactions, and supporting deprescribing initiatives.

Our methodological development is motivated by longitudinal cohort studies of people living with HIV.  
Advances in antiretroviral therapy have transformed HIV from a fatal illness into a chronic condition, and individuals living with HIV now have life expectancies approaching those of the general population \citep{marcus2020comparison}.
As this population ages, comorbid conditions such as cardiovascular disease, metabolic disorders, and neurocognitive impairment become increasingly prevalent, leading to complex and evolving medication regimens that extend beyond antiretroviral therapy alone \citep{zhabokritsky2024non}. 
Polypharmacy has therefore emerged as a growing concern in HIV care, raising clinically important questions about how concurrent non-HIV medications interact with antiretroviral treatment, influence medication adherence, and contribute to health outcomes such as cognitive decline and dementia risk \citep{rubin2022degree}. 
Addressing these questions requires methodology for representing and modeling heterogeneous, high-dimensional longitudinal medication data. 

However, developing such a methodology is complicated by the unique analytical challenges presented by these data.
First, medication records are high-dimensional and sparse: among thousands of possible medications, any given individual typically uses only a small subset at any visit, yielding extremely sparse multivariate profiles. 
Second, medication use is naturally encoded as multivariate binary or low-count data, complicating trajectory modeling relative to continuous longitudinal outcomes. 
Third, clinical encounters occur at irregular and subject-specific time points, violating the regular-grid assumptions underlying many longitudinal models. 
Finally, medications are not exchangeable features: drugs are related through therapeutic class, mechanism of action, and clinical indication. 
The Anatomical Therapeutic Chemical (ATC) classification system \citep{nahler2009anatomical} encodes these relationships in a clinically meaningful hierarchy, but this rich domain knowledge is rarely incorporated directly into statistical models of longitudinal medication data.

Existing approaches address subsets of these challenges but do not provide a unified framework that simultaneously handles them.
Parametric trajectory models \citep{nagin2018group} and latent class mixed models \citep{lin2000latent} have been widely used to identify subgroups with distinct longitudinal patterns, but they are typically designed for low-dimensional continuous outcomes and become unstable or difficult to interpret in high-dimensional sparse binary settings. 
Sequence-based and state-space models \citep{sakoe2003dynamic}, including hidden Markov models \citep{xia2019bayesian}, can accommodate irregular sampling but scale poorly to multivariate medication profiles with hundreds of sparse indicators. 
Recent high-dimensional longitudinal clustering \citep{yang2023model, lu2024sparse} and deep representation learning methods \citep{ zhao2021longitudinal, qiu2025deep} 
address scalability through latent-space modeling and self-supervised learning, but they do not handle sparse multivariate medication trajectories observed at irregular times or incorporate structured pharmacologic hierarchies. 

Furthermore, existing methods are largely task-specific: models are typically developed for a single outcome or inferential objective, such as modeling medication adherence, identifying polypharmacy subgroups, or predicting cognitive decline. 
As a result, analytic pipelines are difficult to generalize across scientific aims, often requiring repeated model development and extensive feature engineering when new questions arise. 
This limitation highlights a broader methodological gap: there is currently no framework for learning representations of longitudinal medication data that accommodates irregular observation times, high-dimensional sparsity, and structured pharmacologic knowledge while remaining broadly applicable across tasks. 
Such representations provide a common statistical foundation for inference and decision-making, enabling analysts to move from raw medication records to interpretable summaries for population-level discovery and individual-level risk stratification.

To address this gap, we propose LOPEL (LOngitudinal Polypharmacy Embedding Learning), a self-supervised learning framework for task-agnostic representation learning from longitudinal medication data.
LOPEL operates in two stages. 
First, we learn visit-level embeddings of multivariate medication profiles using Gaussian process self-supervised learning (GPSSL) \citep{duan2025self}, incorporating an ATC-informed kernel that encodes therapeutic similarity and borrows strength across pharmacologically related drugs. Second, we model the longitudinal trajectories of these visit-level embeddings over irregular observation times using GPSSL defined on distributions under a Wasserstein geometry, yielding compact subject-level representations that capture both trajectory dynamics and posterior uncertainty.
The resulting embeddings are reusable across downstream tasks, including identifying patient subgroups, predicting clinically relevant outcomes, and forecasting future medication patterns.
Through simulation studies and an application to medication data from aging cohorts of people living with HIV (PWH), we demonstrate the effectiveness of LOPEL in identifying clinically interpretable subgroups with distinct medication trajectories.

The remainder of this paper proceeds as follows. Section \ref{sec:motivation} introduces the motivating study and describes the data and preprocessing steps. Section \ref{sec:method} presents the LOPEL framework. Section \ref{sec:simulation} evaluates its performance using simulation studies, and Section \ref{sec:realdata} applies LOPEL to study polypharmacy patterns in PWH. Section \ref{sec:discussion} concludes with a discussion.

\section{Motivating Application and Data Description} \label{sec:motivation}

The motivating application for this work is a harmonized longitudinal medication dataset derived from three cohort studies investigating the clinical and neurocognitive effects of HIV, including PWH and HIV-negative comparison participants. These studies include the CNS HIV Antiretroviral Therapy Effects Research (CHARTER) \citep{heatonHIVassociatedNeurocognitiveDisorders2010a}, the National NeuroAIDS Tissue Consortium (NNTC) \citep{morgello2001national}, and HIV Neurobehavioral Research Program (HNRP) \citep{WelcomeHNRPa}, spanning January 1999 through March 2020. Additional details on each cohort are provided in Supplement Section B. 

In PWH, medication regimens typically include antiretroviral therapy (ART), which targets HIV infection, as well as additional medications used to manage comorbid chronic conditions.
Our scientific goal is to characterize age-dependent trajectories of non-ART polypharmacy to understand how multimorbidity evolves with age.
We therefore focus on non-ART medications, which reflect the emergence and management of comorbid disease and serve as a direct marker of multimorbidity dynamics.

Medication names were normalized using the RxNorm API's \texttt{getApproximateMatch} function \citep{petersApproximateMatchingMethod2011a, liu2005rxnorm}. All drug strings were mapped to standardized ingredient-level RxNorm concepts, with manual verification to ensure accuracy.
Each medication was then annotated with ATC codes, a hierarchical drug classification system \citep{world2017atc}. The ATC system organizes medications into five nested levels: anatomical main group, therapeutic subgroup, pharmacological subgroup, chemical subgroup, and chemical substance. Medications sharing higher-level ATC categories are typically used for related therapeutic purposes, while lower levels capture increasingly specific pharmacological distinctions. Some medications are associated with multiple ATC codes, reflecting use across different therapeutic indications or pharmacological targets. ATC annotations were obtained from RxNorm when available and supplemented using RxClass \citep{rxclass}. Assignments from RxClass were retained only when the corresponding concepts shared identical active ingredient sets with the normalized RxNorm concept, ensuring pharmacological consistency.

For example, the antidepressant sertraline has ATC code \texttt{N06AB06}. The first character, \texttt{N}, denotes the anatomical main group (nervous system), and the next two characters, \texttt{06}, specify the therapeutic subgroup (psychoanaleptics). The fourth character, \texttt{A}, indicates the pharmacological subgroup (antidepressants), and the fifth character, \texttt{B}, identifies the chemical subgroup (selective serotonin reuptake inhibitors, SSRIs) \citep{lochmann2019selective}. The final two digits, \texttt{06}, uniquely identify sertraline as a chemical substance.
Another SSRI, such as fluoxetine (\texttt{N06AB03}), shares the same classification through level 4, reflecting similar mechanisms of action and therapeutic indications.
This hierarchical structure encodes clinically meaningful similarity between drugs and provides domain knowledge that can be leveraged in statistical modeling.

Because our scientific focus is age-related multimorbidity, we retain medications associated with chronic disease management. Standardized medications with ATC codes are further mapped to clinically-defined chronic condition categories to facilitate downstream interpretation and visualization. This mapping is based on RxRisk \citep{prattValidityRxRiskComorbidity2018, widagdoValidityUpdatedRxRisk2025}, a validated pharmacy-based comorbidity index that infers chronic conditions from prescription medication data.

After preprocessing, the dataset contains $M = 694$ distinct non-ART medications, including 49 with multiple ATC codes and 21 mapped to multiple chronic conditions. It includes 1,171 PWH contributing 3,188 visits and 555 PWOH contributing 1,234 visits. Participants have a mean of 2.56 visits over an average follow-up of 3.30 years and use an average of 3.65 non-ART medications per visit. The resulting visit-by-medication matrix is highly sparse, with the vast majority of entries equal to zero.


\section{Method} \label{sec:method}

Assume that we observe longitudinal medication data for $n$ participants. For participant $i$, visits occur at irregular ages $t_{ij}$, $j = 1, \dots, v_i$, where $v_i$ denotes the number of visits. At each visit, medication use is represented by a regimen $x_{ij} \in \mathcal{X}$, where $x_{ij}$ is the set of medications taken at that visit and $\mathcal{X}$ is the space of all possible regimens. 
Our goal is to construct low-dimensional subject-level representations that capture longitudinal patterns of medication use, while incorporating structured pharmacologic similarity among medications.

 \subsection{Overview of LOPEL}

\begin{figure}[htbp!]
     \centering
     \includegraphics[width=0.99\textwidth]{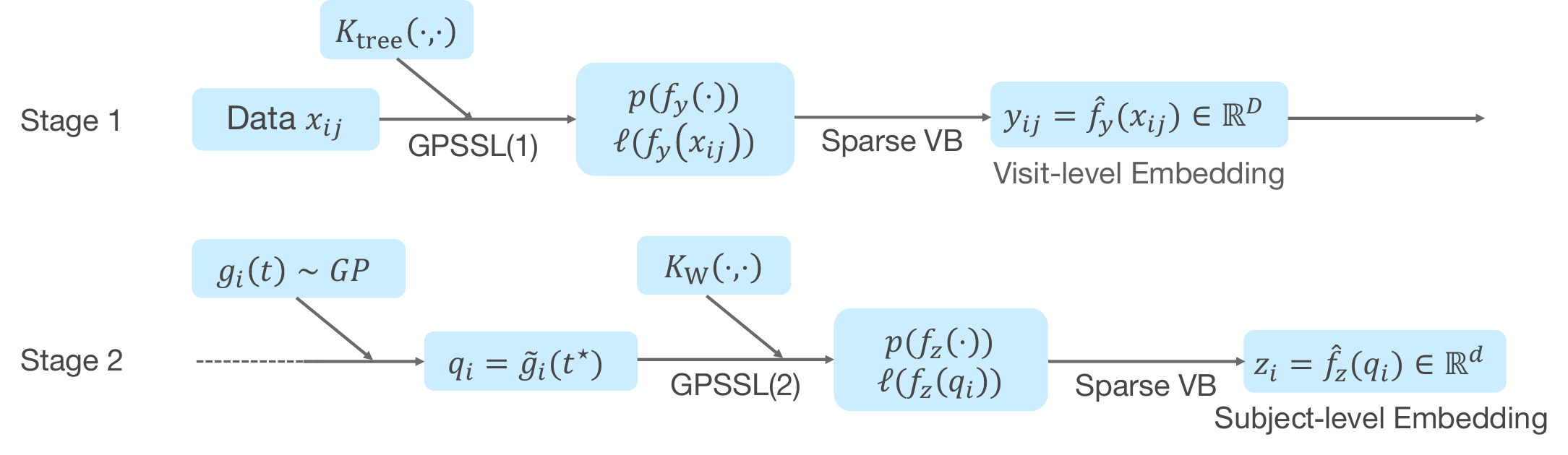}     
     \caption{Flowchart of the two stage construction of LOPEL.}
     \label{fig:flowchart}
\end{figure}

We propose LOPEL, a two-stage self-supervised representation learning framework for longitudinal medication data. 
In the first stage, we learn visit-level embeddings that map high-dimensional medication regimens into a lower-dimensional continuous space, incorporating pharmacologic similarity through the ATC hierarchy. 
In the second stage, we learn subject-level embeddings that summarize the longitudinal trajectories of these visit-level representations over irregular time, capturing both temporal dynamics and estimation uncertainty.
The resulting embeddings provide compact, interpretable summaries of longitudinal polypharmacy patterns and can be used for downstream analyses.
Figure \ref{fig:flowchart} shows a flowchart of the inference pipeline for LOPEL.

\subsection{Visit-level medication embeddings leveraging ATC hierarchy}
\label{sec:ATC-SSL}
The first stage of LOPEL constructs low-dimensional representations of medication regimens at individual visits. Because the number of possible medications $M$ is large while only a small subset is used at any given visit, these regimens are inherently high-dimensional and sparse. At the same time, medications exhibit structured pharmacologic relationships, making it important to preserve clinically meaningful similarities among drugs. We therefore learn visit-level embeddings that incorporate this structure and provide a compact summary of each regimen.

Recall that each visit-level regimen is denoted by $x_{ij} \in \mathcal{X}$. For example, if patient $i$ uses sertraline (SRT) and metformin (MET) at visit $j$, then $x_{ij} = \{ \text{SRT},\text{MET} \}$.  Our goal is to learn a mapping
$
f_y : \mathcal{X} \to \mathbb{R}^D, \;\; D \ll M,
$
such that pharmacologically similar regimens are mapped to nearby points in the embedding space.
To achieve this, we model the embedding function $f_y$ nonparametrically, allowing similarity between medication regimens to be flexibly captured. Specifically, each coordinate of $f_y$ is treated as an unknown function on $\mathcal{X}$ with a Gaussian process prior 
$
f_y^{(k)} ( \cdot) \sim \GP(m(\cdot), K(\cdot,\cdot)), \; k = 1, \dots, D,
$
where the coordinates are assumed independent a priori but share a common kernel $K(\cdot,\cdot)$.
This prior induces smoothness with respect to the similarity structure encoded by $K$, so that pharmacologically related regimens are mapped to nearby embeddings.
See later for detailed construction of $K$. In the SSL context the $y_{ij} = f_x(x_{ij})$ serve as the target labels.

Given the GP prior and the observed regimens, we construct a posterior over the embedding function. In supervised settings, this would be based on a likelihood derived from labeled outcomes. In our setting, where such labels are unavailable, we instead adopt a generalized Bayesian framework using Gaussian process self-supervised learning (GPSSL) \citep{duan2025self}, defining a posterior of the form
\begin{equation} \label{eq:generalized_post}
    \tilde{p}(f_y) \propto p(f_y)\exp\{-\ell(f_y, X)\},
\end{equation} 
where $p(f_y)$ denotes the GP prior on $f_y$, and $\ell(f_y, X)$ is a loss function evaluated over the observed visit-level regimens $X = \{x_{ij}\}_{i,j}$. This formulation follows the generalized Bayesian inference (GBI) framework \citep{bissiri2016general}, in which the loss serves as a surrogate negative log-likelihood, enabling posterior-like inference without explicit probabilistic labels. 

Let $Y = f_y(X)$. 
Let $s_k^2$ denotes the sample variance of the $k-$th dimension of $Y$, and let 
$
\hat{\Sigma} =  \frac{1}{N-1}\sum_{v=1}^N (Y_v - \bar{Y})(Y_v - \bar{Y})^\top
$
denote the empirical covariance matrix, where $N$ denotes the total number of visit-level observations. 
The GPSSL loss is defined as
\begin{equation} \label{eq:GPSSL_loss}
\ell(f_y, X) = c_V \varloss(Y) + c_C \covloss(Y),
\end{equation}
where $c_V$ and $c_C$ are tuning parameters controlling the contributions of the two components. The variance term $
\varloss(Y) = \displaystyle \frac{1}{D}  \sum_{k=1}^D \max\left(0, \gamma - \sqrt{s_k^2+\epsilon}\right)
$ encourages each embedding dimension to exhibit sufficient variability across observations and prevents collapse to a nearly constant representation. 
The covariance component penalizes correlation between embedding dimensions.
The covariance loss
$
\covloss(Y) = \displaystyle \sum_{k \neq \ell} \hat{\Sigma}_{k, \ell}^2
$
shrinks off-diagonal covariance terms toward zero, encouraging different dimensions to capture complementary aspects of the data rather than redundant information. Together, these components encourage the embeddings to distinguish between different medication regimens. 

The GP prior and the loss function jointly balance two objectives: the prior encourages similar regimens encoded by the kernel to map to nearby embeddings, while the loss prevents degenerate solutions and promotes meaningful variation across regimens. 
Following \citet{duan2025self},
we approximate the generalized Bayesian posterior  \eqref{eq:generalized_post} using sparse variational inference \citep{knoblauch2019generalized, hensman2015scalable}.
Specifically, we introduce $M_u$ inducing inputs $U_x$ with corresponding function values $U_y = f_x(U_x)$. 
A variational distribution is specified by way of independent Gaussians
$q(U_y) = N(m, S) = \prod_{k=1}^D N(m^{(k)}, s^{(k)})$. The resulting evidence lower bound (ELBO) is
$\text{ELBO}  =  -\mathbb{E}_{q(f_y)}\!\left[\ell(f_y)\right] - \text{KL}\left(q(U_y) || p(U_y)\right),$
where $q(f_y) = \int p(f_y \mid U_y) q(U_y) d U_y$, and $p(U_y)$ is implied by the GP prior.
The expectation is estimated via Monte Carlo sampling, and the ELBO is optimized jointly over the variational parameters, kernel hyperparameters, and inducing locations using stochastic gradient descent. 
The visit-level embedding for patient $i$ at visit $j$ with regimen $x_{ij}$ is then defined as the generalized posterior mean and  approximated based on interpolation under the GP model:
\[
y_{ij} =  \mathbb{E}_{\tilde{p}}[f_y(x_{ij})] \approx \mathbb{E}_{q(U_y)} \left[ \mathbb{E}(f_y(x_{ij}) \mid U_y) \right ] = 
k(x_{ij}, U_x) K_{U_x, U_x}^{-1} m \in \mathbb{R}^D,
\] 
where $k(x_{ij}, U_x)$ is the $1 \times M_u$ vector of kernel evaluation between $x_{ij}$ and the inducing inputs and 
$K_{U_x, U_x}$ is the $M_u \times M_u$ kernel matrix among inducing inputs. Further technical details are provided in Supplement Section C.1. 

\noindent{\bf ATC-informed tree kernel for regimen similarity.}
The kernel $K(\cdot, \cdot)$ 
is central to this construction, as it determines how similarity between medication regimens is quantified. A key challenge is that medications are not exchangeable features; instead, they are structured through therapeutic class, pharmacologic mechanism, and clinical indication, as encoded by the ATC hierarchy. We incorporate this structure by defining a tree-based kernel that measures similarity between regimens based on shared ATC ancestry.
 
\begin{figure}[t]
     \centering
     \includegraphics[width=0.85\textwidth]{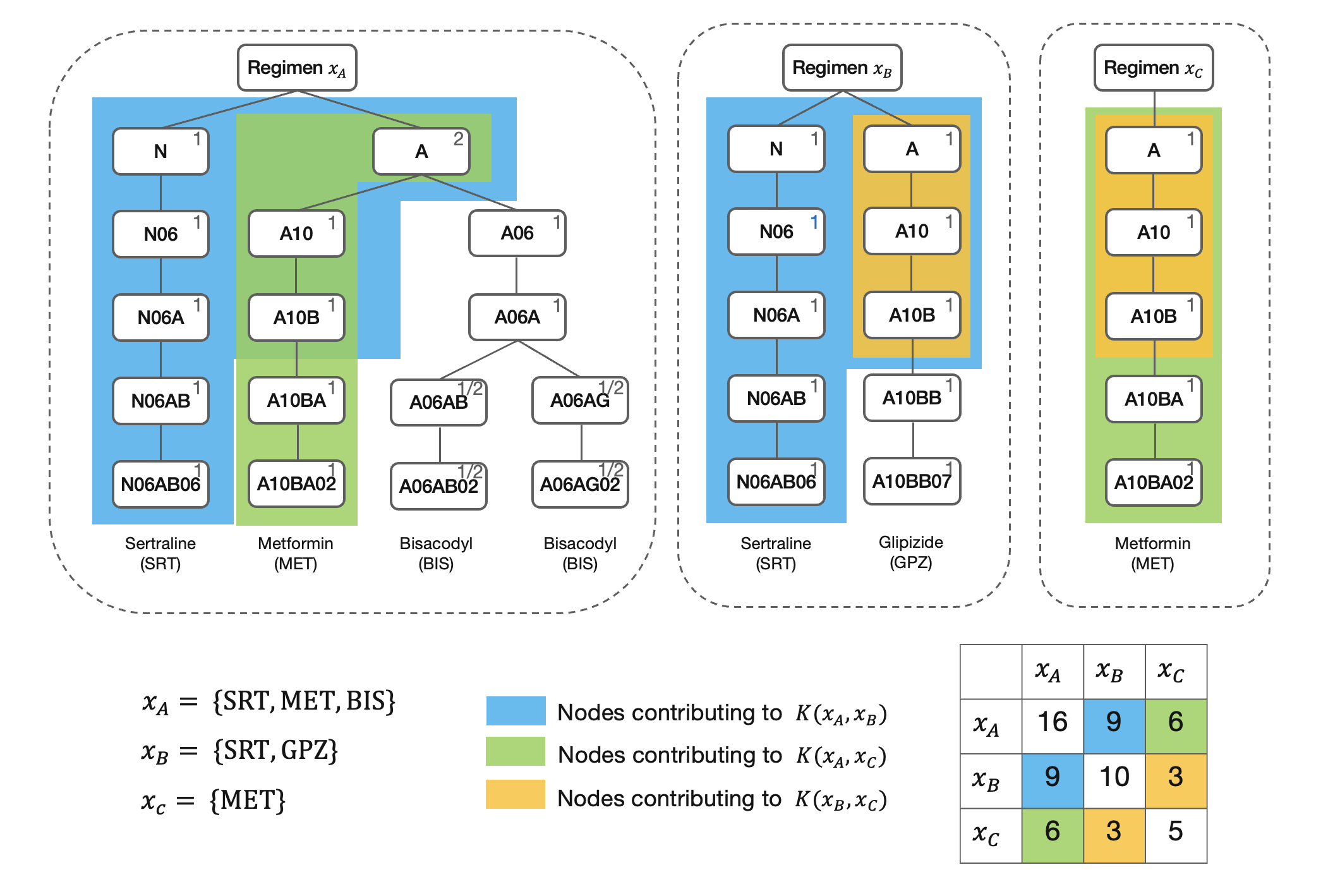}     
     \caption{Illustration of the ATC hierarchy and the proposed tree-feature kernel $K_{\text{tree}}(x, x')$.  Each node represents an ATC prefix, and medications correspond to leaf nodes. 
     The numeric annotation at node $u$ shows the value of $c_u(x_i)$, indicating the contribution of that regimen at the node. Colored boxes highlight shared ATC prefixes contributing to pairwise similarity between regimens. The similarity matrix in the bottom right summarizes the resulting pairwise kernel values when all level weights are set to one.}
     \label{fig:ATC_illustration}
\end{figure}

Let $\mathcal{T}$ denote the ATC tree of depth five corresponding to the five ATC levels. Each medication is represented as a path from the root to a leaf node, where each level corresponds to a progressively more specific ATC category. Internal nodes correspond to shared prefixes of ATC codes, representing common anatomical, therapeutic, or pharmacological groupings (see Figure \ref{fig:ATC_illustration}).
Because of this hierarchical structure, two medications are considered similar if they share common ancestry in the tree, with deeper shared nodes indicating stronger pharmacologic similarity. Accordingly, similarity between regimens can be defined by the extent to which their medications overlap along shared branches of the tree.

We represent each regimen through node-level features that aggregate medication contributions across the tree. For a regimen $x$, we define a feature at each node $u \in \mathcal{T}$ as 
$\phi_u(x) = w_{l(u)} \cdot c_u(x),$
where $l(u)$ denotes the level of node $u$, $c_u(x)$ counts the number of medications in $x$ whose ATC paths pass through node $u$, and $ w_{\ell} \geq 0$ is a level-specific weight controlling the importance of ATC level $l$. 
Thus $\phi_u(x)$ measures the extent to which regimen $x$ occupies the subtree rooted at node $u$.

To handle medications associated with multiple ATC codes, we distribute each medication’s contribution evenly across its codes so that it contributes a total weight of one. Specifically, if $A(m)$ is the set of ATC codes for medication $m$ and $n_m = |A(m)|$, then 
\begin{equation*} \label{eq:cux}
    c_u(x) = \sum_{m \in x} \frac{1}{n_m} \sum_{a \in A(m)} 1\{u \in \text{path}(a)\}. 
\end{equation*}

The similarity between two regimens $x$ and $x'$ is then defined as the inner product of their tree features: 
$K_{\text{tree}}(x, x') = \lambda \sum_{u \in \mathcal{T}}  \phi_u(x) \phi_u(x'),$
where $\lambda > 0$ is a learnable output-scale parameter controlling the overall variance. By construction, this kernel assigns higher similarity to regimens that share more, or deeper, ATC prefixes.

To illustrate how this kernel captures pharmacologic similarity, consider four non-ART medications: sertraline (SRT; N06AB06), a selective serotonin reuptake inhibitor (SSRI); metformin (MET; A10BA02) and glipizide (GPZ; A10BB07), glucose-lowering agents under ATC group A10; and bisacodyl (BIS; A06AB02 and A06AG02), a stimulant laxative. BIS is associated with two ATC codes.  We consider three visit-level regimens:
$x_A = \{ \text{SRT, MET, BIS}\}, \;\; x_B = \{ \text{SRT, GPZ}\}, \;\; x_C = \{ \text{MET}\}.$
The values displayed at each node in Figure \ref{fig:ATC_illustration} correspond to  $c_u(x_i)$, reflecting how medications contribute along their ATC paths. For example, in regimen $x_A$,  each medication contributes to all nodes along its ATC path. For medications associated with multiple ATC codes, such as BIS, the contribution is divided equally across the corresponding paths so that the total contribution remains one.

Similarity between regimens is determined by overlap in these node-level features, with deeper shared prefixes leading to higher similarity. The colored regions in Figure \ref{fig:ATC_illustration} highlight shared ATC prefixes contributing to pairwise similarity.  The similarity between $x_A$ and $x_B$ is relatively large for two reasons. First, both regimens include SRT, leading to overlap along the antidepressant branch (N06). Second, MET and GPZ both belong to the metabolic group A10, contributing additional overlap despite differences at more specific levels. In addition, BIS shares a higher-level anatomical category (A) with these medications, further increasing similarity.
In contrast, $x_A$ and $x_C$ share only the MET branch, so their similarity arises from overlap along a single path. 
Finally, $x_B$ and $x_C$ overlap only at a higher-level ATC category, A10B, leading to the smallest similarity among the three pairs.

In summary, Stage 1 learns visit-level embeddings by combining a Gaussian process prior with an ATC-informed tree kernel and a self-supervised regularization objective. The resulting embeddings map high-dimensional, sparse medication regimens into a structured continuous space that preserves pharmacologic relationships and provides the basis for modeling longitudinal dynamics in Stage 2.

\subsection{Subject-level embeddings capturing longitudinal patterns} \label{sec:method_stage2}

While Stage 1 summarizes medication regimens at individual visits, clinical heterogeneity also arises from how these regimens evolve over time. Two participants may have similar regimens at a given visit but differ substantially in their prior medication history and subsequent trajectories, including differences in onset timing and duration of use. Stage 2 therefore focuses on capturing trajectory-level structure. Our objective is to construct subject-level representations that summarize longitudinal medication dynamics while accounting for irregular observation times and estimation uncertainty. To achieve this, we extend the GPSSL framework to the trajectory level. In this setting, the inputs correspond to individual participants’ longitudinal medication trajectories, and similarity is defined through a kernel over these trajectories.

We begin by defining a representation for each participant’s longitudinal trajectory. Let $\mathcal{G}$ denote the space of longitudinal embedding trajectories. 
For participant $i$, we have visit-level embeddings $y_{ij} \in \mathbb{R}^D$ at times $t_{ij}$, $j = 1, \dots, v_i$ from Stage 1. We model the underlying trajectory as a function $g_i(t)$, 
$
y_{ij} = g_i(t_{ij}) + \varepsilon_{ij}, \; \varepsilon_{ij} \sim N(0, \Sigma_\varepsilon).
$
To jointly model all $D$ embedding dimensions, we place a multitask Gaussian process prior
$
g_i(\cdot) \sim \GP\!\big(m_g(\cdot),\; K_g(\cdot, \cdot)\big), \; K_g(t, t') = B k(t,t'),
$
where $k(t,t')$ governs smoothness over time and $B$ captures dependence across dimensions. This allows different components of the embedding to co-evolve over time, reflecting the fact that multiple medication domains often change together. With this prior and the visit-level embeddings $\{y_{ij}\}_{j=1,\dots, v_i}$, we obtain a posterior Gaussian process, 
$
\tilde g_i(\cdot)\sim \GP\!\big(m_i(\cdot),\, K_i(\cdot,\cdot)\big).
$
The posterior mean function $m_i(t)$ represents the estimated trajectory, while the covariance function $K_i(t,t')$ quantifies uncertainty. 

To stabilize participant specific trajectory estimation, we first fit the GP model to the pooled visit-level data and use the resulting hyperparameter estimates to regularize the participant specific GP fits. The observation noise parameters, task covariance, and correlated observation noise structure are fixed at their pooled estimates, while the remaining parameters are estimated separately for each participant. This partial pooling strategy accommodates participant specific temporal dynamics while borrowing information across individuals for covariance components that would otherwise be weakly identified from sparse trajectories.

Because visit frequency and follow-up duration vary across participants, the uncertainty in the estimated trajectories differs substantially between individuals. Participants with more frequent observations have more precisely estimated trajectories, whereas those with sparse observations exhibit greater uncertainty. Comparing only posterior mean trajectories would ignore these differences and treat all trajectories as equally certain.  To address this, we represent each participant not only by their estimated trajectory but also by the associated uncertainty. Specifically, the Gaussian process posterior for each participant provides both a mean function and a covariance function, which together define a distribution over trajectories. However, participants are observed at different time points, making direct comparison difficult. To enable comparison across individuals, we evaluate each posterior process on a common time grid $t^\star$, yielding a multivariate Gaussian representation
$
q_i = \tilde{g}_i(t^\star) = \mathcal{N}(m_i(t^\star), K_i(t^\star, t^\star)),
$
which summarizes both the estimated trajectory (through the mean) and the associated uncertainty (through the covariance) on a shared time scale.

Having defined $q_i$ for each participant, we next quantify similarity between participants by comparing these trajectory distributions. Specifically, we define a trajectory kernel based on the 2-Wasserstein distance between Gaussian measures, 
$$
K_W(q_i,q_j) = \exp\!\left(-\frac{W_2^2(q_i, q_j)}{2\tau^2}\right).
$$
The Wasserstein distance is well suited for this setting, as it captures differences in trajectory distributions. For Gaussian distributions, it admits a closed-form expression.
However, for Gaussian measures with unrestricted covariance matrices, the squared 2-Wasserstein distance for the covariance component is not, in general, conditionally negative definite. Consequently, exponentiating the negative squared distance does not necessarily produce a positive-definite kernel, as required to define a valid GP prior.

To ensure computational scalability, we approximate each posterior covariance matrix by retaining only its diagonal entries. This corresponds to preserving marginal variances at each time point and embedding dimension while ignoring correlations across time and dimensions. Specifically, let $q_i = N({m}_i, \text{diag}({\sigma}_i^2))$.
For diagonal Gaussian measures, the squared 2-Wasserstein distance reduces to
\begin{equation*} \label{eq:W2-diag}
    W_2^2(q_i,q_j) = \|{m}_i - {m}_j\|^2 +  \|{\sigma}_{i} - {\sigma}_{j}\|^2.
\end{equation*}
The resulting simplification substantially reduces the computational cost of evaluating the Wasserstein distance. In addition to reducing computational cost, this approximation guarantees the validity of the proposed trajectory kernel. This approximation discards the dependence structure in the uncertainty and does not capture how uncertainty evolves jointly over time or across embedding dimensions. However, it retains the overall magnitude and location of uncertainty in each trajectory. In practice, these marginal variances capture the dominant signal needed to distinguish well-estimated from poorly estimated trajectories, and are sufficient to preserve clinically meaningful differences while enabling efficient computation.

We then embed participants into a low-dimensional space by applying GPSSL  at the trajectory level. Treating trajectory distributions as inputs, we learn a mapping $f_z$ from $\mathcal{G}$ to $\mathbb{R}^d$ through evaluating the generalized posterior analog to Stage 1, 
$\tilde{p}(f_z) \propto p(f_z)\exp\{-\ell(f_z, \{q_{i}\}_{i})\},$
where the GP prior $p(f_z)$ is defined using the proposed trajectory kernel and the loss $\ell$ is given by Equation \eqref{eq:GPSSL_loss}. The posterior is approximated via sparse variational inference analog to stage 1. The resulting subject-level embedding for participant $i$ is defined as the posterior mean  $z_i = \mathbb{E}_{\tilde{p}} [f_z(q_i)] \in \mathbb{R}^d$, which provides a compact summary of longitudinal polypharmacy dynamics and serves as the basis for downstream analyses.  In summary, the full procedure consists of per-participant GP fitting, posterior evaluation on a common grid, construction of the trajectory kernel, and GPSSL-based embedding learning. Further technical details are provided in Supplement Section C.2. 

The subject-level embeddings $z_i$ learned by LOPEL provide compact and interpretable summaries of each participant’s longitudinal polypharmacy trajectory. These embeddings are task-agnostic and can be used as inputs to a wide range of downstream analyses.
In this paper, we focus on unsupervised clustering as an illustrative example and apply Gaussian mixture models to $\{z_i\}$ to identify subgroups of participants with distinct longitudinal medication patterns. More broadly, the same embeddings can be used for supervised prediction of clinical outcomes, forecasting of future medication use, or as low-dimensional covariates in regression models relating polypharmacy trajectories to health outcomes.


\subsection{Selection of embedding dimensions and loss weights}
\label{sec:selectdim}
The embedding dimensions $D$ and $d$ are user-specified hyperparameters that control the complexity of the visit-level and subject-level representations, respectively. Larger values allow for more expressive representations but increase model complexity. In practice, these dimensions can be selected based on downstream task performance, model selection criteria, or stability considerations.
Each stage additionally includes loss weights that determine the relative contributions of its component objectives. We therefore treat both the embedding dimensions and the loss weights as hyperparameters to be selected.

In our experiments, we determine the embedding dimensions by evaluating how well the learned representations preserve the underlying similarity structure. For each candidate dimension, we compute the Gram matrix of the embeddings and compare it to the corresponding similarity kernel, defined by the ATC-informed kernel in Stage 1 and the trajectory kernel in Stage 2, using centered kernel alignment \citep{kornblith2019similarity}. 
This criterion quantifies how well the inner-product structure of the embeddings matches that of the target kernel, ensuring that the learned representations preserve the similarity geometry encoded by LOPEL. We then select the smallest dimension that achieves maximal alignment, yielding a parsimonious representation capturing the dominant structure in the data.

For each stage, we first evaluate candidate embedding dimensions using an initial loss-weight specification. We then retain the two dimensions with the highest centered kernel alignment and conduct a refined search over these dimensions and the candidate loss-weight combinations.
This procedure is applied separately in Stage 1 and Stage 2 using the corresponding stage-specific loss weights and target kernel. The embedding dimension and loss-weight combination achieving the highest centered kernel alignment is selected for each stage.

\section{Simulation study} \label{sec:simulation}

In this section, we conduct simulation studies to evaluate the performance of LOPEL for longitudinal polypharmacy data and to compare it with several benchmark methods, using clustering as the downstream task.
The aim of the simulation study is to validate that for realistic sample sizes and data structure LOPEL can indeed meaningfully encode clinically
relevant structure. 
Using different comparisons we also explore whether the stage 1 part of LOPEL provides additional value beyond available methods based on time warping, and similarly whether stage 2 of LOPEL improves upon a longitudinal clustering (for the specific example of the clustering as desired downstream analysis).


\subsection{Simulation setup}

We generate synthetic data designed to mimic longitudinal polypharmacy data in our motivating study. In each setting, we simulate 300 participants, each assigned to one of $K = 3$ latent clusters with equal probability, where clusters represent distinct condition-driven medication trajectories over age. For each participant $i$, the number of visits is sampled from $[3, 20]$ and visit ages lie in $[50, 90]$.

Medication use is generated through a latent condition-level structure.  
We consider four underlying clinical conditions: alcohol dependency, depression, diabetes, and hypertension, each representing a distinct dimension of disease burden. Each medication is assigned to a single condition, defining a mapping from medications to conditions, while each condition may be associated with multiple medications.

For each cluster $k$ and condition $c$, we define an age-dependent prevalence function $p_{kc}(t)$, representing the probability that condition $c$ is present at age $t$. These functions are constructed on the logit scale to represent increasing, decreasing, or non-monotonic trajectories over age. Specifically,
\[
\mathrm{logit}\big(p_{kc}(t)\big) = \ell_{\min} + (\ell_{\max} - \ell_{\min}) \cdot g_{kc}(t),
\]
where $\ell_{\min} = \mathrm{logit}(p_{\min})$ and $\ell_{\max} = \mathrm{logit}(p_{\max})$ determine the lower and upper bounds of the trajectory, and $g_{kc}(t)$ is a smooth function controlling its shape. We consider three types of trajectories: increasing curves defined by $g_{kc}(t) = \sigma(\lambda (t - \tau))$, decreasing curves defined by $g_{kc}(t) = 1 - \sigma(\lambda (t - \tau))$, and non-monotonic (bump-shaped) curves defined by $g_{kc}(t) = \min\{\sigma(\lambda (t - \tau_s)), 1 - \sigma(\lambda (t - \tau_e))\}$, where $\sigma(x) = (1 + e^{-x})^{-1}$, $\lambda > 0$ controls the steepness of transitions, and $\tau$ (or $\tau_s, \tau_e$) determines the location of change points.

\begin{figure}[htbp!]
    \centering
    \includegraphics[width=0.99\linewidth]{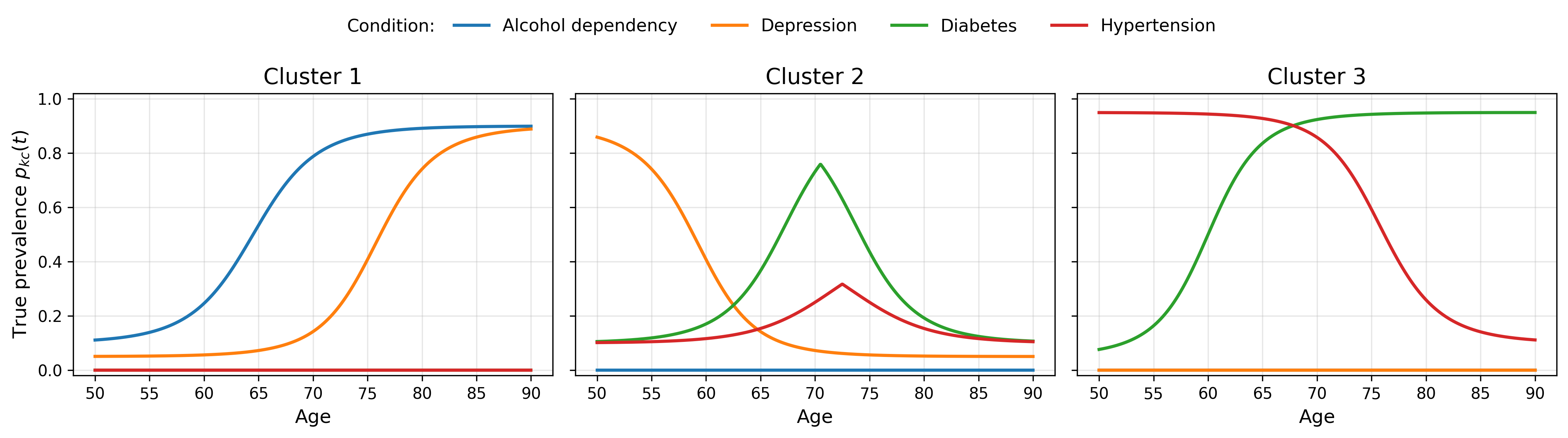}
    \caption{True cluster-specific condition prevalence curves $p_{kc}(t)$ over age $t$.}
    \label{fig:sim_condition_p}
\end{figure}

The ground truth cluster structure is defined by how trajectory types are assigned across conditions and clusters. Cluster 1 is characterized by increasing prevalence of alcohol dependency and depression over age. Cluster 2 exhibits decreasing depression prevalence together with bump-shaped trajectories for diabetes and hypertension. Cluster 3 is characterized by increasing diabetes and decreasing hypertension prevalence. Conditions not explicitly assigned within a cluster are assumed to have zero prevalence throughout. Figure~\ref{fig:sim_condition_p} illustrates the resulting cluster-specific condition prevalence curves, and all parameter values are provided in Supplementary Table S1. 

Medication-level probabilities are derived from the condition-level trajectories. For participant $i$ in cluster $k$, consider a medication $m$ associated with condition $c$. We define 
$
p_{im}(t) = \sigma\!\left\{\sigma^{-1}\!\big(p_{kc}(t)\big) + \delta_i \right\}, 
\quad \delta_i \sim \mathcal{N}(0, \sigma_\delta^2),
$
where $\sigma(\cdot)$ denotes the logistic function and $\delta_i$ is a subject-specific random effect that introduces individual-level heterogeneity while preserving the temporal pattern of the underlying condition.
Binary medication usage indicators are then generated as
$e_{ijm} \sim \mathrm{Bernoulli}(p_{im}(t_{ij}))$, for each participant $i$, visit $j$, and medication $m$.
Medications not selected as active for a given participant are assigned zero usage across all visits.

For each condition, a pool of candidate medications is constructed from the ATC-to-condition mapping in the motivating study, with sizes 10, 3, 10, and 10 for alcohol dependency, depression, diabetes, and hypertension, respectively. 

We evaluate performance across four simulation scenarios. Scenarios 1–3 share the condition-level prevalence structure described above and illustrated in Figure~\ref{fig:sim_condition_p}, but vary in medication multiplicity per condition and visit-time sampling scheme. \textbf{Scenario 1} assigns one active medication per condition per participant and samples visit times uniformly over $[50,90]$, providing a clean baseline. \textbf{Scenario 2} allows two concurrent medications per condition, while retaining uniform visit times. 
It introduces medication multiplicity and within-condition heterogeneity, although the presence of multiple condition-related medications may also provide redundant information about the underlying condition trajectory.
\textbf{Scenario 3} further introduces irregular, cluster-dependent visit times alongside two medications per condition:
$t_{ij} = 50 + 40 B_{ij}, \; B_{ij} \sim \mathrm{Beta}(\alpha_k, \beta_k),$ where Cluster 1 has visit times skewed toward younger ages $(\alpha_1 = 2, \beta_1 = 5)$, Cluster 2 has approximately symmetric visit ages $(\alpha_2 = 5, \beta_2 = 5)$, and Cluster 3 has visit times skewed toward older ages $(\alpha_3 = 5, \beta_3 = 2)$. Together, Scenarios 1–3 assess performance as observational complexity increases while the underlying clusters remain well-separated in trajectory shape. 

\textbf{Scenario 4} targets a more challenging regime by modifying only Cluster 3, while Clusters 1 and 2 retain the same condition prevalence trajectories as in Scenarios 1–3. Specifically, Cluster 3 is modified to exhibit bump-shaped diabetes and hypertension trajectories qualitatively similar to those of Cluster 2, differing only in peak age and peak hypertension prevalence. This deliberately reduces the trajectory-based separation between Clusters 2 and 3. Cluster 3 is further distinguished by substantially sparser follow-up, with 5–10 visits per participant compared to 15–20 in Cluster 2. This setting evaluates whether LOPEL's Wasserstein-based kernel can recover cluster structure when separation must be inferred in part from differential observation density rather than trajectory shape alone. The corresponding prevalence curves and full parameter specifications are provided in Supplementary Figure S1 and Table S2. 

For comparison, we consider three alternative methods that span a range of modeling strategies, from purely distance-based approaches to hybrid pipelines that incorporate the LOPEL visit-level embedding framework. 

\begin{enumerate} 
    \item \textbf{KmeansDTW}: K-means clustering for time series data using Dynamic Time Warping (DTW) \citep{sakoe2003dynamic} as the similarity metrics, implemented via the \textit{tslearn} Python package. 
    This method evaluates whether flexible temporal alignment is sufficient to recover cluster structure in high-dimensional, sparse medication trajectories.

    \item \textbf{LOPEL1-KmeansDTW}: A hybrid approach that first learns visit-level embeddings using the LOPEL stage 1 framework described in Section \ref{sec:ATC-SSL}, and then applies K-means clustering with DTW distance to the resulting longitudinal embedding trajectories using the \textit{tslearn} Python package. 
    This comparator isolates the contribution of the second-stage trajectory representation learning component of LOPEL.
    
    \item \textbf{LOPEL1-clusterMLD}:  A second hybrid strategy that uses the same ATC-informed visit-level embeddings, followed by hierarchical clustering for multivariate longitudinal data (clusterMLD) via the \textit{clusterMLD} R package \citep{zhou2023clustermld}.  This method evaluates whether hierarchical trajectory clustering can recover cluster structure once pharmacologic similarity is incorporated at the visit level.
    
\end{enumerate}

A common set of tuning parameters is adopted across all scenarios to ensure comparability. For LOPEL, visit-level embeddings of dimension $D$ and loss weights $c_V, c_C$ are learned using the ATC-based kernel with level specific weights $\bm{w}^2 = (0.05, 0.30, 0.30, 0.15, 0.10)$, and a learning rate of 0.02 for 500 iterations. These ATC kernel weights are chosen to place greater emphasis on similarity at the therapeutic and pharmacological subgroup levels, reflecting the clinical relevance of the corresponding levels in the ATC hierarchy. Subject-level embeddings of dimension $d$ and loss weights $c_V, c_C$ are subsequently learned using $\gamma=10$ and and a learning rate schedule of 0.05 for the first 200 iterations and 0.01 for the remaining 300 iterations. Embedding dimensions are selected from $D, d \in \{1,2,3,4,5\}$ and loss weights are selected from $(c_V, c_C) \in \{ (100,10), (50,10), (10,10), (10,50), (10,100) \}$ using the centered kernel alignment criterion described in Section~\ref{sec:selectdim}; the resulting alignment curves and selected dimensions across scenarios are provided in Supplementary Figure S2, S3, and Table S3. 
For all methods, the true number of clusters is supplied  to the clustering algorithms, so that evaluation focuses on clustering quality rather than model selection.

\subsection{Simulation results}

\begin{figure}[t]
     \centering
    \includegraphics[width=0.99\linewidth]{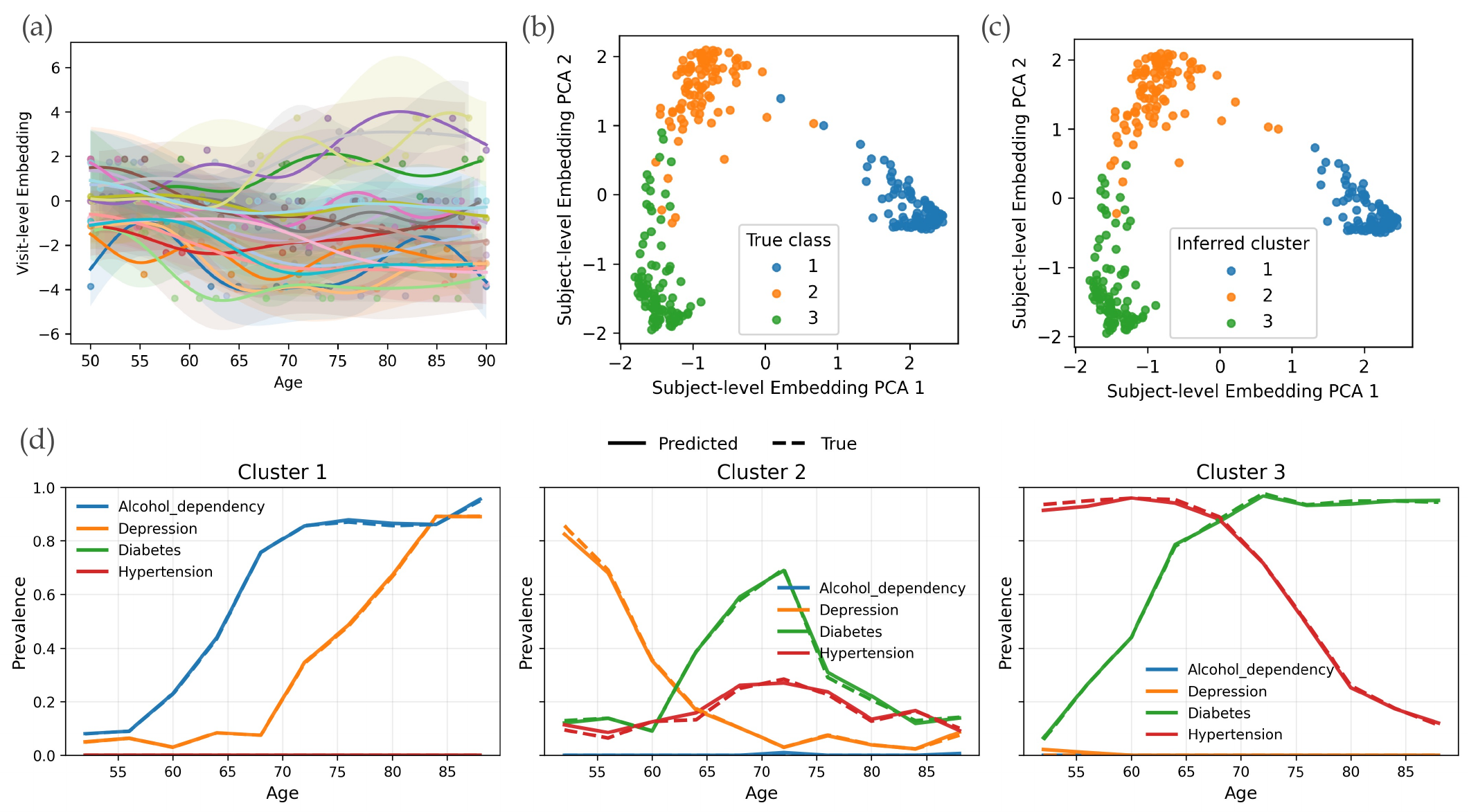}
        \caption{Simulation results under Scenario 1 using LOPEL.
(a) Gaussian process regression of the first embedding dimension for 20 participants.
(b–c) Subject-level embeddings projected onto the first two principal components, colored by true (b) and inferred (c) cluster labels.
(d) Condition-level medication prevalence within inferred clusters (solid) versus true clusters (dashed), aggregated in five-year age bins.
        }
        \label{fig:sim_results_sc1}
\end{figure}

We first illustrate LOPEL under Scenario 1. Figure \ref{fig:sim_results_sc1}(a) shows Gaussian process regression of the first dimension of the visit-level embeddings over time for 20 representative participants, corresponding to the trajectory modeling step in Stage 2. The resulting trajectories are smooth and exhibit clear age-dependent structure, indicating that LOPEL effectively captures underlying temporal patterns in high-dimensional medication regimens at the visit level.
Figure \ref{fig:sim_results_sc1}(b) and \ref{fig:sim_results_sc1}(c) display the corresponding subject-level embeddings projected onto their first two principal components, colored by true and inferred cluster labels, respectively. The two panels exhibit nearly identical separation, demonstrating that LOPEL preserves the latent cluster geometry and achieves accurate recovery of cluster membership in the learned embedding space.
Figure \ref{fig:sim_results_sc1}(d) further examines the interpretability of the inferred clusters by comparing condition-level medication prevalence within inferred clusters (solid lines) to that obtained using true cluster labels (dashed lines). Prevalence is computed as the proportion of participants within each five-year age bin using any medication for the corresponding condition. The close agreement across conditions indicates that individuals are correctly clustered with distinct longitudinal medication patterns, and that the recovered clusters correspond to the true underlying trajectory structure, demonstrating that LOPEL produces accurate and interpretable representations of longitudinal medication data.

To assess clustering performance quantitatively, we compute the Adjusted Rand Index (ARI), which measures agreement between inferred and true cluster labels. Table~\ref{tab:sim_results} reports ARI for LOPEL and all three competing methods across the four scenarios. LOPEL achieves the highest ARI in every scenario, with performance remaining consistently strong as data complexity increases. 

\begin{table}[htbp!]
    \centering
    \begin{tabular}{c|c|c|c|c}
    \hline
Scenario   & KmeansDTW & LOPEL1-KmeansDTW & LOPEL1-clusterMLD & LOPEL  \\ \hline
      1 &  0.03    &    0.92       &     0.57      &      \textbf{0.93}\\
      2 & 0.60    &     0.99       &     0.98       &     \textbf{1.00}  \\
      3 & 0.07    &     0.92      &     0.50       &     \textbf{0.94} \\
      4 & 0.54 & 0.59 & 0.60 & \textbf{0.63} \\
      \hline
    \end{tabular}
    \caption{Adjusted Rand Index (ARI) for LOPEL and competing methods across scenarios. }
    \label{tab:sim_results}
\end{table}

In Scenarios 1 and 2, where visit times are regularly sampled, LOPEL achieves near-perfect clustering (ARI = 0.93 and 1.00, respectively). The introduction of two concurrent medications per condition in Scenario 2 weakens the direct correspondence between observed medications and the underlying condition signal, yet LOPEL continues to recover the true clusters with perfect accuracy, and the inferred prevalence trajectories remain closely aligned with the ground truth (Supplementary Figure S4). This demonstrates the robustness of the ATC-informed visit-level representation to medication heterogeneity. 
In contrast, KmeansDTW applied directly to raw medication indicators performs poorly in Scenario 1 (ARI = 0.03), reflecting its inability to share information across pharmacologically related medications in a high-dimensional, sparse setting. Its partial recovery in Scenario 2 (ARI = 0.60) is likely attributable to the increased redundancy from concurrent medications, which incidentally stabilizes the raw indicator space. Replacing DTW with ATC-informed visit-level embeddings (LOPEL1--KmeansDTW) markedly improves performance in both scenarios, confirming that pharmacologic similarity encoding at the visit level is critical. LOPEL1--clusterMLD, despite using the same embeddings, performs less favorably across both scenarios, suggesting that parametric trajectory clustering lacks the flexibility to capture the nonlinear temporal patterns present in the data.

Scenario 3 additionally introduces irregular, cluster-dependent visit times, inducing temporal misalignment across participants. LOPEL maintains high ARI (0.94) and coherent trajectory recovery (Supplementary Figure S5). KmeansDTW and LOPEL1--clusterMLD show particularly low ARI (0.07 and 0.50, respectively), as neither is designed to handle irregular observation times. LOPEL1--KmeansDTW remains competitive but also shows some degradation (ARI = 0.92). Although DTW permits flexible sequence alignment, it operates on discretized visit indices rather than continuous time, and can spuriously align observations from different ages when visit schedules are irregular and cluster-dependent.
In contrast, LOPEL models trajectories as continuous stochastic processes and compares participants through distributional similarity in the Wasserstein sense, naturally accommodating both irregular timing and uncertainty in trajectory estimation. 

Scenario 4 presents the most challenging setting, where Cluster 3 exhibits trajectory shapes qualitatively similar to Cluster 2 and is further characterized by substantially sparser follow-up (5--10 visits versus 15--20 in Cluster 2), so that cluster separation must be inferred in part from differences in trajectory uncertainty rather than shape alone. All competing methods show reduced performance in this setting. KmeansDTW, LOPEL1--KmeansDTW, and LOPEL1--clusterMLD achieve ARIs of 0.54, 0.59, and 0.60, respectively, whereas LOPEL attains the highest ARI of 0.63 (Supplementary Figure S6).
The result suggests that representing each participant's trajectory as a posterior distribution and incorporating uncertainty through the Wasserstein-based kernel provides additional information beyond point-estimated trajectory distances. This advantage is particularly relevant when follow-up intensity varies across participants, as commonly occurs in real-world longitudinal studies.

Overall, LOPEL's advantage over competing methods becomes more pronounced as data complexity increases. By integrating ATC-informed visit-level representations with continuous, uncertainty-aware trajectory modeling, LOPEL is robust to medication multiplicity, irregular observation schedules, and heterogeneous follow-up.

\section{Application to Polypharmacy in Aging PWH} \label{sec:realdata}

We apply LOPEL to characterize heterogeneity in longitudinal polypharmacy among aging PWH, with the goal of identifying clinically meaningful subgroups that differ in medication burden, comorbidity structure, and age-dependent medication trajectories. Understanding such heterogeneity is critical for risk stratification and targeted clinical management in aging HIV populations, where multimorbidity and polypharmacy are increasingly prevalent. 

We analyze the polypharmacy dataset described in Section~\ref{sec:motivation}, restricting to participants whose first study visit occurred after 2010 and who contributed at least three visits, so that the estimated trajectories reflect contemporary prescribing practices and support reliable longitudinal modeling. The resulting analytic dataset includes 427 PWH contributing 2,224 visits across 694 distinct medications, and 155 PWOH contributing 773 visits. 
The analysis focuses on PWH, since the primary scientific objective is to identify subgroups with distinct longitudinal polypharmacy profiles within this population, and PWOH are retained as a descriptive reference group.
The most prevalent chronic condition categories and their prevalence among PWH and PWOH, along with the number of medications mapped to each condition, are summarized in Supplementary Table S4.

\begin{figure}[htbp!]
    \centering
    \includegraphics[width=0.95\linewidth]{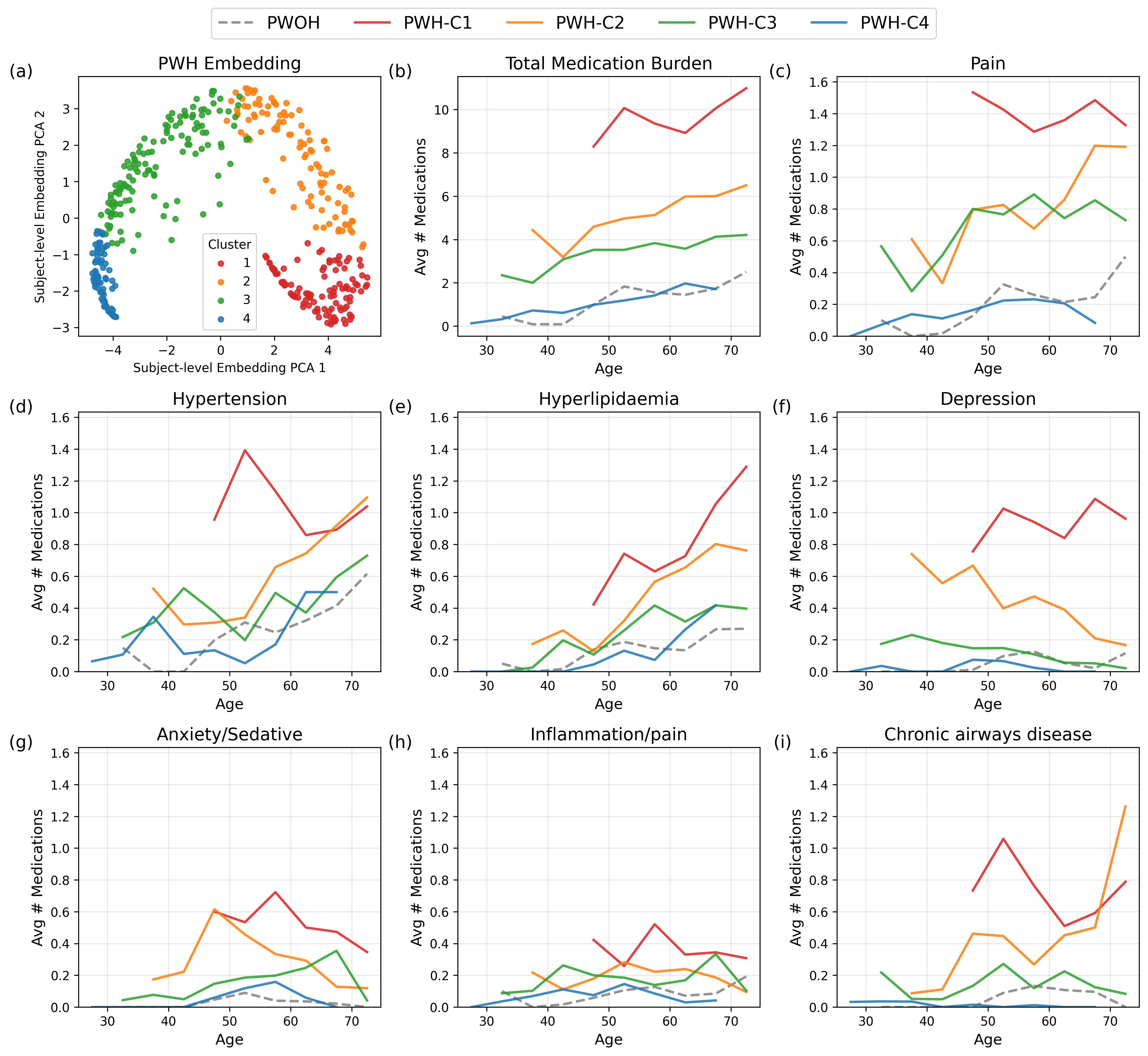}
        \caption{Longitudinal medication patterns for PWH subgroups identified by LOPEL.
(a) Subject-level embeddings projected onto the first two principal components, showing four identified PWH clusters.
(b) Mean total medication burden over age, averaged within five-year age bins.
(c–i) Condition-specific medication use over age for the seven most prevalent chronic conditions.
Solid lines represent the four PWH subgroups, and the gray dashed line represents PWOH.}
    \label{fig:realdata_med_condition}
\end{figure}

LOPEL is implemented as described in Section~\ref{sec:method}. 
Both stages are trained for 800 iterations, with all other hyperparameters set to the values used in the simulation study. The loss weights are fixed at $(50, 10)$ based on the simulation results, which indicate that training performance is relatively insensitive to the choice of loss weights when the embedding dimensions are appropriately selected.
The embedding dimensions $D$ and $d$ are selected from $\{1,2,3, 4, 5\}$ using the centered kernel alignment criterion (Supplementary Figure S7), yielding $D = 2$ and $d = 3$. 
Each participant is thus represented by a three-dimensional embedding summarizing their longitudinal medication trajectory. 
Clustering these embeddings using a Gaussian mixture model, with the number of clusters selected by the silhouette criterion \citep{rousseeuw1987silhouettes}, identifies four subgroups among PWH (PWH-C1 through PWH-C4), with sample sizes 106, 102, 130, and 89, respectively.  
Figure~\ref{fig:realdata_med_condition}(a) shows the first two principal components of the subject-level embeddings, where the four clusters are well-separated in the learned representation space.

Supplementary Table S5 summarizes baseline demographic, clinical, and HIV-related characteristics across the four PWH clusters and PWOH. PWH-C1 has the oldest median age (57 years) and the highest prevalence of several comorbid conditions, including hyperlipidemia (69\%) and hypertension (68\%). In contrast, PWH-C4 is the youngest subgroup (median age 49 years) with substantially lower comorbidity prevalence, reflecting a lower-risk population at an earlier stage of disease accumulation. The degree of polypharmacy further reinforces this stratification: 72\% of PWH-C1 fall into the high polypharmacy category ($\geq$10 medications), whereas 72\% of PWH-C4 are in the low polypharmacy category (0--3 medications). These patterns demonstrate that the learned clusters correspond to clinically meaningful differences in disease burden and treatment complexity.

We next examine longitudinal medication burden over age (Figure~\ref{fig:realdata_med_condition}(b)). 
For visualization, medication use is averaged within five-year age bins, and summaries are shown only for bins containing at least 20 participants to ensure stability. Across nearly all age groups, PWH exhibit higher overall burden than PWOH, consistent with elevated comorbidity and treatment complexity in this population.  
More importantly, the four PWH clusters show different burden trajectories over age. 
PWH-C1 exhibits the highest burden, with rapid accumulation after midlife reaching over 10 medications in later years. PWH-C2 and PWH-C3 show intermediate burden but differ in trajectory shape, with PWH-C2 increasing more sharply beginning around age 45 and PWH-C3 exhibiting a more gradual accumulation. In contrast, PWH-C4 maintains a consistently low and relatively stable burden over time. These findings indicate that heterogeneity in polypharmacy is not captured solely by overall medication count, but also by differences in the timing and progression of medication use.

We further examine age-specific medication use across the seven most prevalent chronic conditions (Figure~\ref{fig:realdata_med_condition}(c–i)), providing a detailed view of how comorbidity patterns differ across subgroups.
PWH-C1 is characterized by elevated medication use across multiple domains, including pain, hypertension, hyperlipidemia, and depression, consistent with a multisystem multimorbidity phenotype requiring complex medication management \citep{zhabokritsky2024non}. 
PWH-C2 exhibits elevated use of depression and anxiety/sedative medications at earlier ages followed by decline, alongside a marked increase in chronic airway disease at older ages, suggesting a transition from a mental health–dominant profile in midlife to a later-life respiratory burden \citep{do2014excess,crothers2006increased}. 
PWH-C3 shows more moderate but steadily increasing use of cardiometabolic medications, reflecting a more gradual trajectory driven by age-related cardiometabolic risk. In contrast, PWH-C4 maintains low levels across conditions, indicating a relatively stable and low-burden comorbidity profile.

These patterns indicate that LOPEL-derived clusters capture differences in the timing, composition, and progression of comorbidity, beyond overall medication burden. This distinction is further supported by differences in condition co-occurrence (Supplementary Figure S8), where PWH-C1 exhibits dense co-occurrence across multiple domains, PWH-C2 shows stronger clustering among mental health conditions, and PWH-C3 shows co-occurrence primarily within cardiometabolic domains. These results indicate that subgroup differences reflect not only individual conditions but also how conditions accumulate jointly over time.


Lastly, we compare LOPEL with KmeansDTW applied to (i) total medication counts and (ii) raw medication indicators. Clustering based on total medication counts (Supplementary Figure S9) primarily separates individuals by overall burden, yielding subgroups that differ in magnitude but show limited differentiation across clinical domains. Correspondingly, condition-specific trajectories differ mainly in overall levels and fail to capture domain-specific comorbidity patterns.
Applying KmeansDTW to raw medication indicators (Supplementary Figure S10) introduces additional variability but produces poorly interpretable clusters, lacking sufficient resolution to distinguish subgroups with distinct condition-specific medication patterns. 
This reflects the challenges of clustering high-dimensional, sparse medication data without incorporating pharmacologic structure.
In contrast, LOPEL produces subgroups that are both well-separated in representation space and clinically interpretable, demonstrating the importance of integrating pharmacologic similarity, temporal dynamics, and trajectory-level structure in modeling longitudinal polypharmacy.

Overall, LOPEL identifies four clinically distinct longitudinal polypharmacy trajectories among PWH that reflect different patterns of disease accumulation and management complexity. These findings have direct implications for clinical care. Individuals in PWH-C1 may benefit from coordinated, multidisciplinary care and proactive medication review to mitigate risks of adverse drug interactions and cumulative treatment burden \citep{greene2014polypharmacy}. PWH-C2 suggests the importance of early mental health intervention followed by monitoring for emerging respiratory disease \citep{do2014excess}. In contrast, PWH-C3 may benefit from targeted cardiometabolic risk management, while PWH-C4 highlights a group where maintaining stability and early detection of trajectory shifts may be important.

\section{Discussion} \label{sec:discussion}

We introduced LOPEL, a representation learning framework for longitudinal medication data that addresses high dimensionality, sparsity, irregular observation times, and structured pharmacologic relationships. By combining ATC-informed similarity at the visit level with distributional modeling of individual trajectories over time, LOPEL transforms complex medication records into low-dimensional subject-level representations that preserve clinically meaningful longitudinal patterns and support downstream statistical analysis.

A key contribution of this work is the explicit separation of representation learning into visit-level embedding and trajectory-level modeling. This structure allows LOPEL to borrow strength across pharmacologically related medications while capturing temporal dynamics at the subject level. Simulation studies demonstrate that this combination is critical for robust performance under realistic conditions. In the real data application, LOPEL reveals clinically meaningful heterogeneity in longitudinal polypharmacy that is not captured by traditional summary measures such as total medication count. The identified subgroups differ not only in overall burden, but also in the timing, composition, and progression of comorbid conditions over age. These findings highlight that differences in polypharmacy arise not only from the number of medications used, but also from how medication use accumulates and evolves over time.

There are several directions for future work. First, medication-to-condition mapping relies on the RxRisk classification system, which may misclassify medications used off-label or for indications outside their primary mapping. More flexible or data-driven approaches to condition assignment could improve the fidelity of the visit-level representation.  
Second, the current implementation models each participant's trajectory independently before comparing individuals at the distributional level. Extensions that jointly model trajectories across individuals through shared latent processes or hierarchical GP priors could enable borrowing of strength across participants with sparse observations. 
Third, although we use Gaussian process regression to model embedding trajectories in Stage 2, this choice can be adapted to the data and computational setting. For regularly sampled data, simpler approaches such as spline or polynomial regression may suffice, while for very large datasets, approximate Gaussian process methods or scalable sequence models (e.g., recurrent neural networks or transformers) may provide more efficient alternatives.
Fourth, the current implementation propagates only the Stage 1 posterior mean embeddings into Stage 2 and does not explicitly account for uncertainty in the visit-level representations. In principle, Stage 1 posterior variances, or other summaries of the embedding uncertainty, could be incorporated as additional inputs or propagated through a fully probabilistic hierarchical model. In our experiments, however, the posterior mean embeddings provided sufficiently stable representations for the downstream trajectory analysis, and we therefore adopted this simpler approach. 
Future work should evaluate when uncertainty propagation materially affects trajectory estimation or subject-level clustering, particularly for participants with sparse or highly irregular observations.
For settings in which computational speed is prioritized, a computationally cheaper visit-level embedding method could also be considered, such as kernel principal component analysis using the proposed ATC-informed kernel. Such an approach would not reproduce all components of the GPSSL objective, but could provide a useful approximation when scalability is the primary concern. We retain GPSSL in the present framework because it provides a probabilistic representation and maintains methodological consistency across the two stages.

Beyond clustering, the learned representations have broader applications. Because LOPEL produces task-agnostic subject-level embeddings, they can be used for downstream analyses such as predicting clinical outcomes, identifying high-risk trajectories, and evaluating treatment effects in longitudinal settings. More broadly, the framework extends beyond medication data to other domains with high-dimensional, sparse, and irregularly observed longitudinal measurements that exhibit structured relationships, including laboratory measurements, patient-reported symptom measures, and healthcare utilization records.

\bibliographystyle{unsrtnat}
\bibliography{bibfile}

\end{document}


\maketitle

\renewcommand{\thesection}{\Alph{section}}

\renewcommand{\thefigure}{S\arabic{figure}} 
\setcounter{figure}{0} 
\renewcommand{\thetable}{S\arabic{table}} 
\setcounter{table}{0} 

\section{Supplementary Tables and Figures}

\begin{table}[htbp!]
\begin{tabular}{cllcccc}
\toprule
Cluster & Condition & Shape & $\bar{\tau}$ & $\sigma_\tau$ & $p_{\min}$ & $p_{\max}$ \\
\midrule
1 & Alcohol dependency & $\uparrow$ & 64.5 & 1.5 & 0.10 & 0.90 \\
1 & Depression         & $\uparrow$ & 75.0 & 1.5 & 0.05 & 0.90 \\
\midrule
2 & Depression         & $\downarrow$ & 60.0 & 0.5 & 0.05 & 0.90 \\
2 & Diabetes           & bump  & $(68.0,\, 73.0)$ & 0.5 & 0.10 & 0.95 \\
2 & Hypertension       & bump  & $(70.0,\, 75.0)$ & 0.5 & 0.10 & 0.50 \\
\midrule
3 & Diabetes           & $\uparrow$ & 60.0 & 1.5 & 0.05 & 0.95 \\
3 & Hypertension       & $\downarrow$ & 75.0 & 1.5 & 0.10 & 0.95 \\
\bottomrule
\end{tabular}
\centering
\caption{Cluster-specific trajectory parameters for scenarios 1-3. Each row specifies an active (cluster, condition) pair. ``Shape'' denotes the type of age-dependent curve: increasing ($\uparrow$), decreasing ($\downarrow$), or non-monotonic (bump). $\bar{\tau}$ is the mean change-point. 
$\sigma_\tau$ is the standard deviation of the change-point. $p_{\min}$ and $p_{\max}$ are the lower and upper probability bounds. The logistic transition slope governing the steepness of all prevalence curves is $\lambda = 0.25$.}
\label{tab:sim_trajectory_params}
\end{table}

 \begin{table}[htbp!]
\begin{tabular}{cllcccc}
\toprule
Cluster & Condition & Shape & $\bar{\tau}$ & $\sigma_\tau$ & $p_{\min}$ & $p_{\max}$ \\
\midrule
1 & Alcohol dependency & $\uparrow$ & 64.5 & 1.5 & 0.10 & 0.90 \\
1 & Depression         & $\uparrow$ & 75.0 & 1.5 & 0.05 & 0.90 \\
\midrule
2 & Depression         & $\downarrow$ & 60.0 & 0.5 & 0.05 & 0.90 \\
2 & Diabetes           & bump  & $(60.0,\, 70.0)$ & 0.5 & 0.10 & 0.95 \\
2 & Hypertension       & bump  & $(60.0,\, 75.0)$ & 0.5 & 0.10 & 0.50 \\
\midrule
3 & Diabetes           & bump  & $(68.0,\, 78.0)$ & 0.5 & 0.10 & 0.95 \\
3 & Hypertension       & bump  & $(70.0,\, 80.0)$ & 0.5 & 0.10 & 0.90 \\
\bottomrule
\end{tabular}
\centering
\caption{Cluster-specific trajectory parameters for scenario 4. Each row specifies an active (cluster, condition) pair. ``Shape'' denotes the type of age-dependent curve: increasing ($\uparrow$), decreasing ($\downarrow$), or non-monotonic (bump). $\bar{\tau}$ is the mean change-point. 
$\sigma_\tau$ is the standard deviation of the change-point. $p_{\min}$ and $p_{\max}$ are the lower and upper probability bounds. The logistic transition slope governing the steepness of all prevalence curves is $\lambda = 0.25$.}
\label{tab:sim_trajectory_params_sc4}
\end{table}


\begin{figure}[htbp!]
    \centering
    \includegraphics[width=0.99\linewidth]{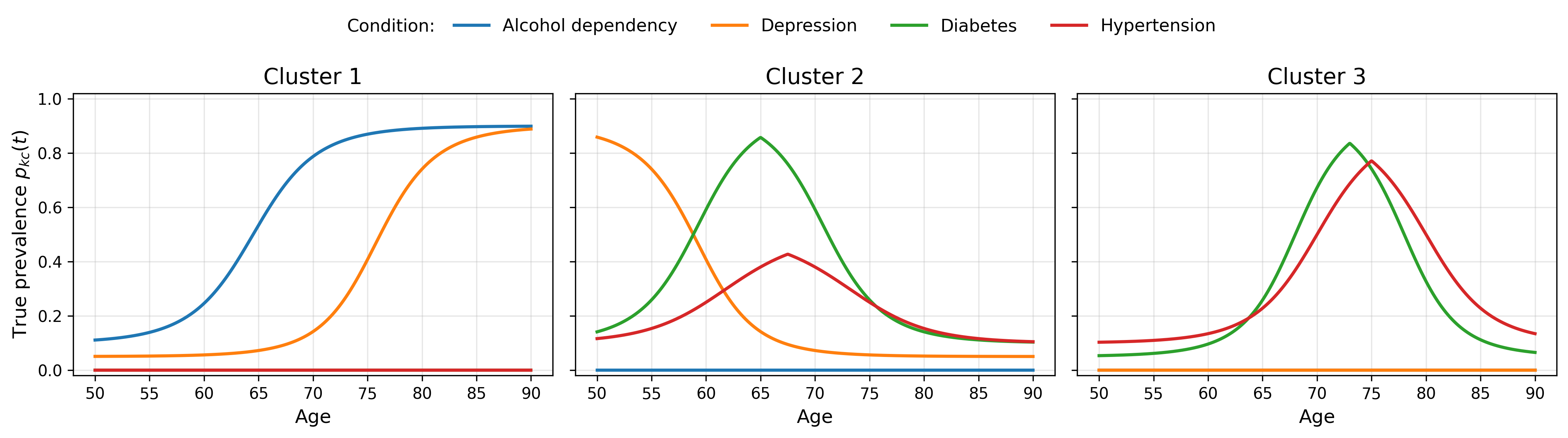}
    \caption{True cluster-specific condition prevalence curve $p_{kc}(t)$ over age $t$ under scenario 4.}
    \label{fig:sim_condition_p_sc4}
\end{figure}



\begin{table}[htbp!]
\centering
\begin{tabular}{lcccc}
\hline & \multicolumn{2}{c}{Stage 1} & \multicolumn{2}{c}{Stage 2} \\ \cline{2-3} \cline{4-5}
\hline
Scenario & $D$ & $(c_V, c_C)$ & $d$ & $(c_V, c_C)$\\
\hline
Scenario 1 & 2 & (50, 10) & 3 & (10, 10) \\
Scenario 2 & 1 & (10, 10) & 5 & (50, 10) \\
Scenario 3 & 1 & (10, 10) & 4 & (100, 10) \\
Scenario 4 & 2 & (100, 10) & 2 & (100, 10) \\
\hline
\end{tabular}
\caption{Selected embedding dimensions and loss weights for Stage 1 ($D, c_V, c_C$) and Stage 2 ($d, c_V, c_C$) across simulation scenarios, based on centered kernel alignment.}
\label{tab:dim_selection}
\end{table}

\begin{figure}[htbp!]
     \centering
     \begin{subfigure}[b]{0.33\textwidth}
         \centering
         \includegraphics[width=\textwidth]{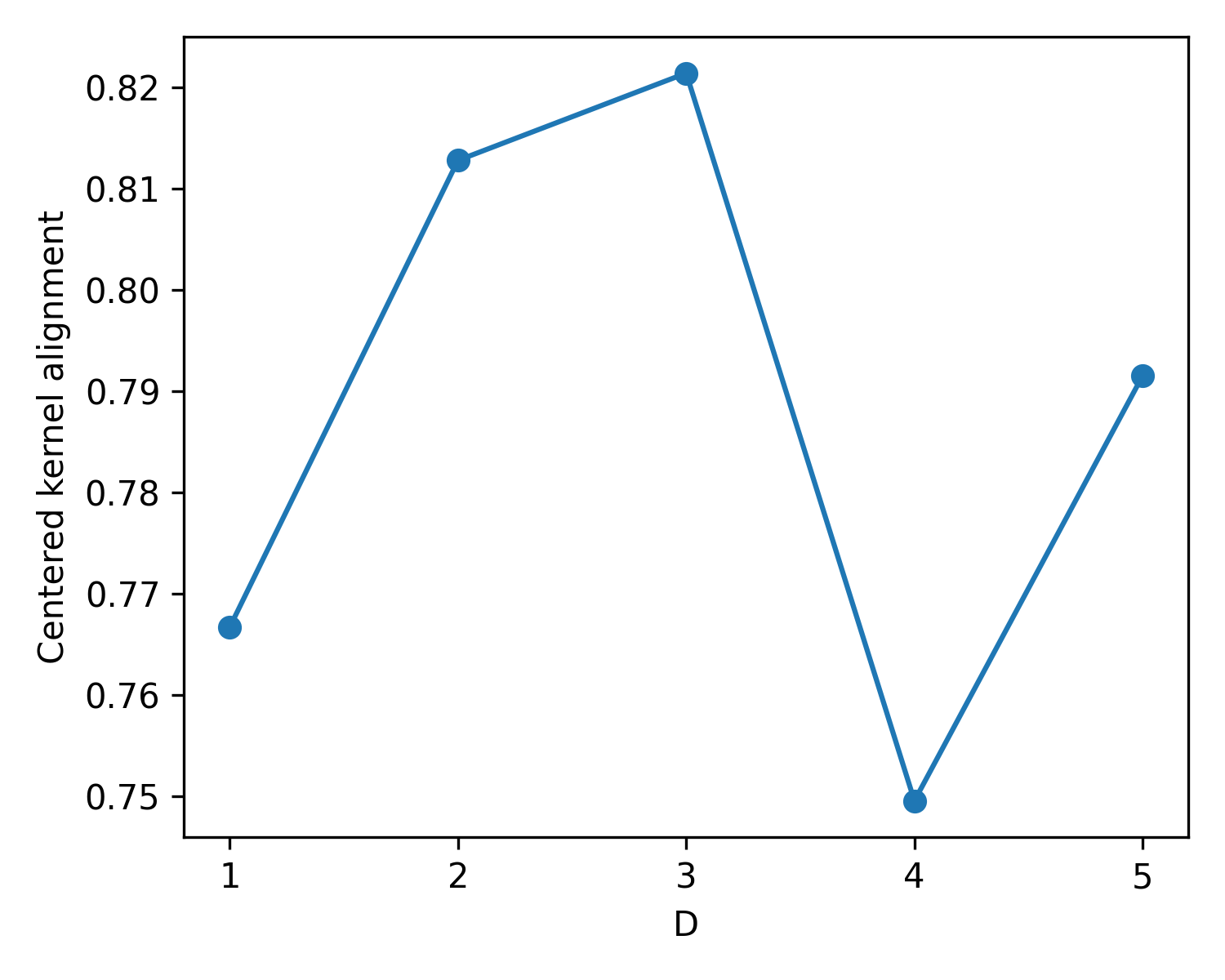}
     \end{subfigure}
     \begin{subfigure}[b]{0.52\textwidth}
         \centering
         \includegraphics[width=\textwidth]{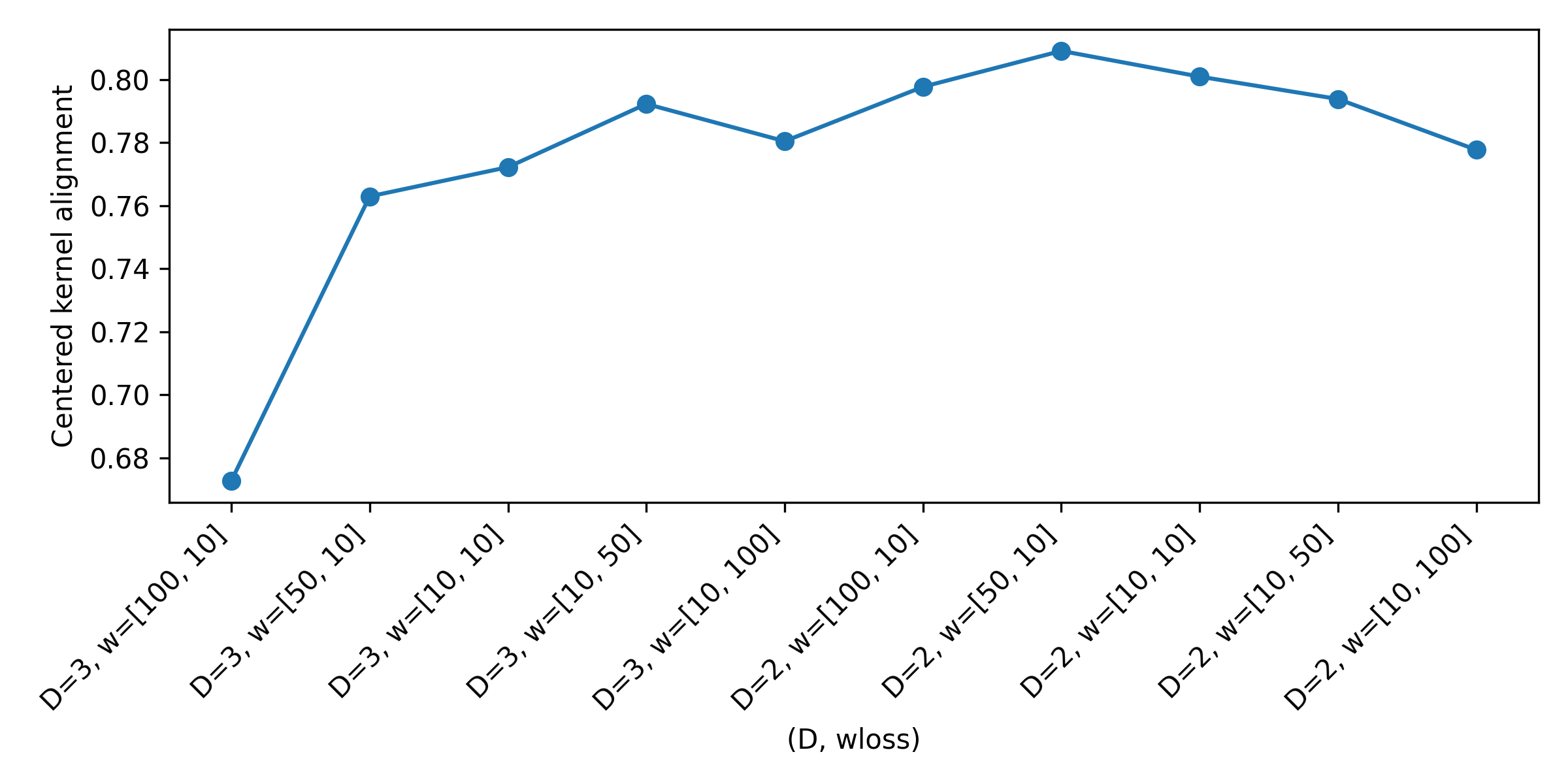}
     \end{subfigure} \\
     \begin{subfigure}[b]{0.33\textwidth}
         \centering
         \includegraphics[width=\textwidth]{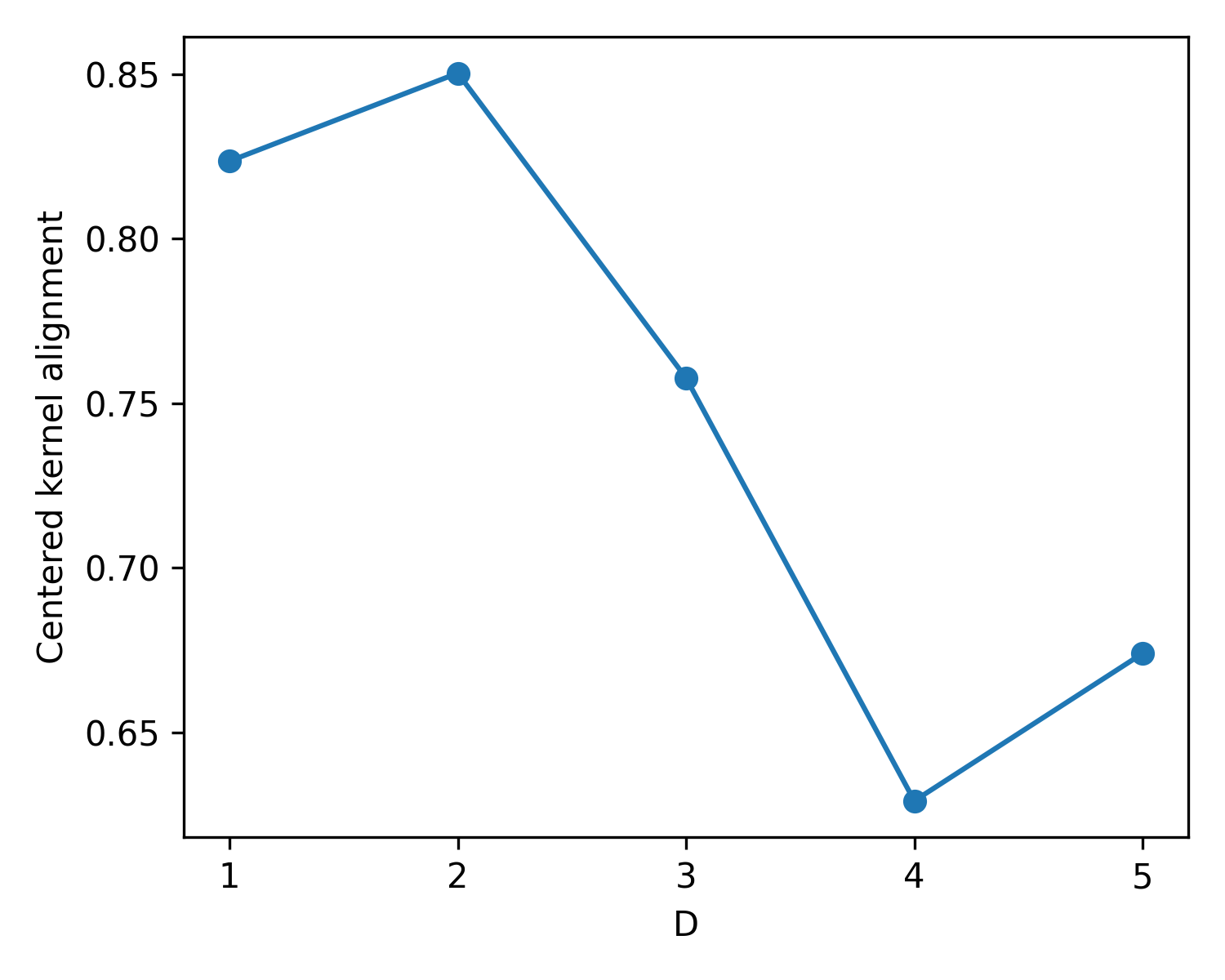}
     \end{subfigure}
     \begin{subfigure}[b]{0.52\textwidth}
         \centering
         \includegraphics[width=\textwidth]{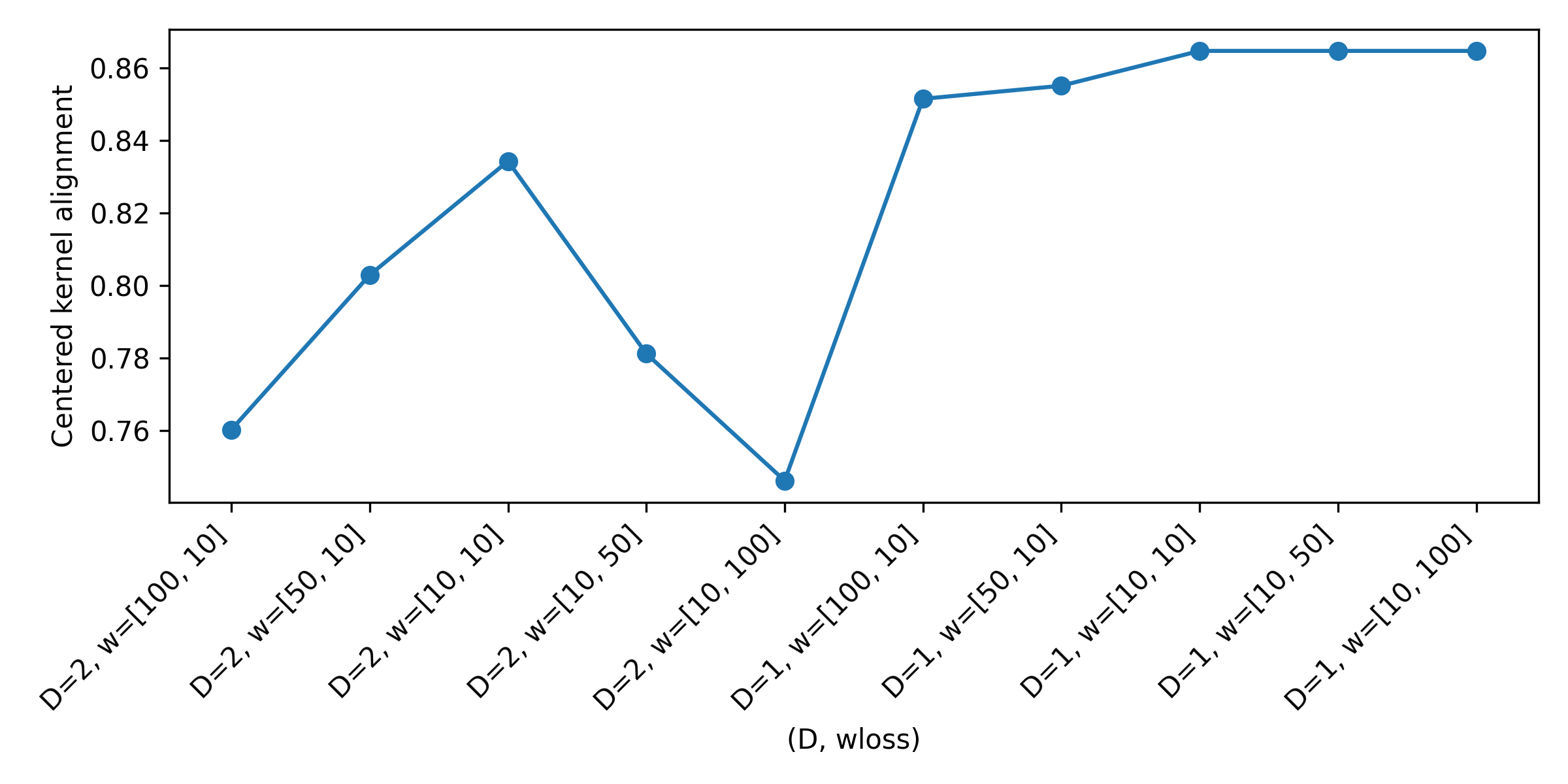}
     \end{subfigure} \\
     \begin{subfigure}[b]{0.33\textwidth}
         \centering
         \includegraphics[width=\textwidth]{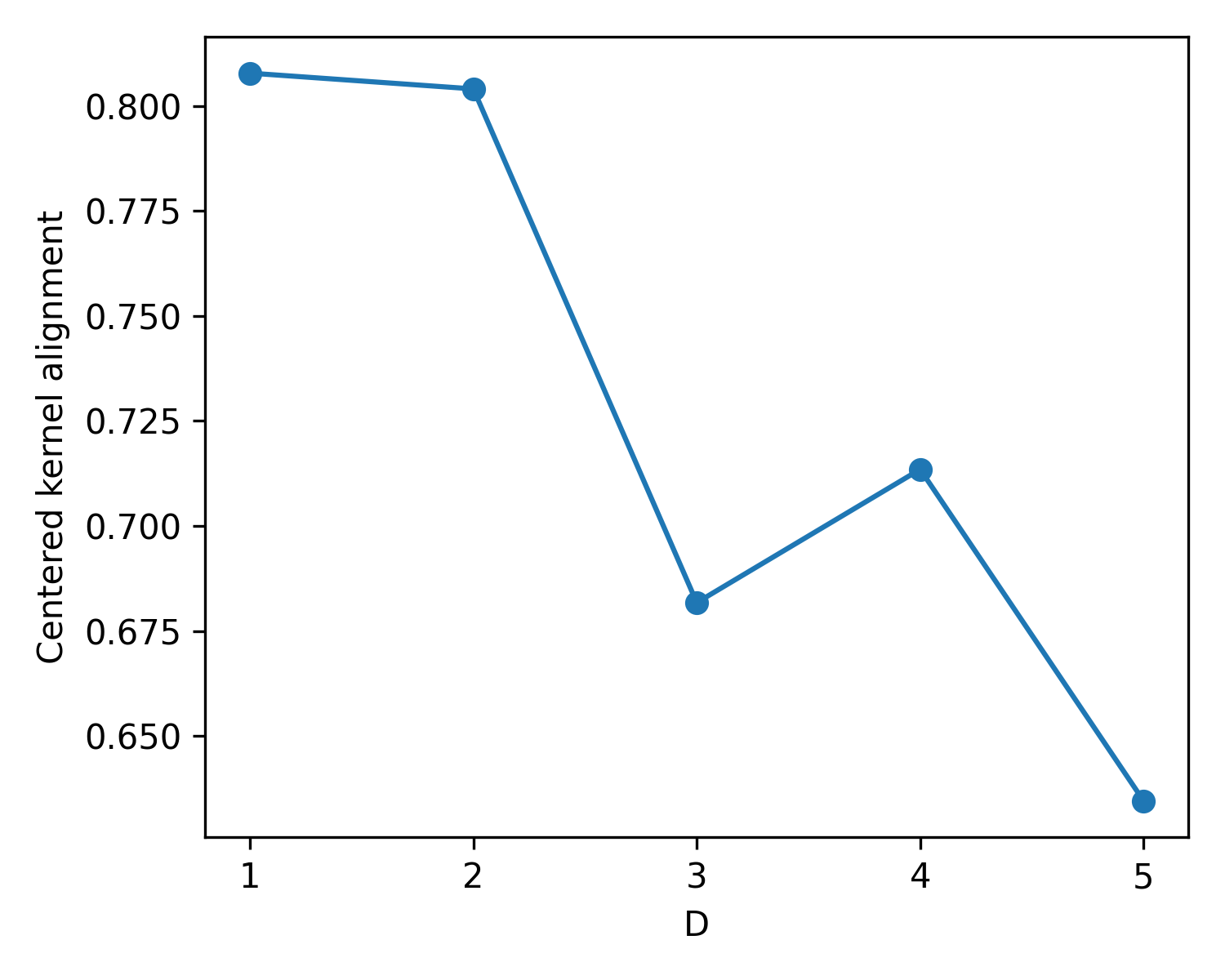}
     \end{subfigure}
     \begin{subfigure}[b]{0.52\textwidth}
         \centering
         \includegraphics[width=\textwidth]{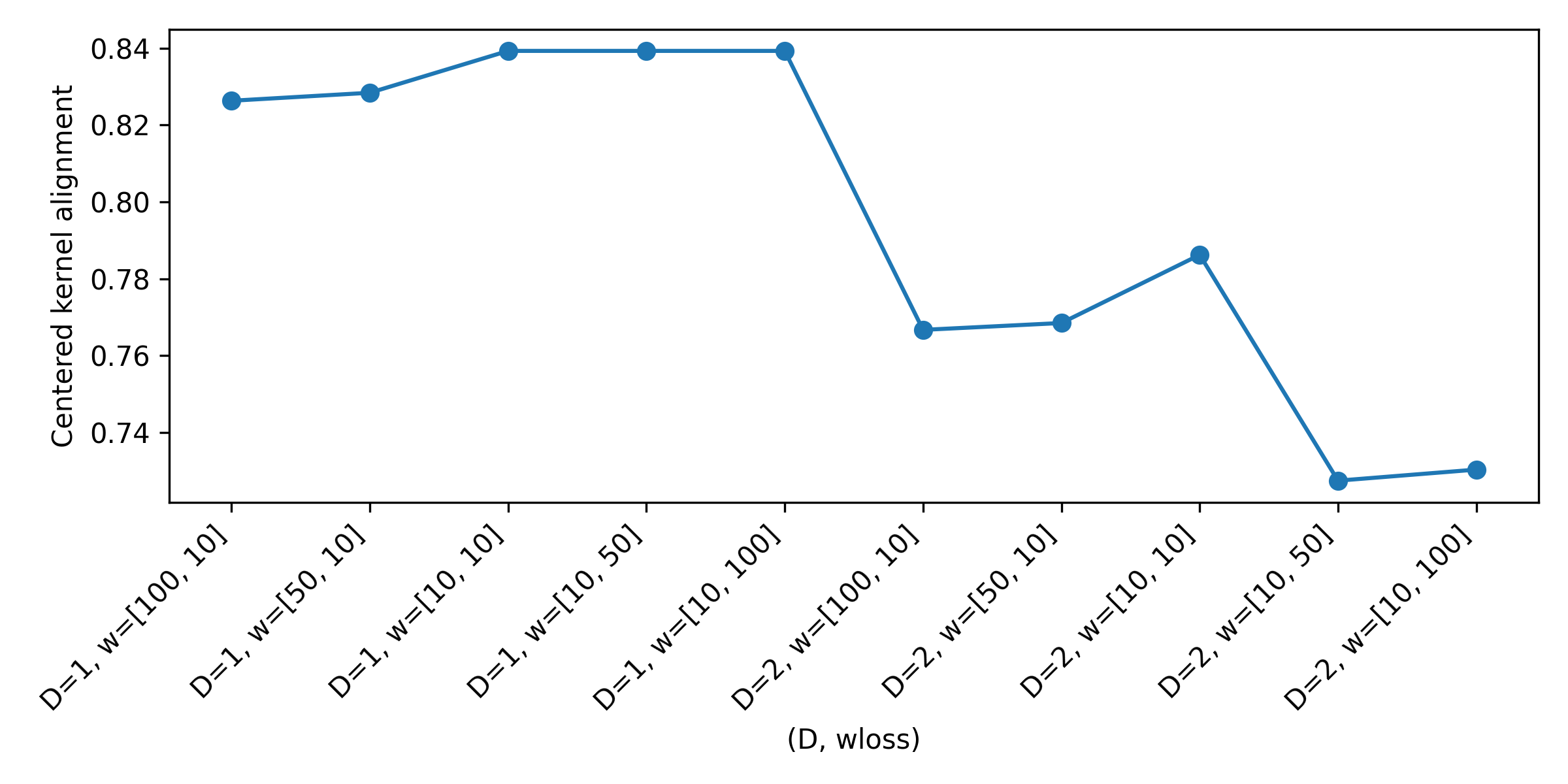}
     \end{subfigure} \\
     \begin{subfigure}[b]{0.33\textwidth}
         \centering
         \includegraphics[width=\textwidth]{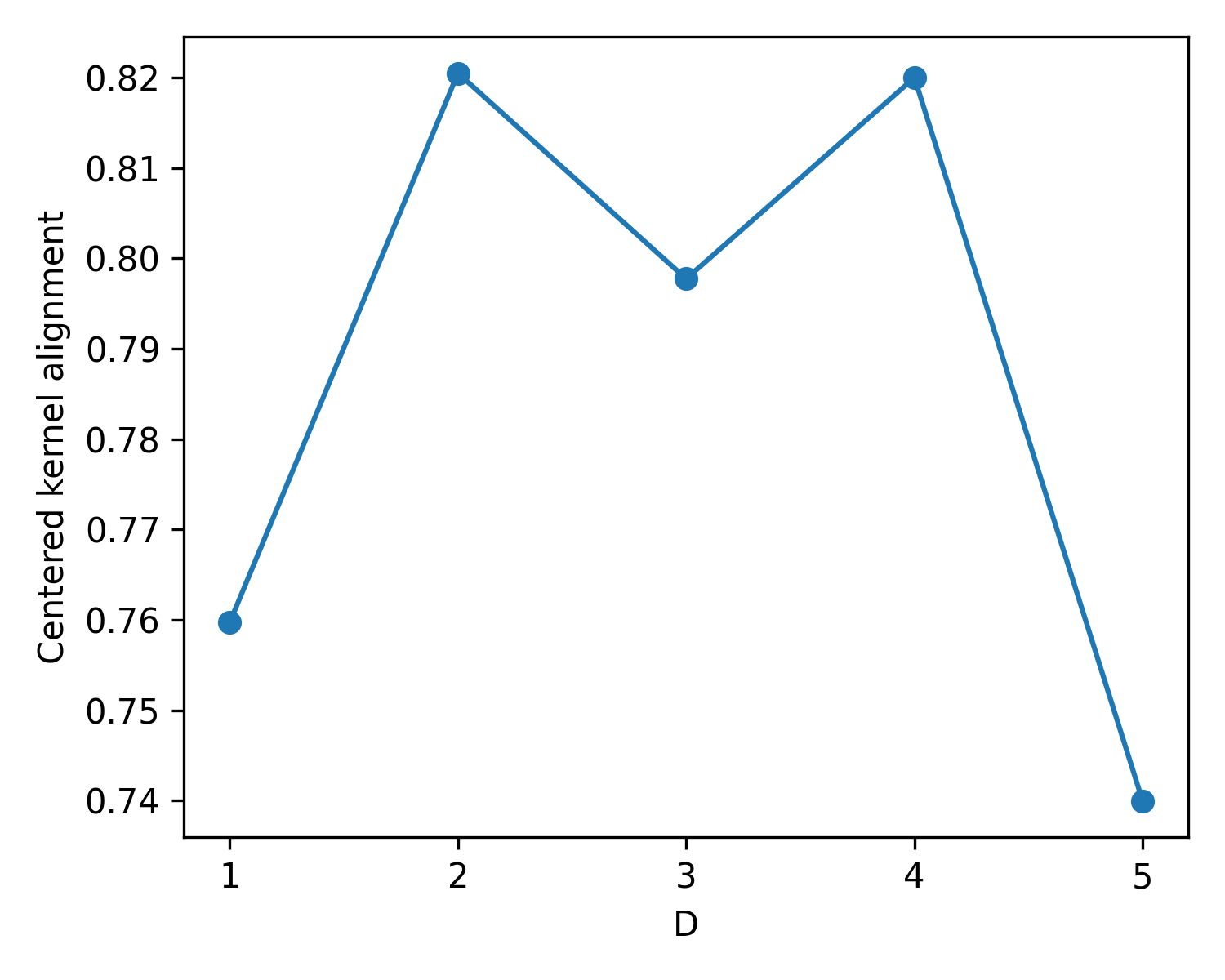}
     \end{subfigure}
     \begin{subfigure}[b]{0.52\textwidth}
         \centering
         \includegraphics[width=\textwidth]{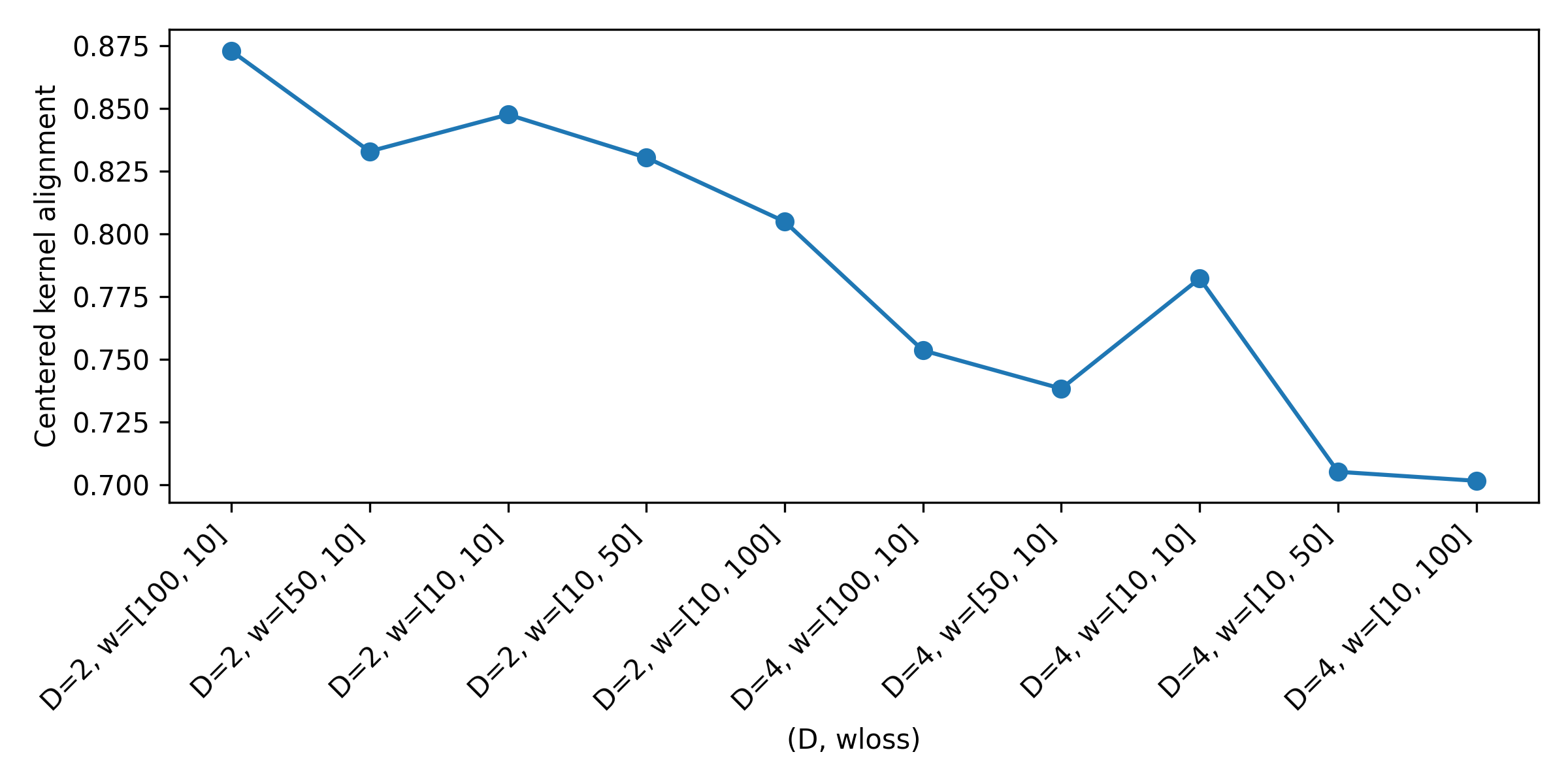}
     \end{subfigure}
        \caption{\small Centered kernel alignment used to select the Stage 1 embedding dimension $D$ and loss weights $w = (c_V, c_C)$ in simulation scenarios 1-4. Rows correspond to scenarios 1-4, respectively. For each scenario, the left panel shows the alignment obtained from an initial search over $D = 1, .\dots, 5$, using $w = (50, 10)$ and $100$ training epochs. The right panels shows the subsequent refined search over all candidate loss weight combinations for the two highest performing values of $D$ from the initial search. Centered kernel alignment quantifies how well the learned embeddings preserve the target similarity structure, with higher values indicating better agreement. The combination of $D$ and $w$ achieving the highest alignment was selected for each scenario.}
        \label{fig:sim_Dd}
\end{figure}

\begin{figure}[htbp!]
     \centering
     \begin{subfigure}[b]{0.33\textwidth}
         \centering
         \includegraphics[width=\textwidth]{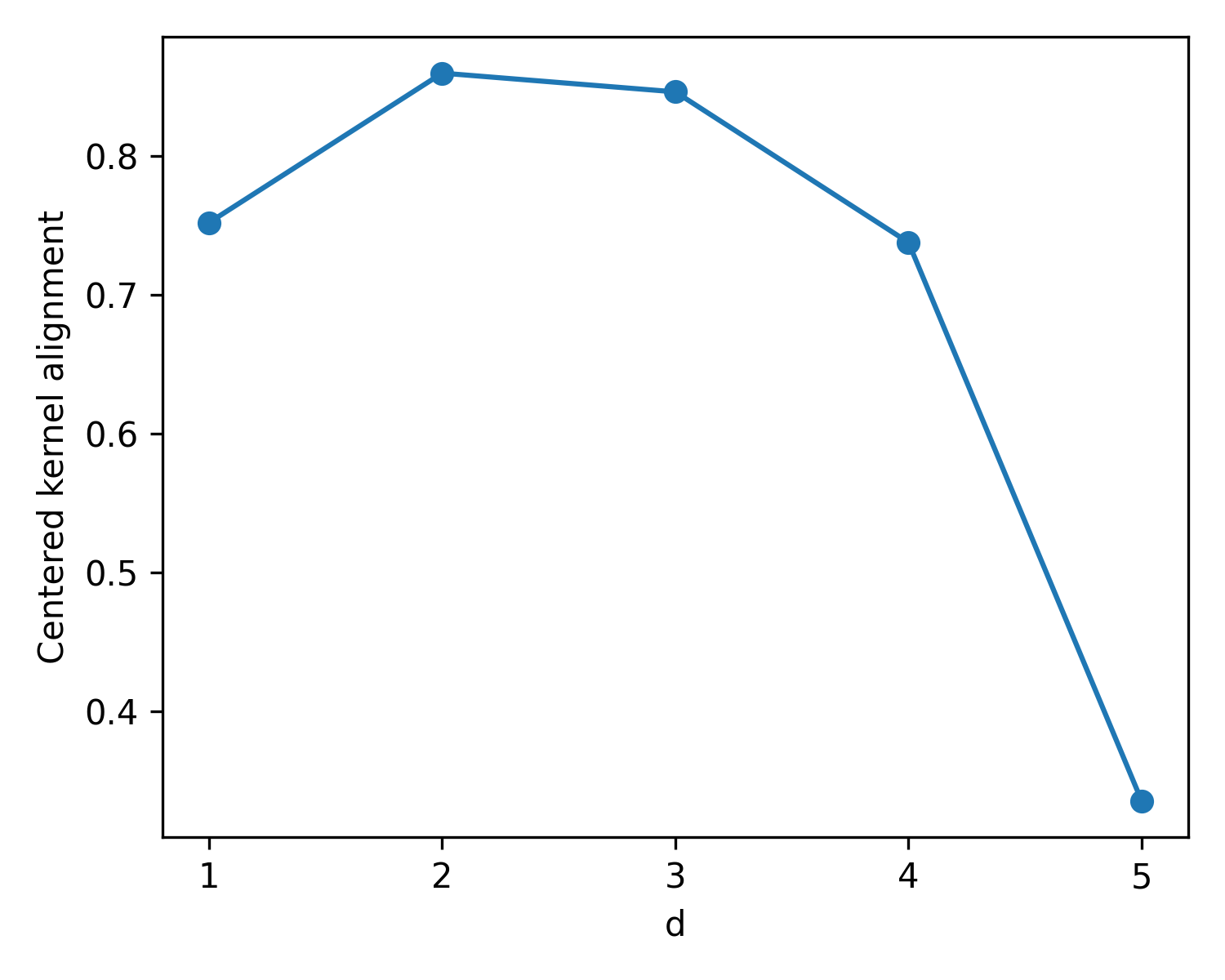}
     \end{subfigure}
     \begin{subfigure}[b]{0.47\textwidth}
         \centering
         \includegraphics[width=\textwidth]{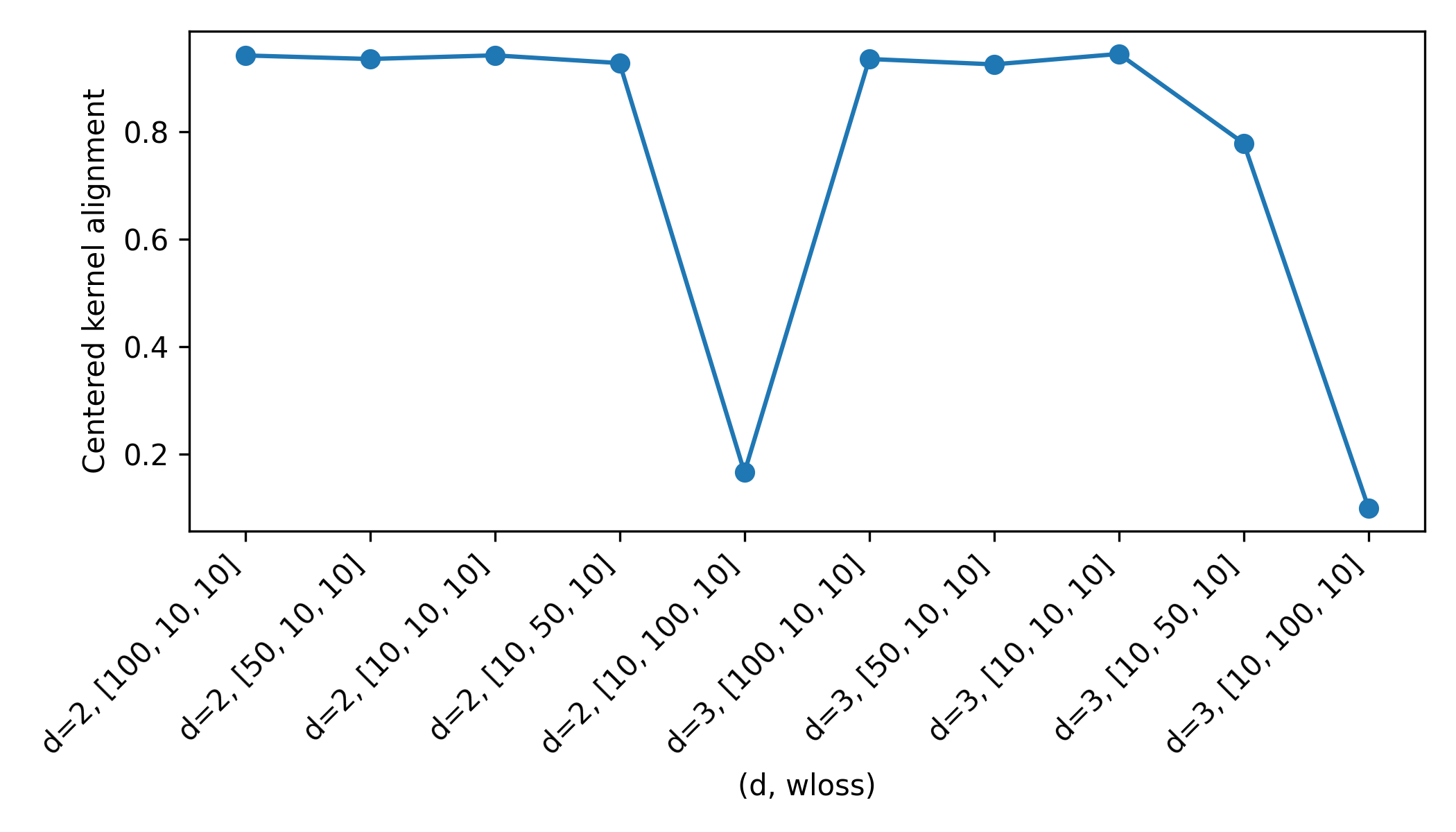}
     \end{subfigure} \\
     \begin{subfigure}[b]{0.33\textwidth}
         \centering
         \includegraphics[width=\textwidth]{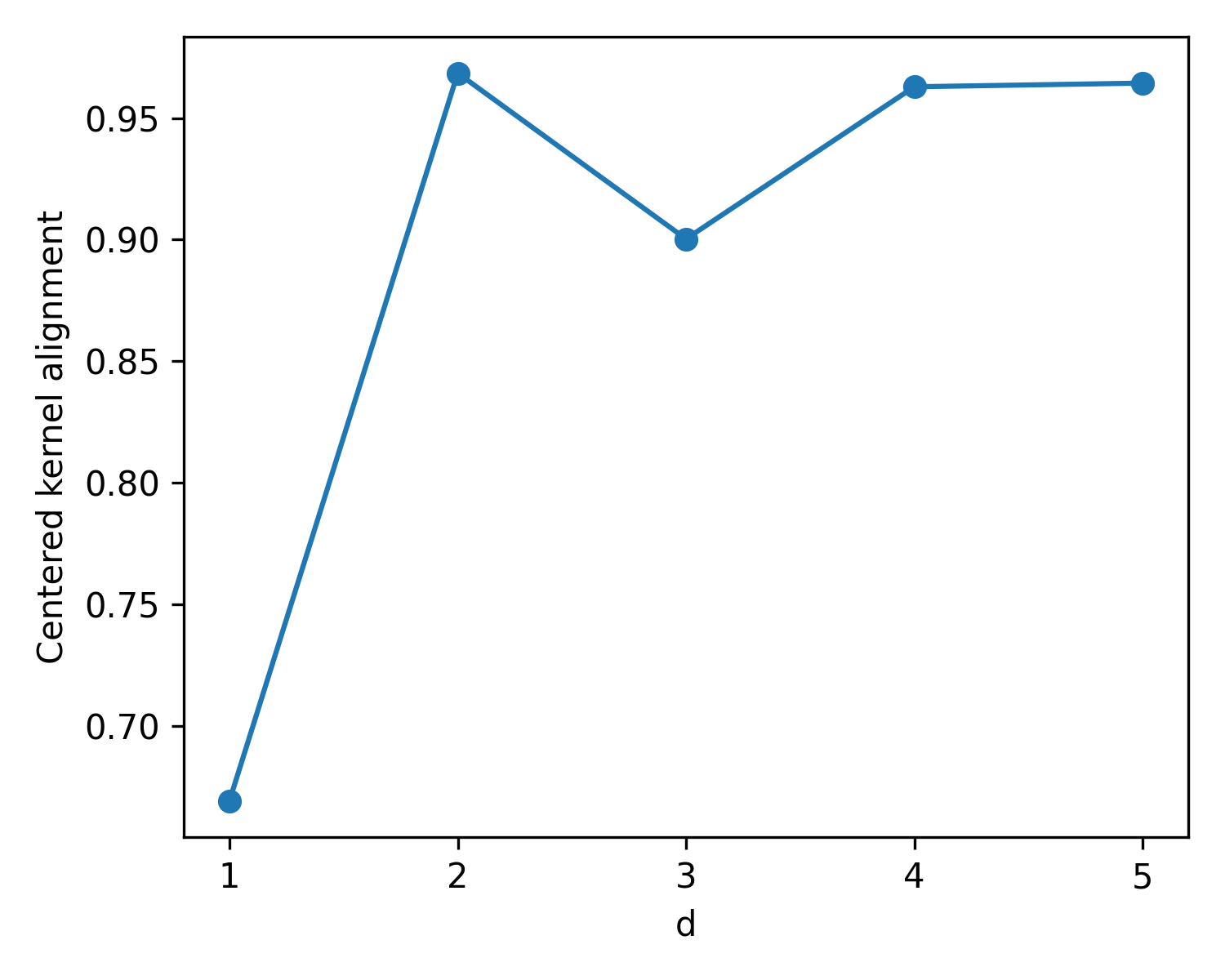}
     \end{subfigure}
     \begin{subfigure}[b]{0.47\textwidth}
         \centering
         \includegraphics[width=\textwidth]{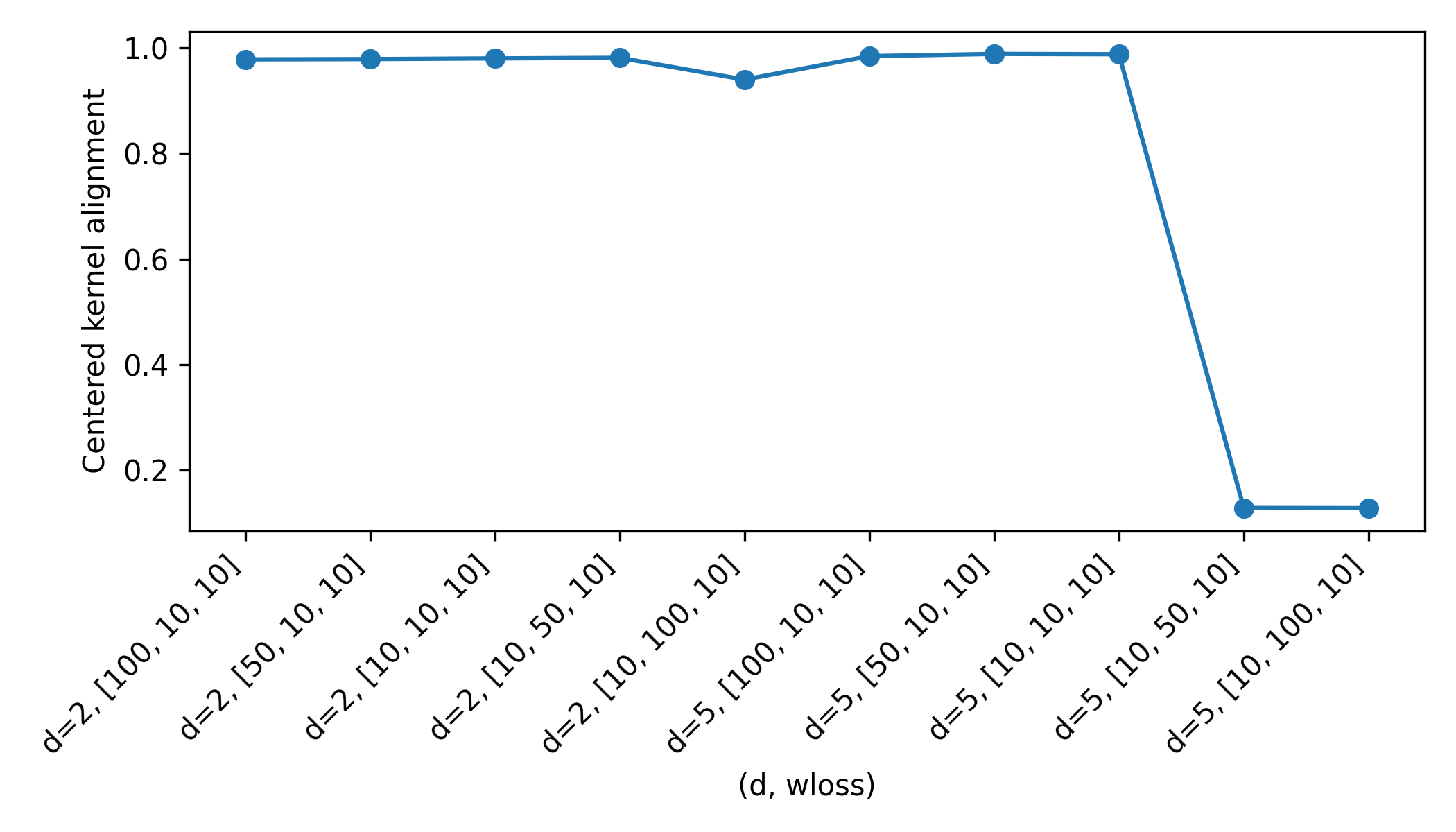}
     \end{subfigure} \\
     \begin{subfigure}[b]{0.33\textwidth}
         \centering
         \includegraphics[width=\textwidth]{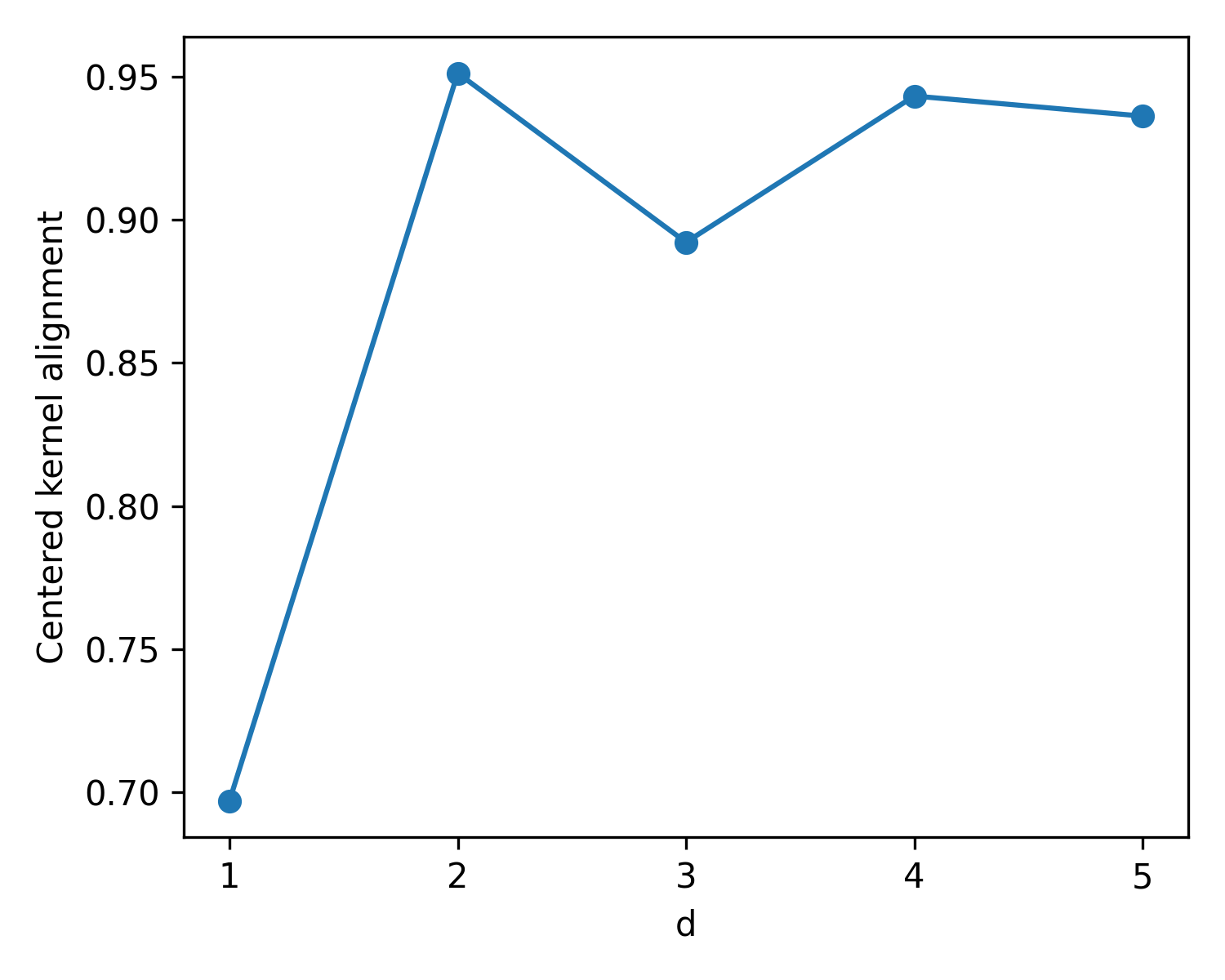}
     \end{subfigure}
     \begin{subfigure}[b]{0.47\textwidth}
         \centering
         \includegraphics[width=\textwidth]{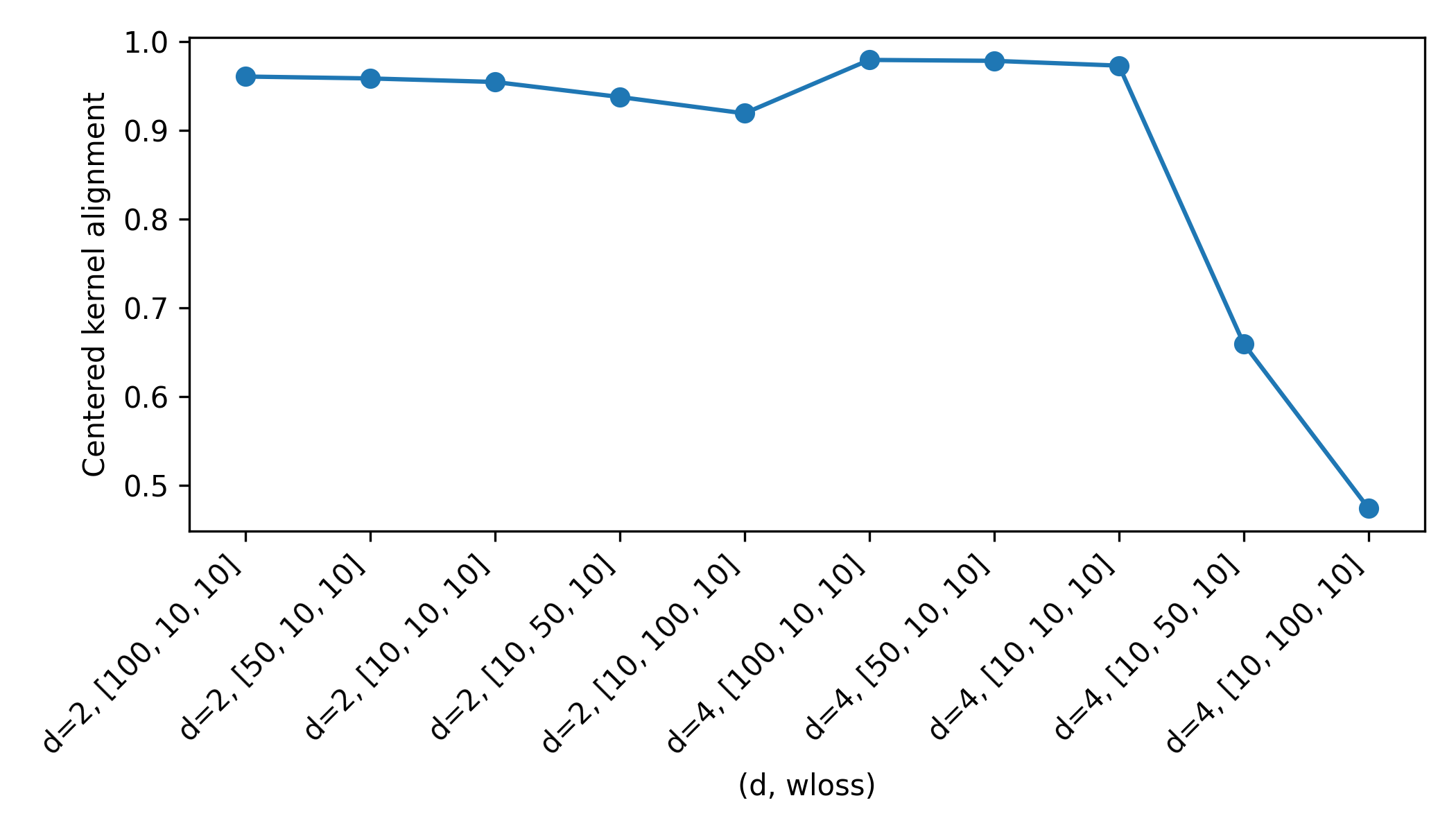}
     \end{subfigure} \\
     \begin{subfigure}[b]{0.33\textwidth}
         \centering
         \includegraphics[width=\textwidth]{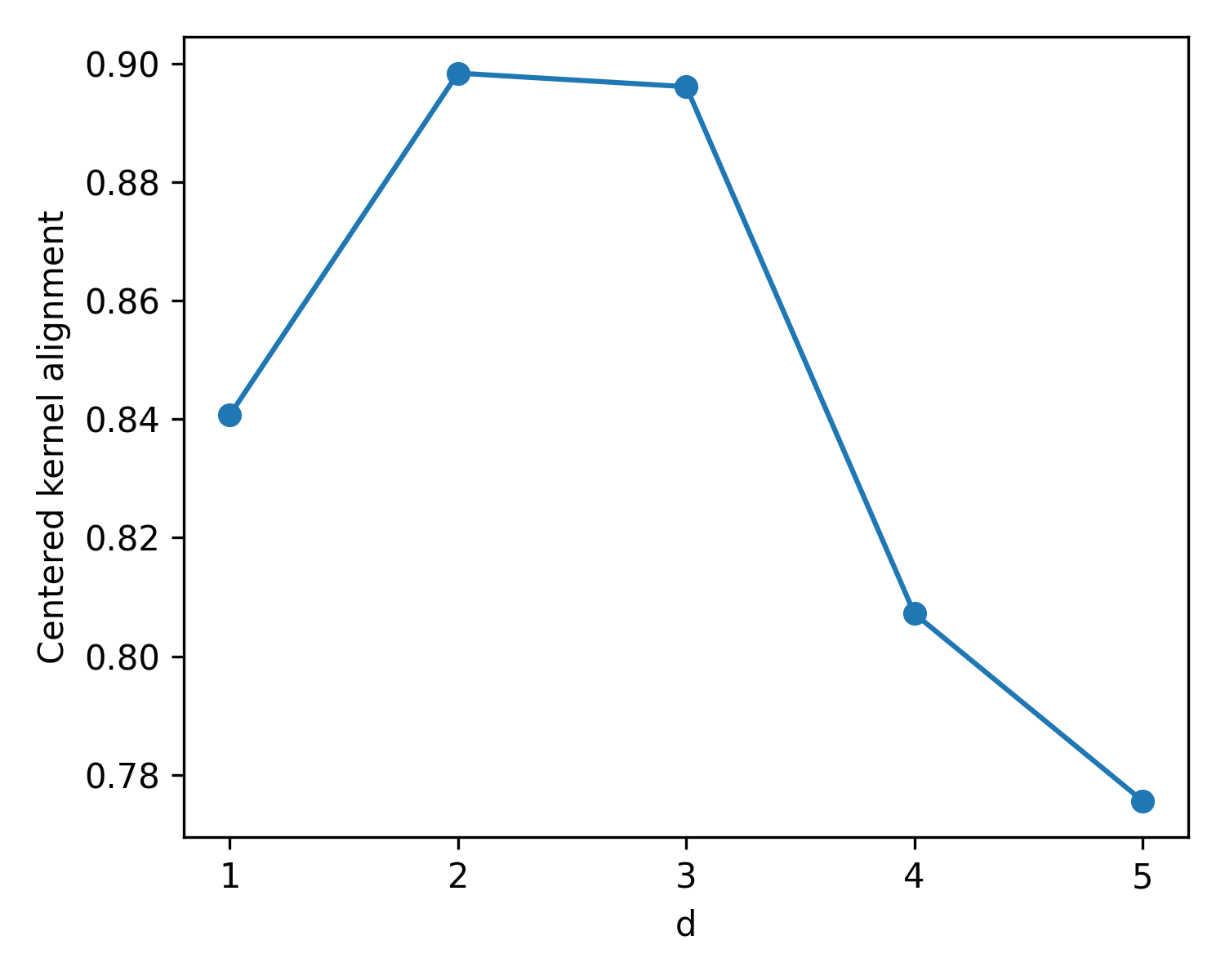}
     \end{subfigure}
     \begin{subfigure}[b]{0.47\textwidth}
         \centering
         \includegraphics[width=\textwidth]{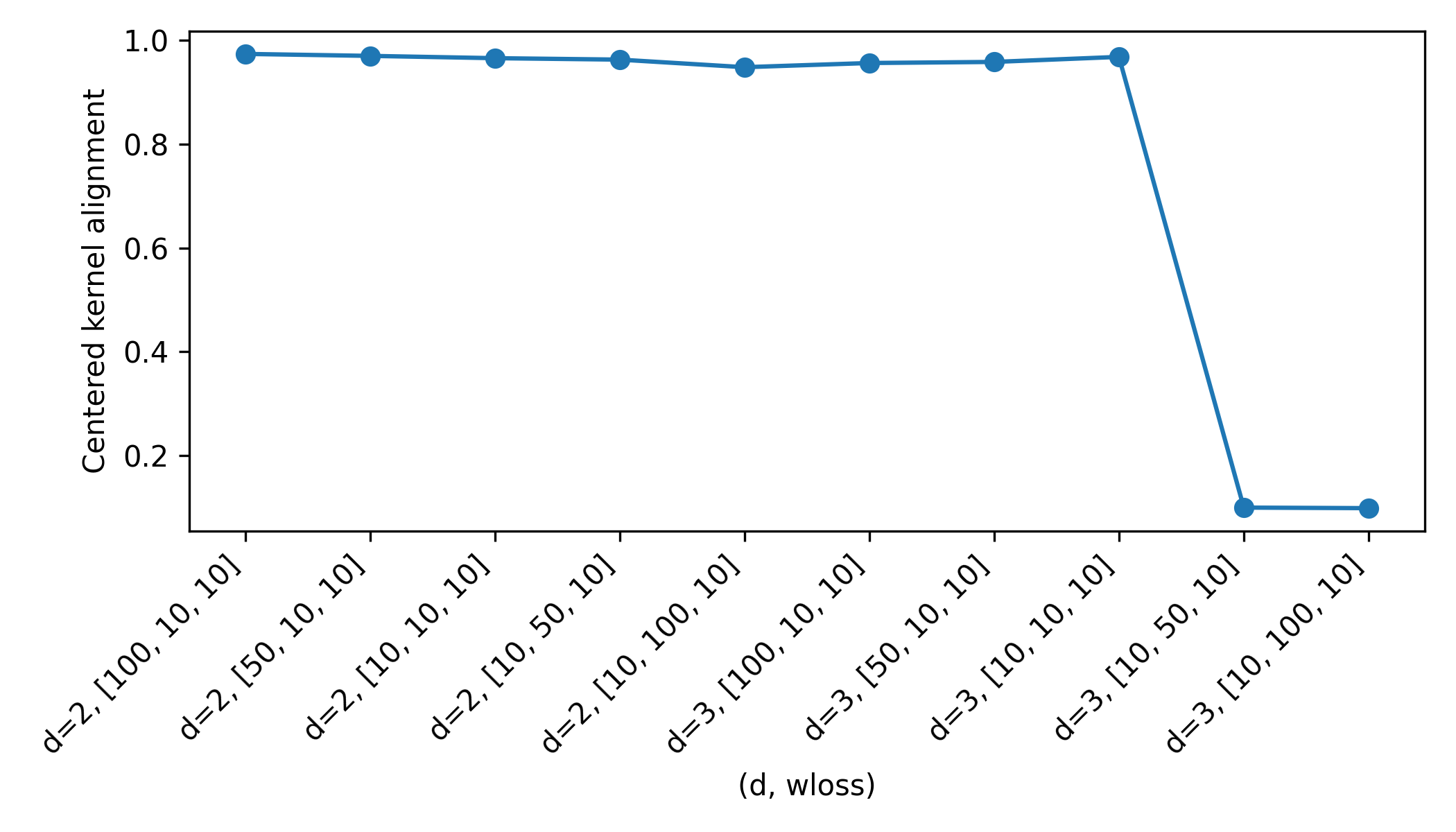}
     \end{subfigure}
        \caption{\small Centered kernel alignment used to select the Stage 2 embedding dimension $d$ and loss weights $w = (c_V, c_C)$ in simulation scenarios 1-4. Rows correspond to scenarios 1-4, respectively. For each scenario, the left panel shows the alignment obtained from an initial search over $d = 1, .\dots, 5$, using $w = (50, 10)$ and $100$ training epochs. The right panels shows the subsequent refined search over all candidate loss weight combinations for the two highest performing values of $d$ from the initial search. Centered kernel alignment quantifies how well the learned embeddings preserve the target similarity structure, with higher values indicating better agreement. The combination of $d$ and $w$ achieving the highest alignment was selected for each scenario.}
        \label{fig:sim_Dd}
\end{figure}


\clearpage \newpage

\begin{figure}[htbp!]
     \centering
     \begin{subfigure}[b]{0.99\textwidth}
         \centering
         \includegraphics[width=\textwidth]{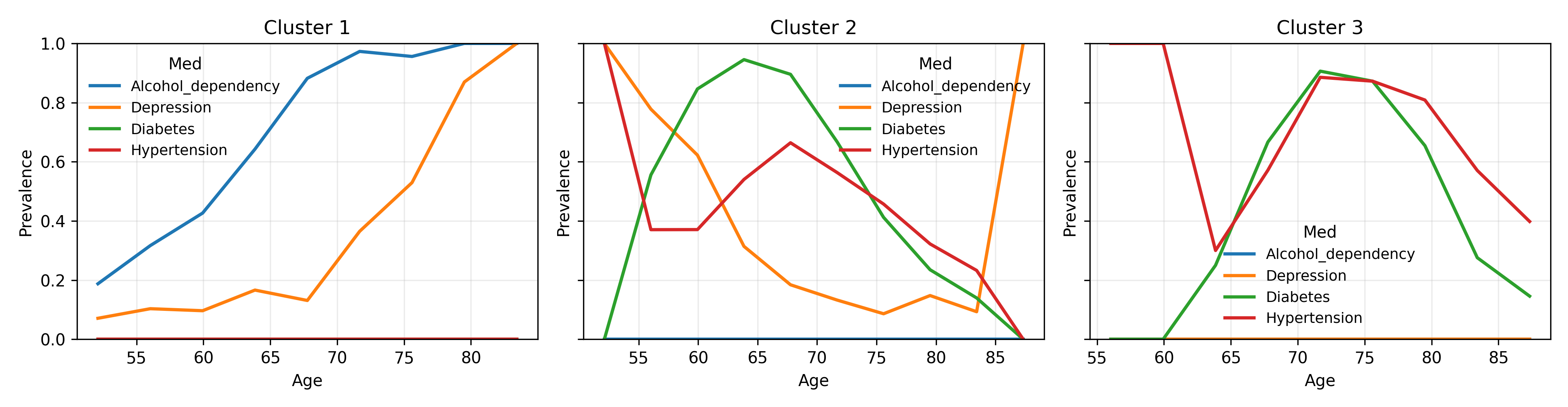}
         \caption{}
         \label{fig:sim_set_sc4}
     \end{subfigure}
     \hfill
     \begin{subfigure}[b]{0.38\textwidth}
         \centering
         \includegraphics[width=\textwidth]{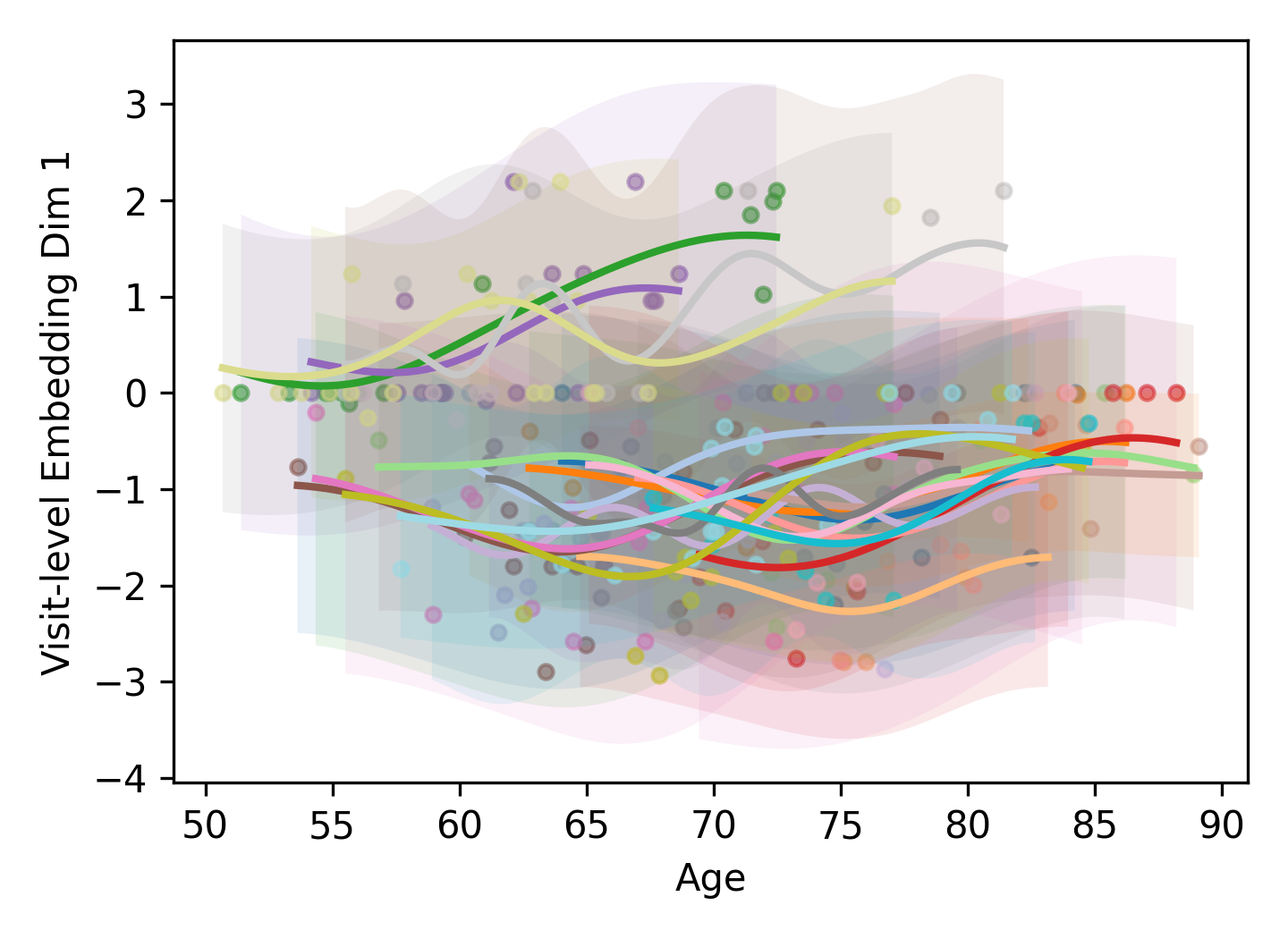}
         \caption{}
         \label{fig:sim_med_ebd_sc4}
     \end{subfigure}
     \begin{subfigure}[b]{0.30\textwidth}
         \centering
         \includegraphics[width=\textwidth]{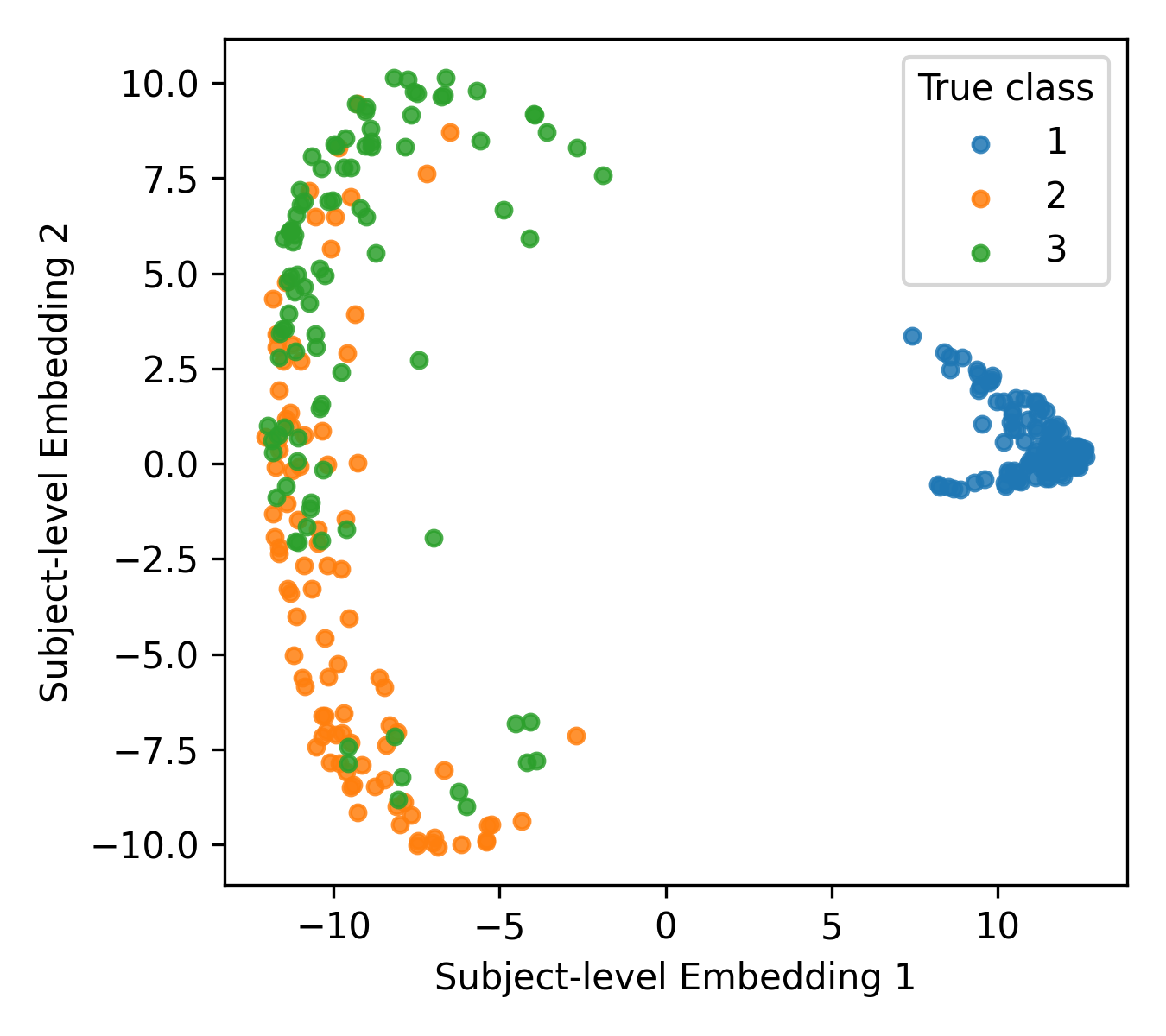}
         \caption{}
         \label{fig:sim_embeddings_plot_sc5}
     \end{subfigure}
     \begin{subfigure}[b]{0.30\textwidth}
         \centering
         \includegraphics[width=\textwidth]{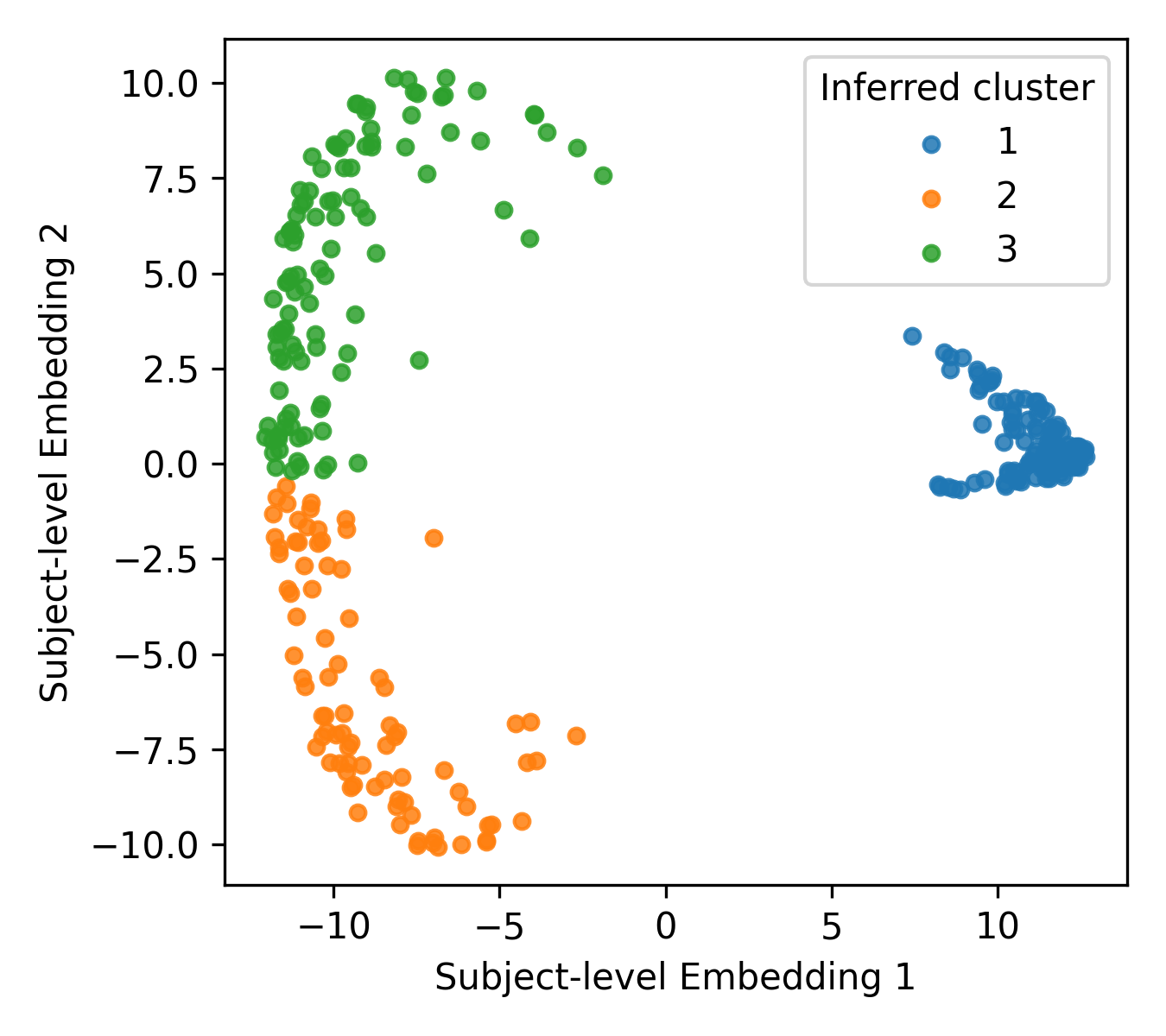}
         \caption{}
         \label{fig:sim_embeddings_plot_sc4}
     \end{subfigure}
     \hfill
     \begin{subfigure}[b]{0.99\textwidth}
         \centering
         \includegraphics[width=\textwidth]{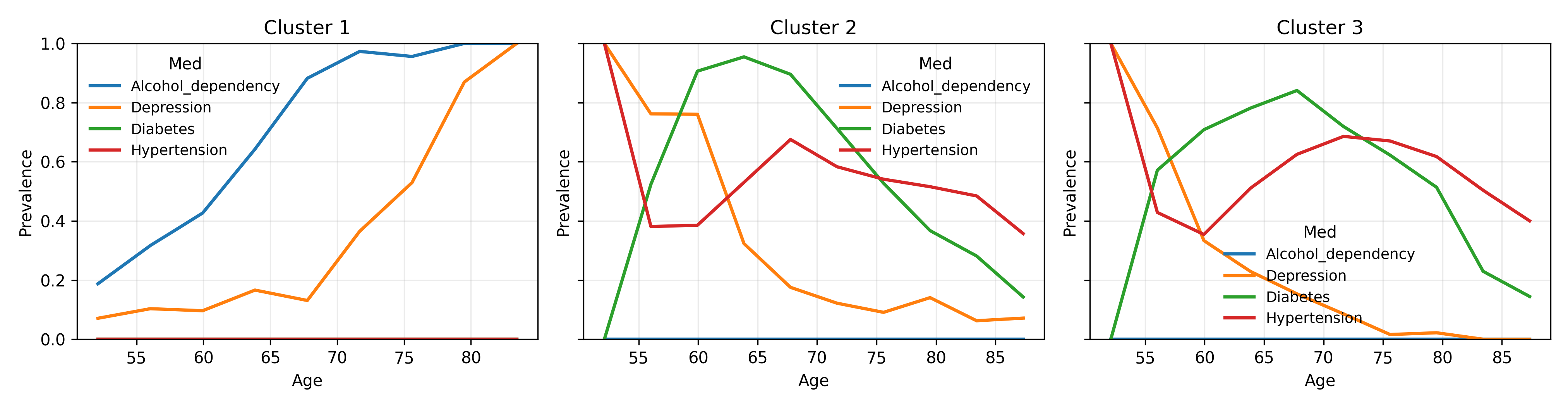}
         \caption{}
         \label{fig:sim_prevalence_per_cluster_sc4}
     \end{subfigure}
        \caption{Simulation result of scenario 2.
        (a) Empirical condition-level medication prevalence over age within each true cluster. For each condition, prevalence is computed as the proportion of participants in each five-year age bin using any medication for that condition.
        (b) Gaussian process regression of the learned medication embedding, showing 20 participants.
        (c) Subject-level embeddings colored by true clusters labels.
        (d) Subject-level embeddings colored by inferred clusters by LOPEL method.
        (e) Empirical condition-level medication prevalence over age within each LOPEL inferred cluster.
        }
        \label{fig:sim_sc4}
\end{figure}

\begin{figure}[htbp!]
     \centering
     \begin{subfigure}[b]{0.99\textwidth}
         \centering
         \includegraphics[width=\textwidth]{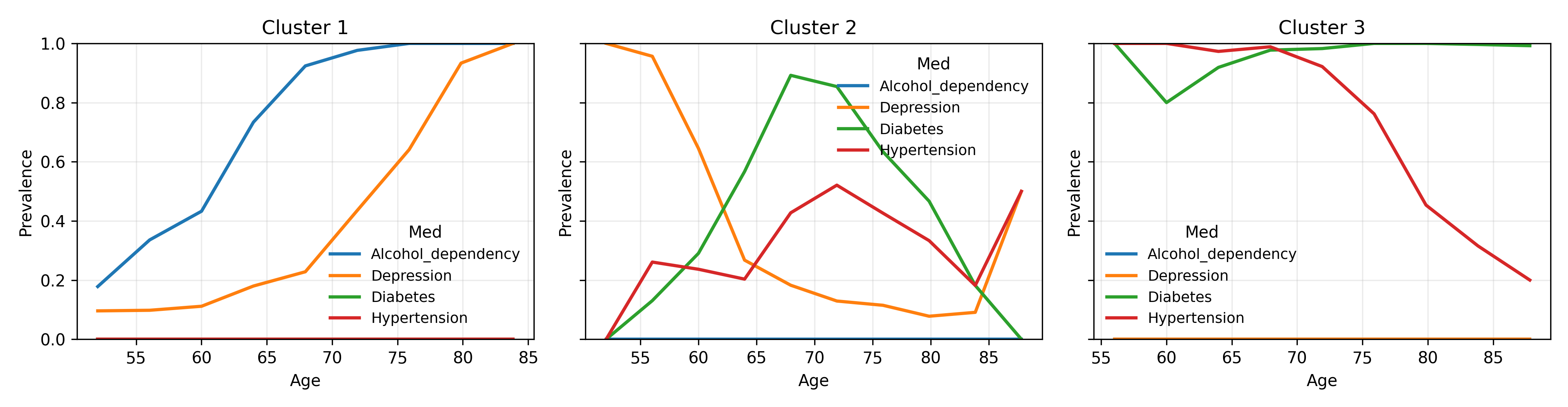}
         \caption{}
         \label{fig:sim_set_sc4}
     \end{subfigure}
     \hfill
     \begin{subfigure}[b]{0.38\textwidth}
         \centering
         \includegraphics[width=\textwidth]{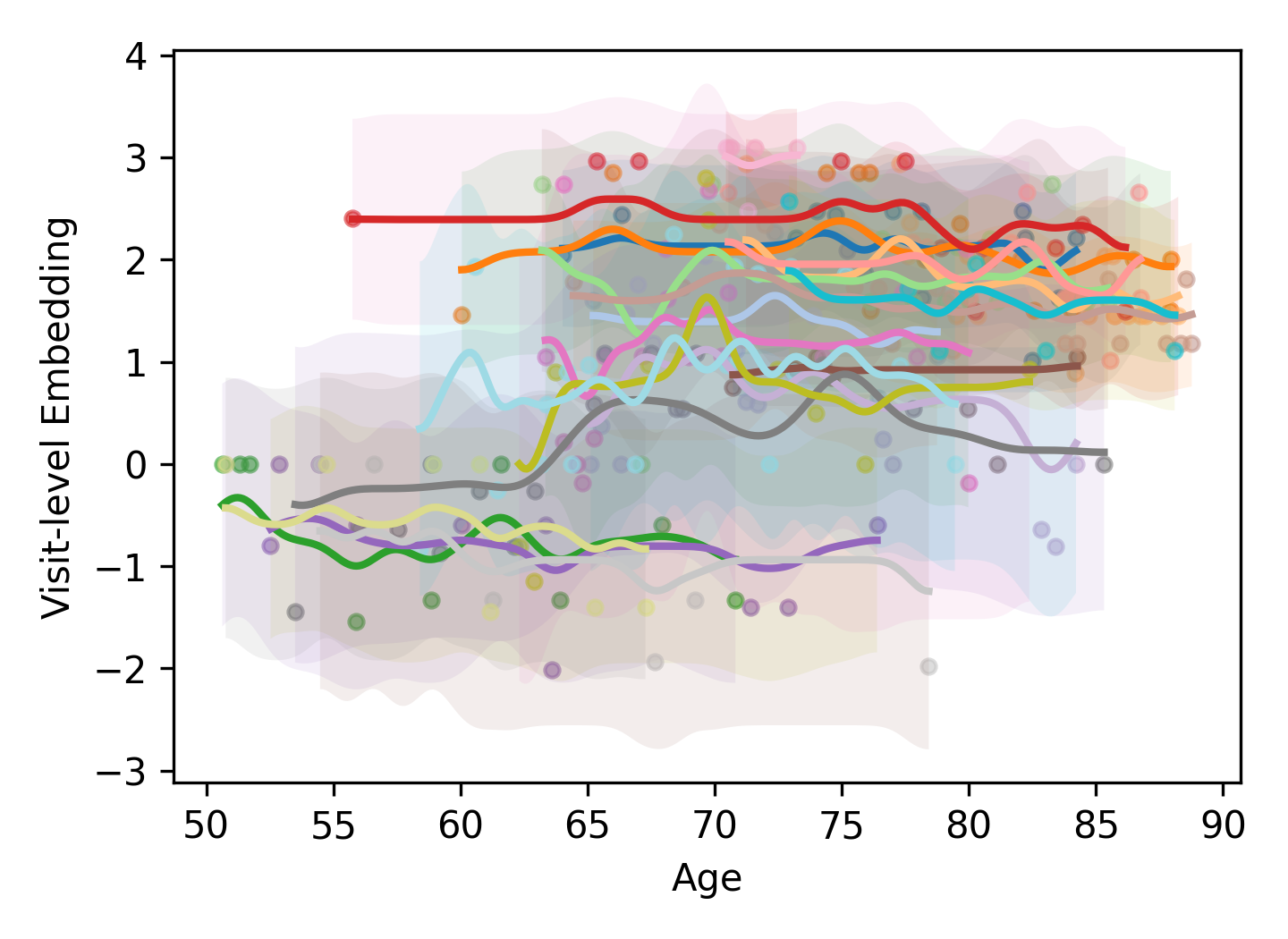}
         \caption{}
         \label{fig:sim_med_ebd_sc5}
     \end{subfigure}
     \begin{subfigure}[b]{0.30\textwidth}
         \centering
         \includegraphics[width=\textwidth]{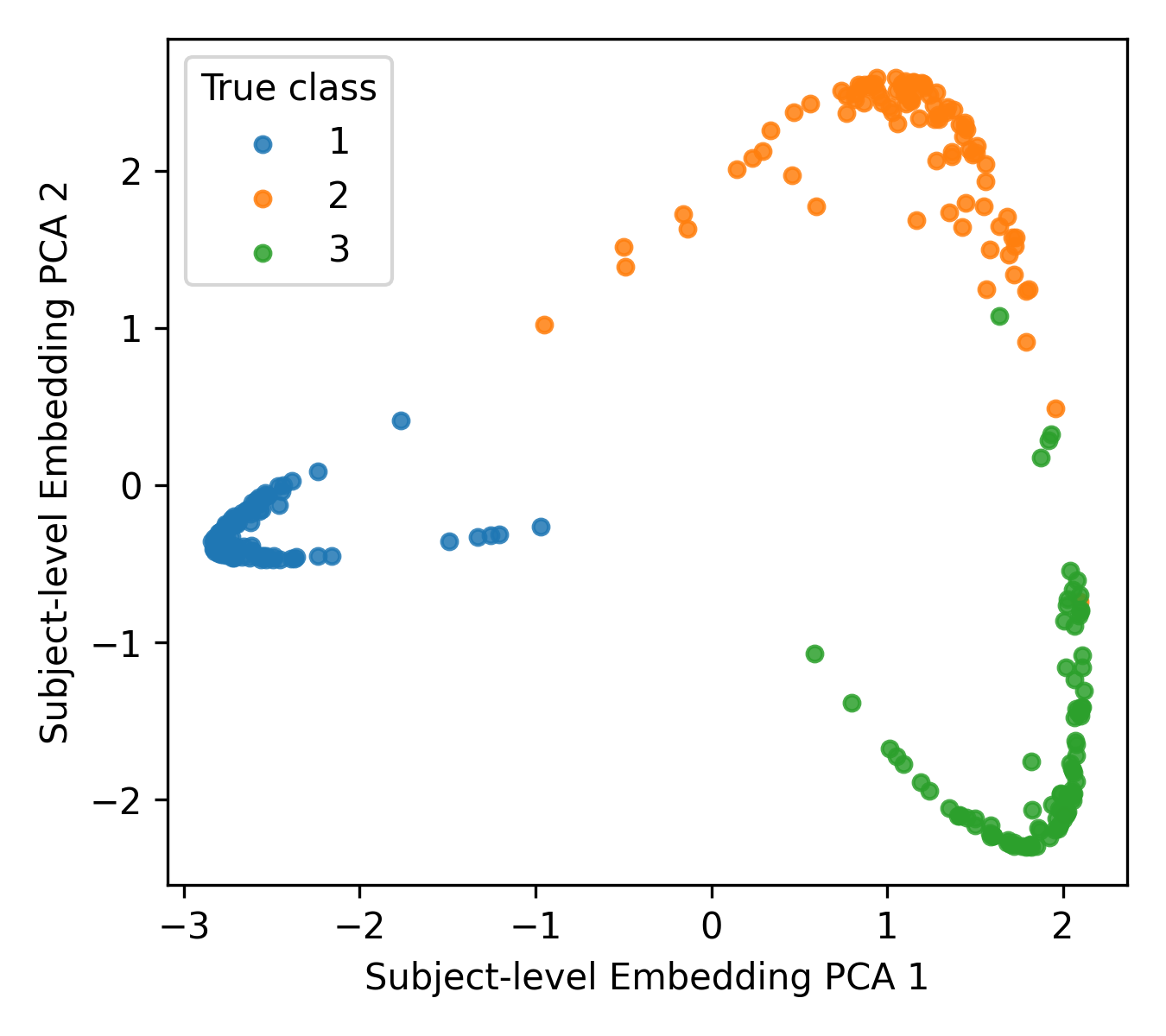}
         \caption{}
         \label{fig:sim_embeddings_plot_sc5_true}
     \end{subfigure}
     \begin{subfigure}[b]{0.30\textwidth}
         \centering
         \includegraphics[width=\textwidth]{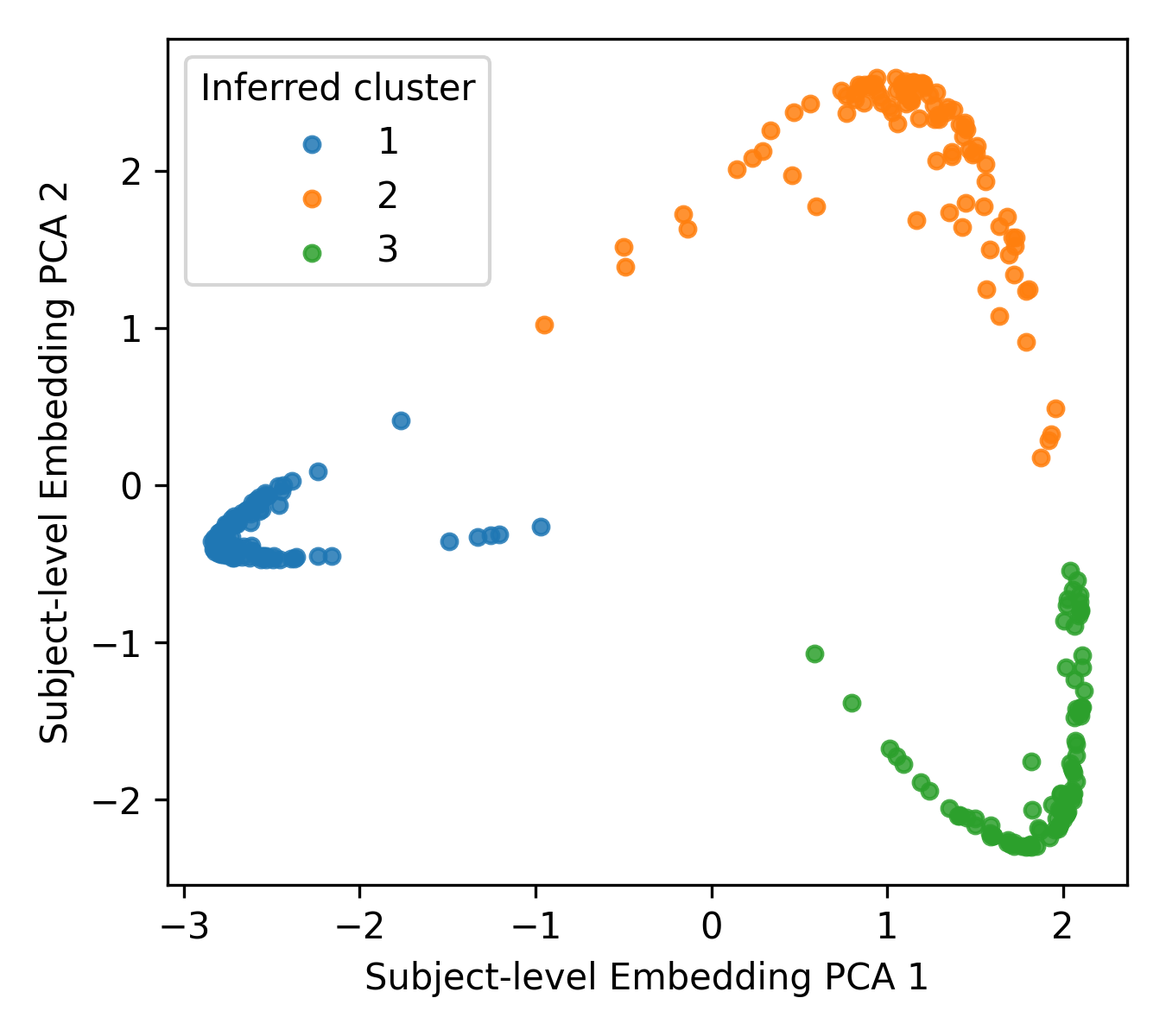}
         \caption{}
         \label{fig:sim_embeddings_plot_sc5}
     \end{subfigure}
     \hfill
     \begin{subfigure}[b]{0.99\textwidth}
         \centering
         \includegraphics[width=\textwidth]{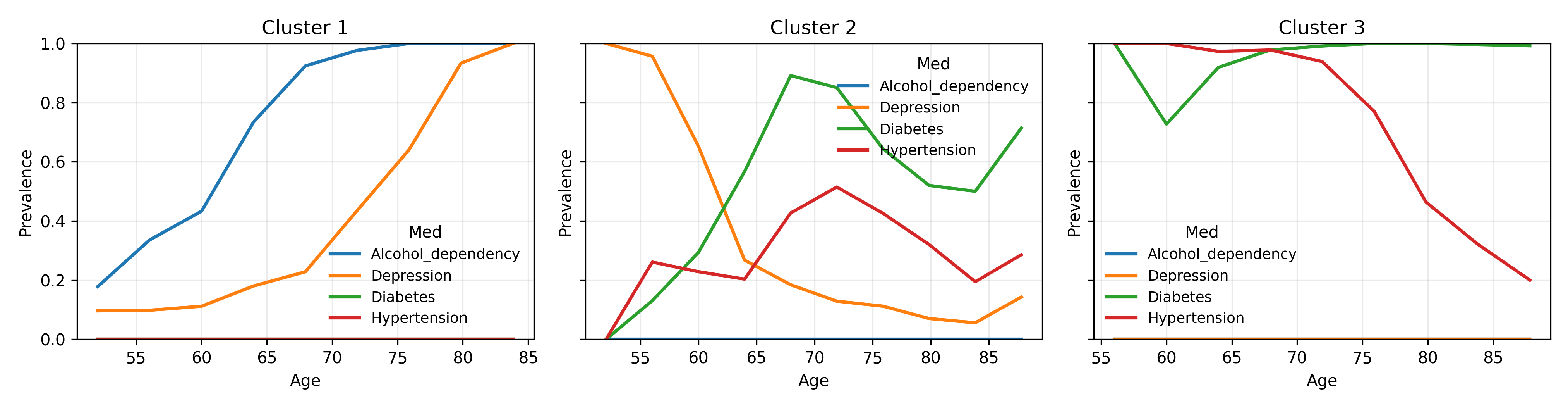}
         \caption{}
         \label{fig:sim_prevalence_per_cluster_sc5}
     \end{subfigure}
        \caption{Simulation result of scenario 3.
        (a) Empirical condition-level medication prevalence over age within each true cluster. For each condition, prevalence is computed as the proportion of participants in each five-year age bin using any medication for that condition.
        (b) Gaussian process regression of the learned medication embedding, showing 20 participants.
        (c) Subject-level embeddings colored by true clusters labels.
        (d) Subject-level embeddings colored by inferred clusters by LOPEL method.
        (e) Empirical condition-level medication prevalence over age within each LOPEL inferred cluster.
        }
        \label{fig:sim_sc5}
\end{figure}

\begin{figure}[htbp!]
     \centering
     \begin{subfigure}[b]{0.99\textwidth}
         \centering
         \includegraphics[width=\textwidth]{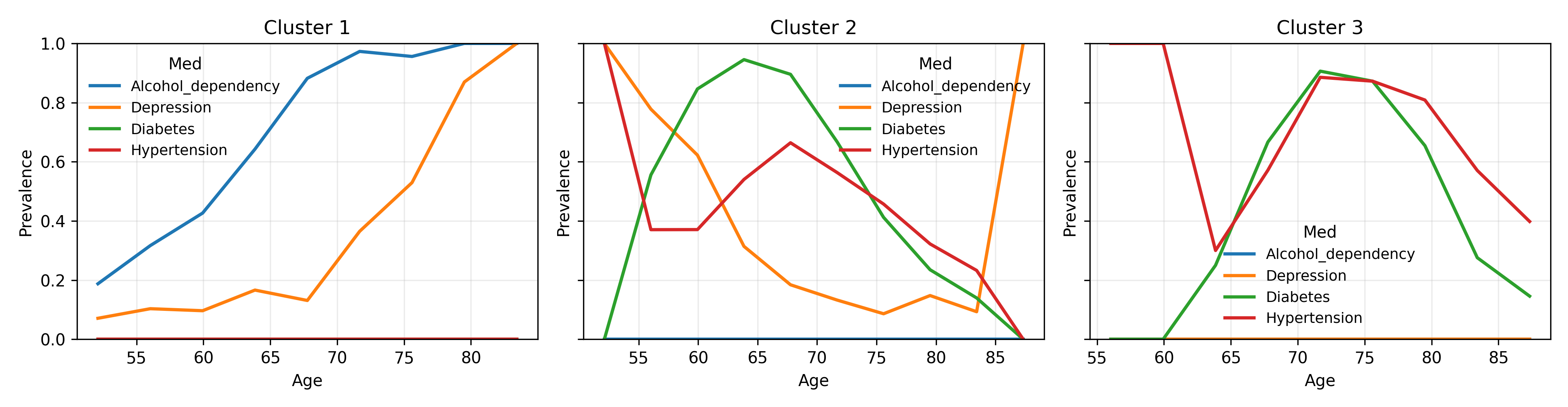}
         \caption{}
         \label{fig:sim_set_sc6}
     \end{subfigure}
     \hfill
     \begin{subfigure}[b]{0.38\textwidth}
         \centering
         \includegraphics[width=\textwidth]{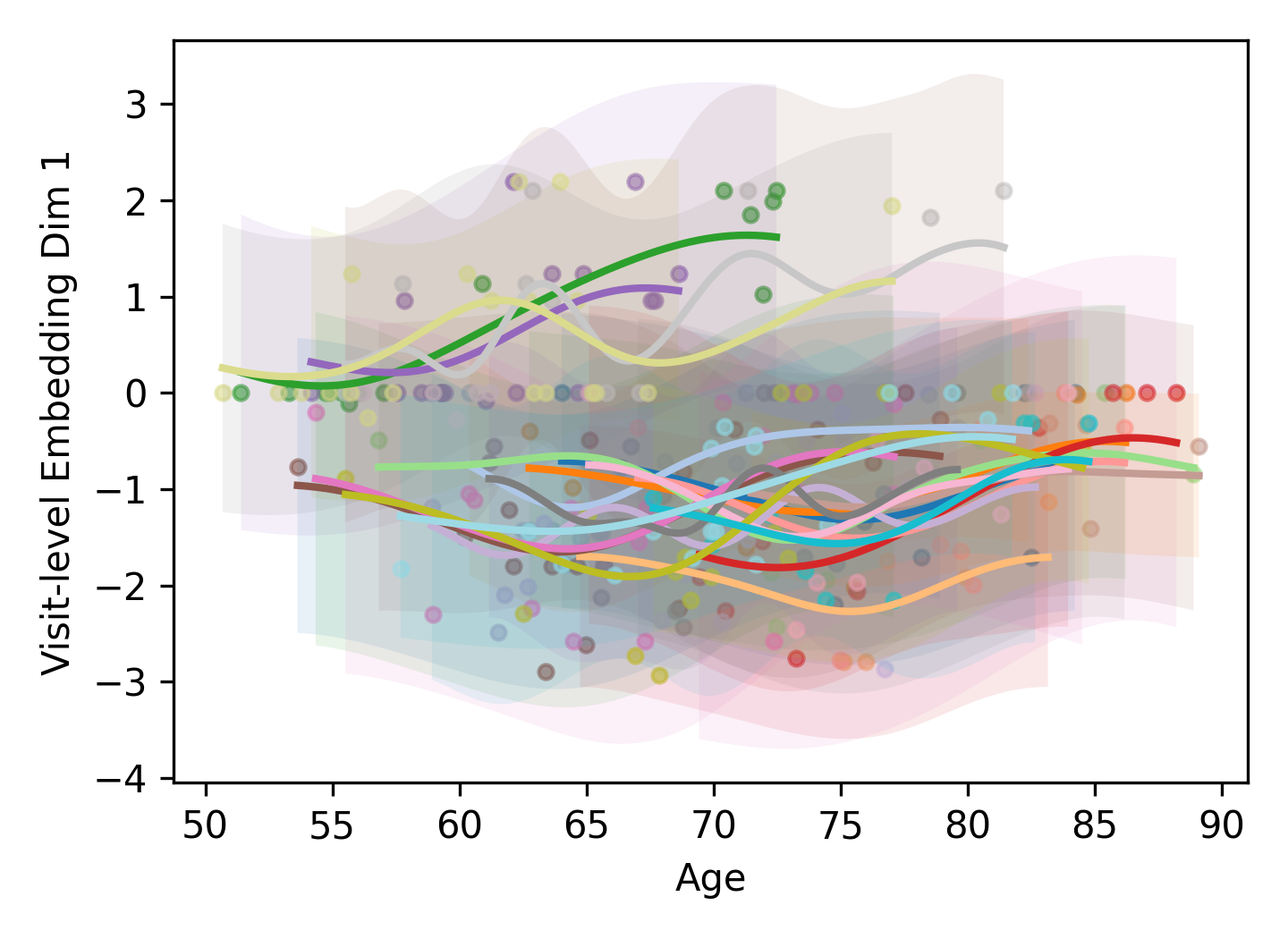}
         \caption{}
         \label{fig:sim_med_ebd_sc6}
     \end{subfigure}
     \begin{subfigure}[b]{0.30\textwidth}
         \centering
         \includegraphics[width=\textwidth]{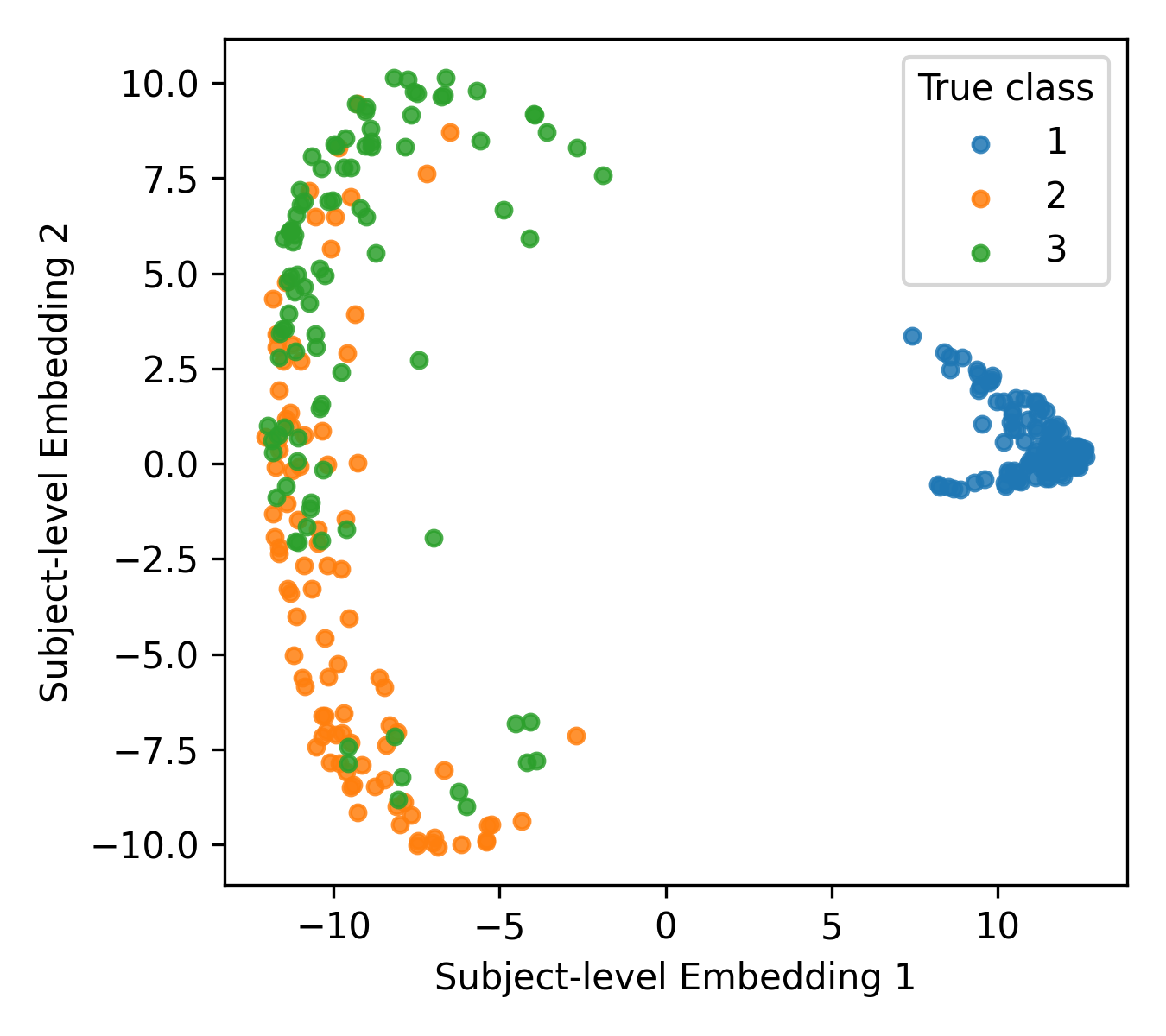}
         \caption{}
         \label{fig:sim_embeddings_plot_sc6_true}
     \end{subfigure}
     \begin{subfigure}[b]{0.30\textwidth}
         \centering
         \includegraphics[width=\textwidth]{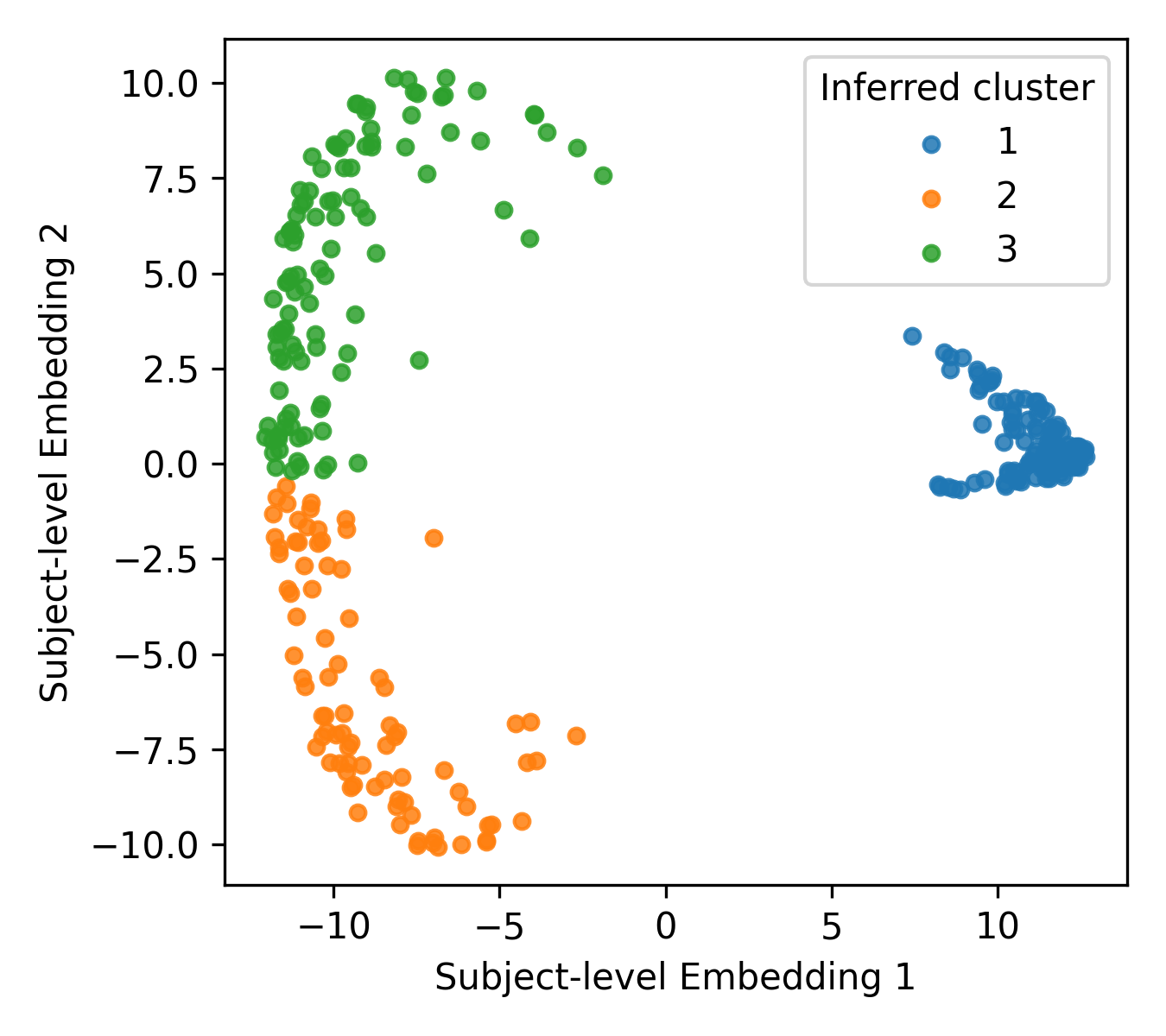}
         \caption{}
         \label{fig:sim_embeddings_plot_sc6}
     \end{subfigure}
     \hfill
     \begin{subfigure}[b]{0.99\textwidth}
         \centering
         \includegraphics[width=\textwidth]{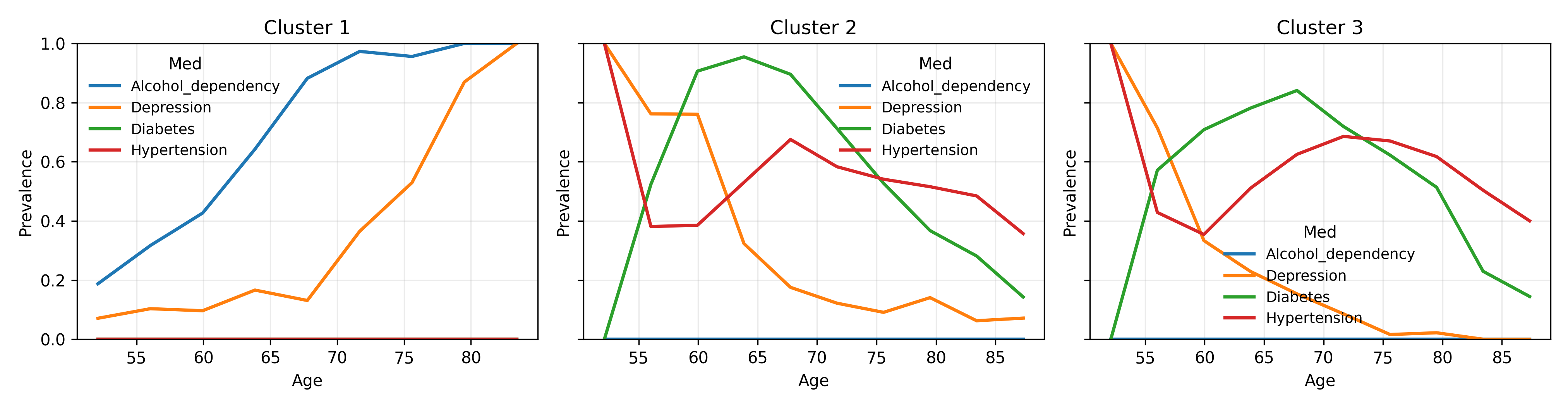}
         \caption{}
         \label{fig:sim_prevalence_per_cluster_sc6}
     \end{subfigure}
        \caption{Simulation result of scenario 4.
        (a) Empirical condition-level medication prevalence over age within each true cluster. For each condition, prevalence is computed as the proportion of participants in each five-year age bin using any medication for that condition.
        (b) Gaussian process regression of the learned medication embedding, showing 20 participants.
        (c) Subject-level embeddings colored by true clusters labels.
        (d) Subject-level embeddings colored by inferred clusters by LOPEL method.
        (e) Empirical condition-level medication prevalence over age within each LOPEL inferred cluster.
        }
        \label{fig:sim_sc6}
\end{figure}

\clearpage \newpage


\begin{table}[htbp!]
\centering
\renewcommand{\arraystretch}{0.75}
\begin{tabular}{l|c|c|c}
\hline
Condition & No. Meds & Prev. (PWH) & Prev. (PWOH) \\
\hline
Pain & 44 & 70.7\% & 28.4\% \\
Hypertension & 72 & 54.3\% & 25.2\% \\
Hyperlipidaemia & 30 & 51.1\% & 21.3\% \\
Depression & 32 & 49.9\% & 9.0\% \\
Anxiety/Sedative & 32 & 38.2\% & 5.8\% \\
Inflammation/pain & 26 & 35.6\% & 16.8\% \\
Chronic airways disease & 38 & 35.4\% & 9.0\% \\
Gastroesophageal reflux disease & 17 & 35.1\% & 7.7\% \\
Allergies & 28 & 29.0\% & 3.2\% \\
Smoking cessation & 6 & 23.7\% & 5.2\% \\
Antiepileptics & 18 & 19.9\% & 1.9\% \\
Congestive heart failure & 7 & 19.9\% & 6.5\% \\
Constipation & 26 & 19.0\% & 2.6\% \\
Benign prostatic hyperplasia & 6 & 18.3\% & 2.6\% \\
Antipsychotics & 24 & 18.0\% & 3.9\% \\
Diabetes & 46 & 17.1\% & 9.7\% \\
Pulmonary hypertension & 2 & 12.4\% & 3.9\% \\
Inflammatory glucocorticoids & 18 & 8.9\% & 3.2\% \\
Glaucoma & 21 & 8.4\% & 3.2\% \\
Hypothyroidism & 4 & 6.8\% & 4.5\% \\
\hline
\end{tabular}
\caption{Prevalence of the 20 most common non-antiretroviral (non-ART) chronic condition categories among PWH ($n$ = 427) and the corresponding prevalence of the conditions in PWOH ($n$ = 155). ``Prev." denotes the percentage of participants ever using medications under each condition and ``No. Meds" denotes the number of distinct medications mapped to each condition.}
\label{tab:condition_prevalence}
\end{table}

\begin{figure}[htbp!]
     \centering
     \begin{subfigure}[b]{0.33\textwidth}
         \centering
         \includegraphics[width=\textwidth]{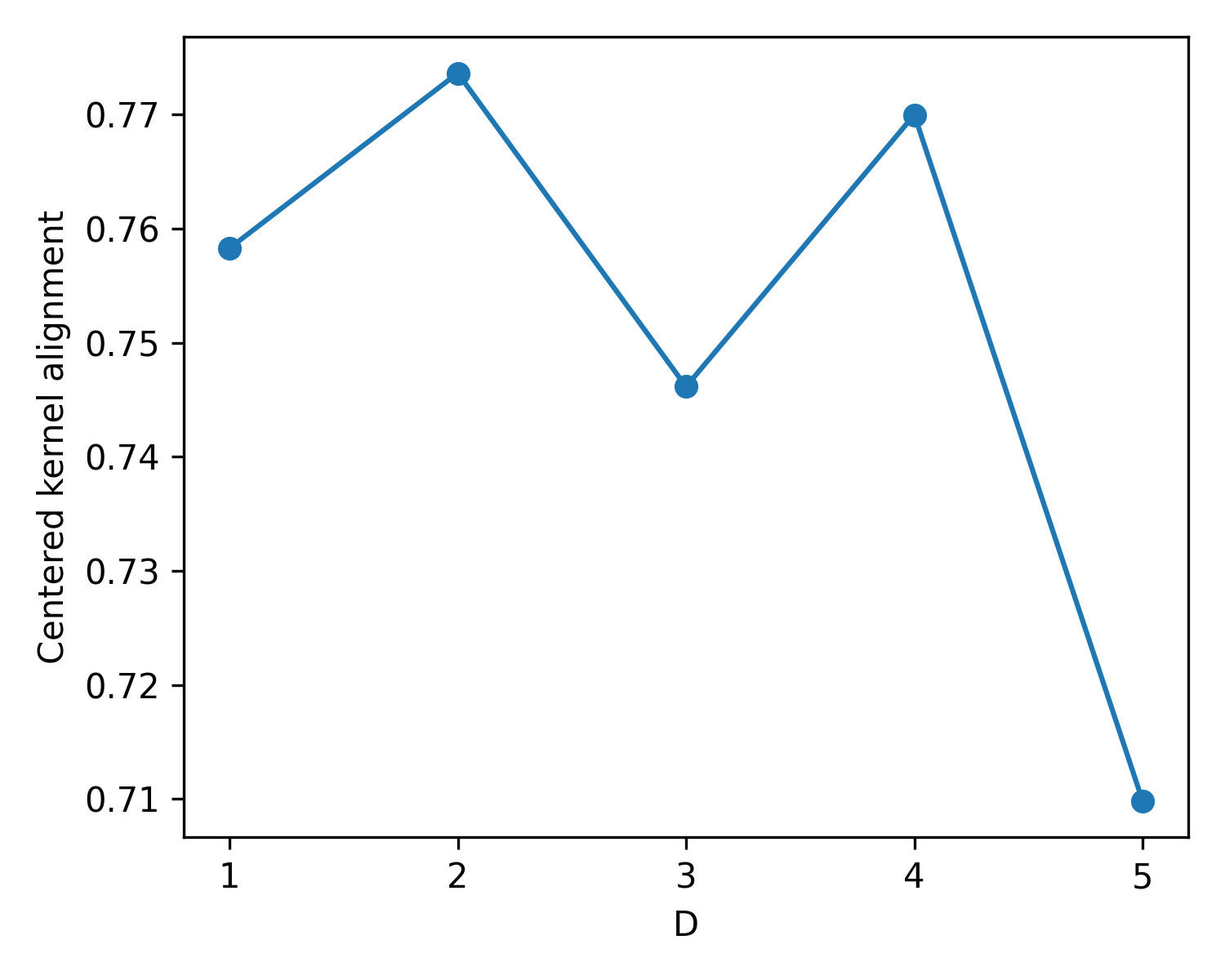}
     \end{subfigure}
     \begin{subfigure}[b]{0.33\textwidth}
         \centering
         \includegraphics[width=\textwidth]{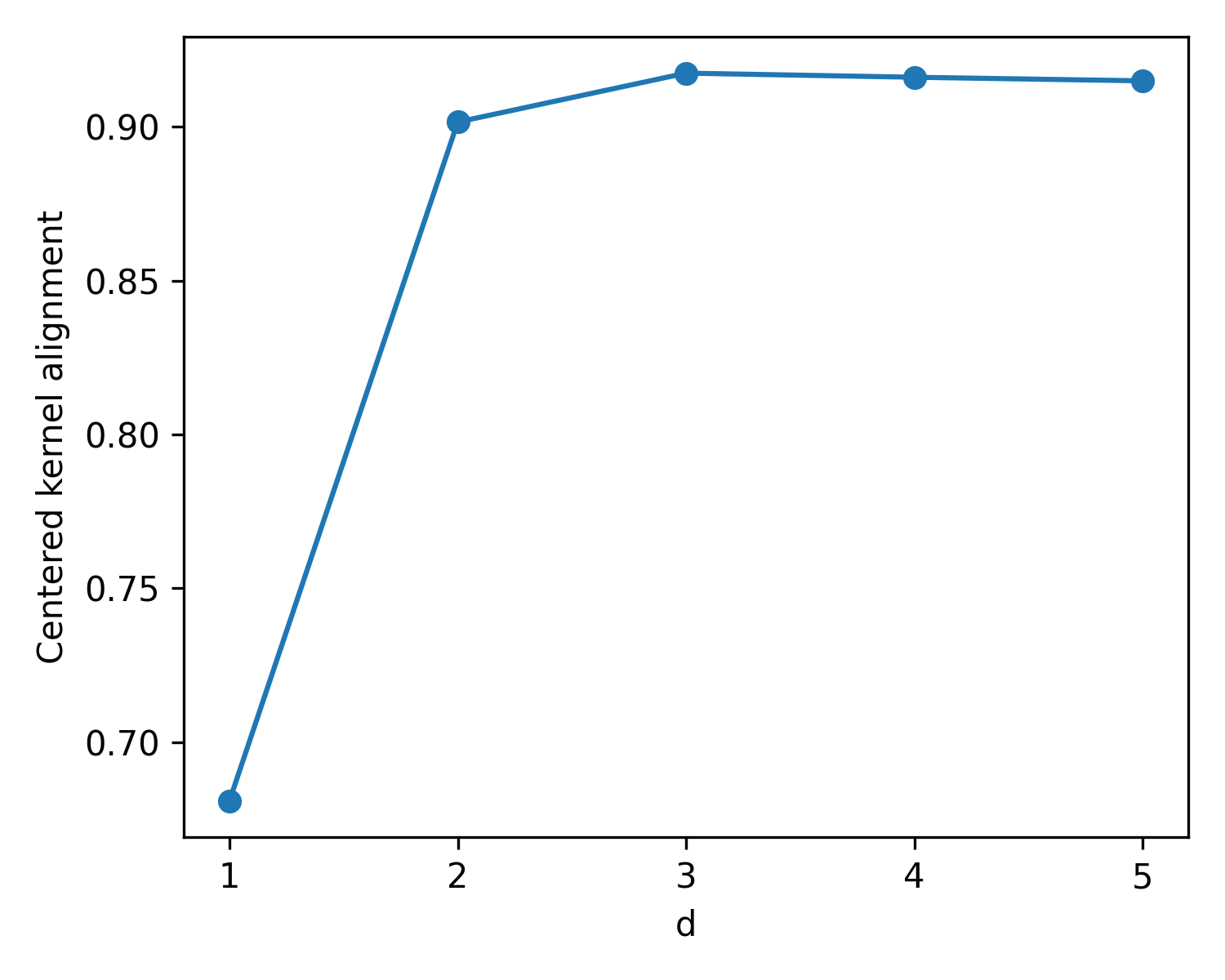}
     \end{subfigure} 
     \caption{Centered kernel alignment values used for selecting the embedding dimensions $D$ (Stage 1) and $d$ (Stage 2) in real data analysis via LOPEL. The selected dimensions correspond to the smallest values of $D$ and $d$ achieving maximal alignment.}
        \label{fig:realdata_dimension_selection}
\end{figure}

\input{table1}

\begin{figure}[htbp!]
    \centering
    \includegraphics[width=0.95\linewidth]{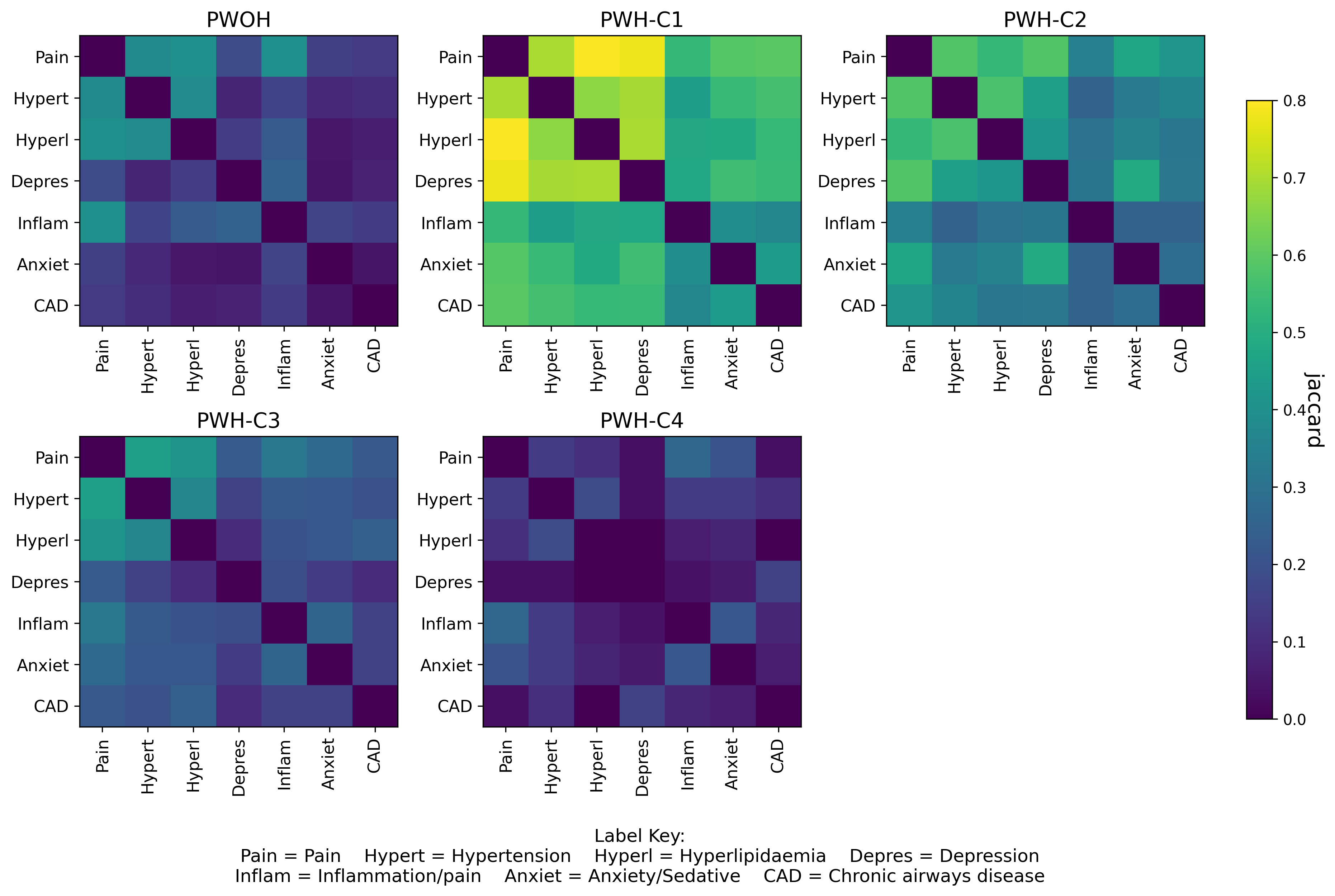}
        \caption{Pairwise co-occurrence of the seven most prevalent medication conditions within each PWH cluster and PWOH.
        Co-occurrence is quantified using the Jaccard index, $J(i,j) = |P_i \cap P_j| / |P_i \cup P_j|$, where $P_i$ denotes participants with at least one medication prescribed for condition $i$.}
    \label{fig:realdata_med_cooccurrence}
\end{figure}

\begin{figure}[htbp!]
    \centering
        \centering
        \includegraphics[width=0.9\linewidth]{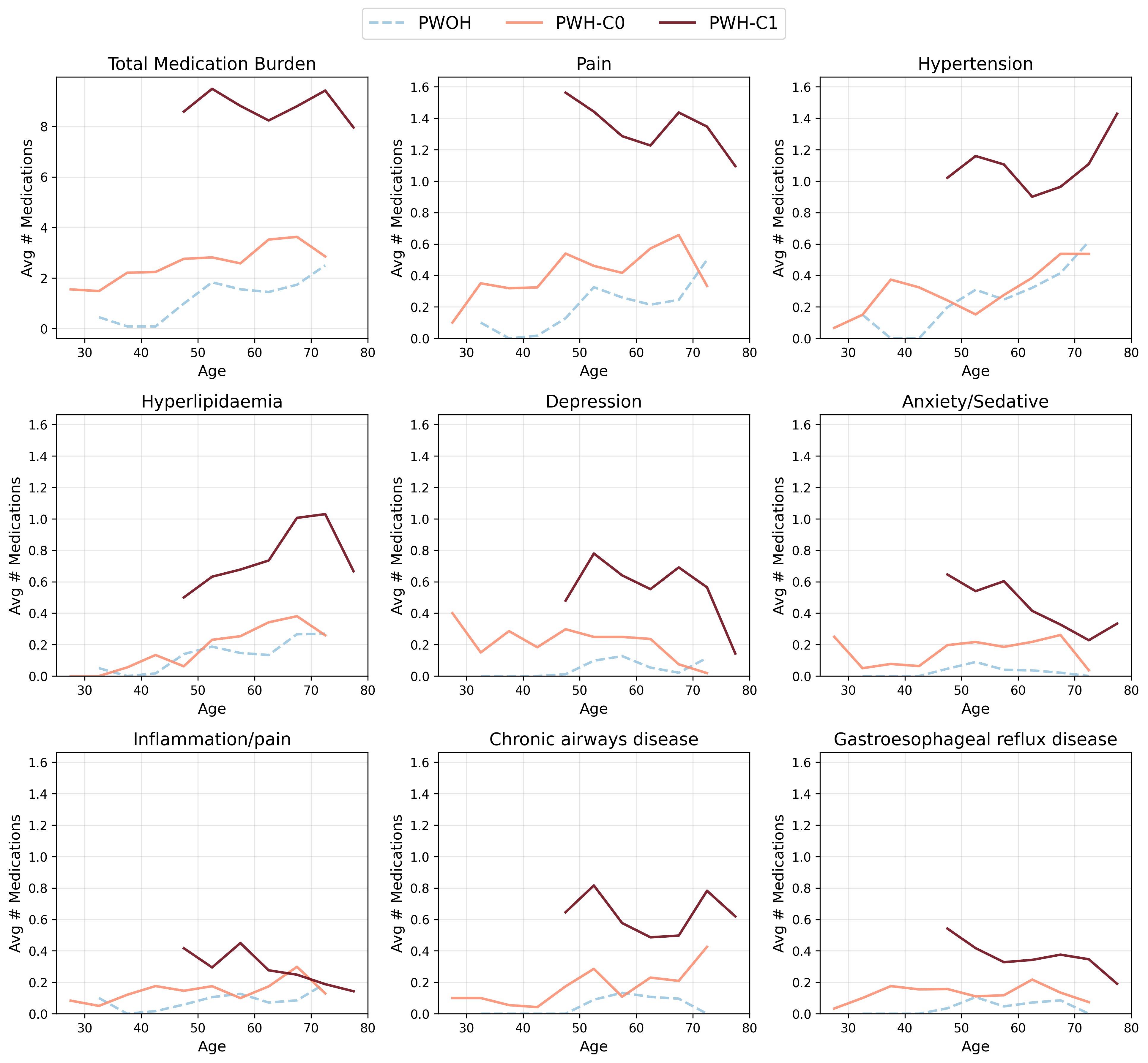}
          \caption{Result by KmeansDTW on total medication counts. Overall medication burden and condition-specific medication burden over age for each PWH cluster and PWOH, averaged within five-year age bins. The red solid lines represent two identified PWH subgroups and the blue dashed line represents PWOH.}
    \label{fig:realdata_nummed_dtw}
\end{figure}

\begin{figure}[htbp!]
    \centering
        \centering
        \includegraphics[width=0.9\linewidth]{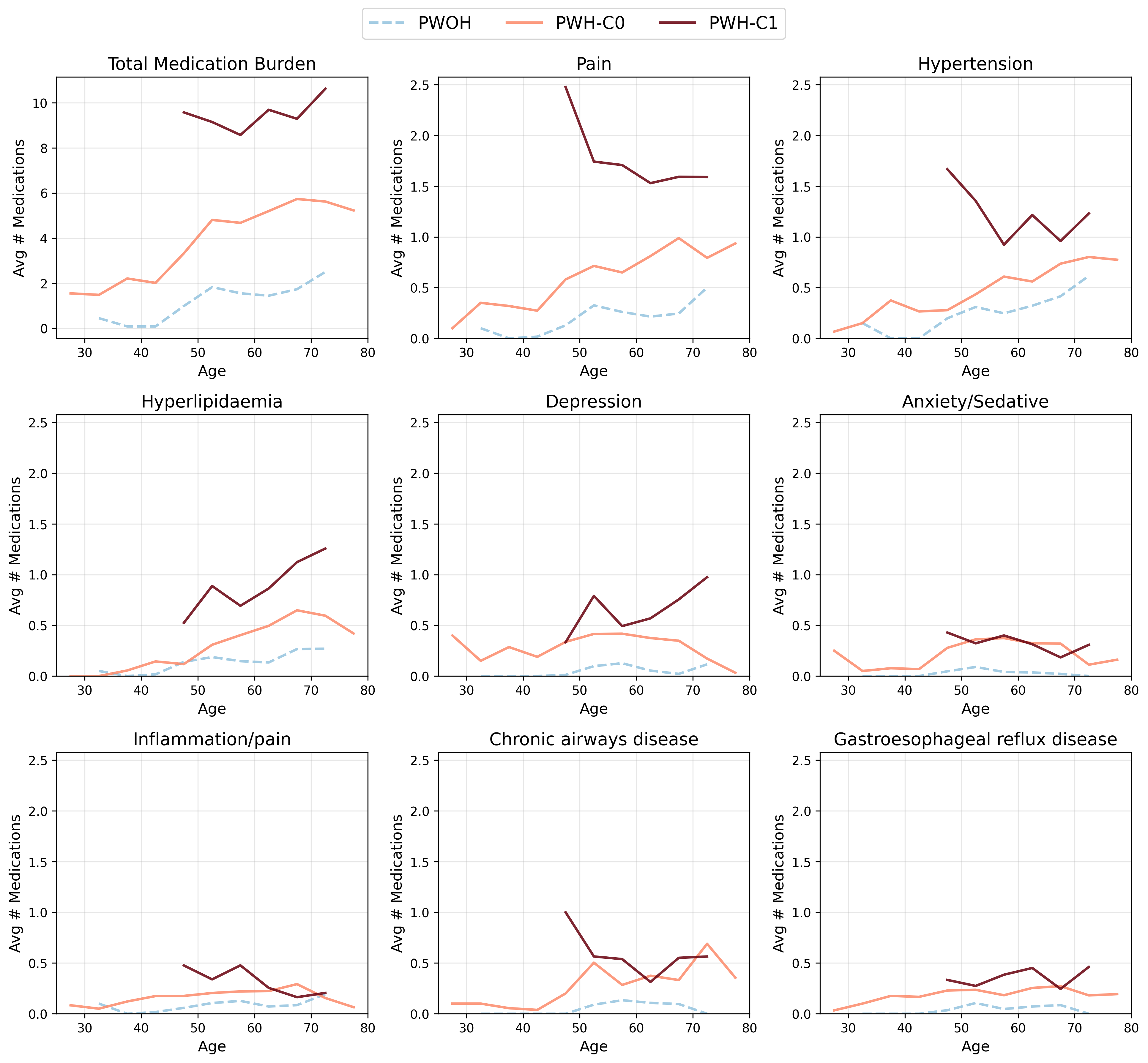}
         \caption{Result by KmeansDTW on raw medication indicators. Overall medication burden and condition-specific medication burden over age for each PWH cluster and PWOH, averaged within five-year age bins. The red solid lines represent two identified PWH subgroups and the blue dashed line represents PWOH.}
    \label{fig:realdata_dtw}
\end{figure}




\clearpage \newpage

\section{Cohort description and data sources} \label{sec:s-data}
\paragraph{CNS HIV Antiretroviral Therapy Effects Research (CHARTER):}  
The CHARTER study is a multi-site cohort of people living with HIV (PWH) receiving care at university-affiliated clinics across the United States. Participants were recruited from six sites (New York, NY; Baltimore, MD; St. Louis, MO; Galveston, TX; Seattle, WA; and San Diego, CA) and followed longitudinally with repeated clinical assessments at intervals of 6 to 12 months. The study provides rich longitudinal data on medication use and comorbid conditions across a diverse population of PWH. 

\paragraph{National NeuroAIDS Tissue Consortium (NNTC):}  
Established in 1998, the NNTC is a multi-site program that collects longitudinal clinical data, including medication records, from PWH across four sites. Participants are followed approximately every six months, providing extended follow-up from May 1999 through March 2020. These data contribute substantial longitudinal depth for studying evolving medication use and comorbidity patterns.

\paragraph{HIV Neurobehavioral Research Program (HNRP):}  
The HNRP integrates data from the HIV Neurobehavioral Research Center (HNRC) and more than 30 affiliated studies. It provides additional longitudinal data on PWH, including medication use and clinical characteristics.




\section{Additional details in LOPEL method} \label{sec:s-method}

\subsection{Posterior Inference via Sparse Variational Approximation} \label{sec:s-VI}

Let $f_y: \mathcal{X} \to \mathbb{R}^D$ denote the GP mapping from the input space to the $D$-dimensional embedding space. 
The generalized Bayesian posterior takes the form
$$\tilde{p}(f_y) \propto p(f_y)\exp\{-\ell(f_y, X)\},$$ 
where $p(\cdot)$ is the GP prior and $\ell(\cdot)$ is the self-supervised loss defined in the main text.
To enable scalable inference, we introduce $M_u$ inducing inputs $U_x = \{u_m\}_{m=1}^{M_u} \subset \mathcal{X}$, initialized by sampling uniformly from the training inputs, with $M_u$ pre-specified. The corresponding inducing variables are $U_y^{(k)} = f_y^{(k)}(U_x) \in \mathbb{R}^{M_u}$ for each output dimension~$k$. The variational distribution factorizes across output dimensions:
\[
  q(U_y) = \prod_{k=1}^{D} \mathcal{N}\!\left(U_y^{(k)} \mid m^{(k)},\, S^{(k)}\right),
\]
where $m^{(k)} \in \mathbb{R}^{M_u}$ and $S^{(k)} \in \mathbb{R}^{M_u \times M_u}$ is parametrized via its Cholesky factor $L^{(k)}$ such that $S^{(k)} = L^{(k)} (L^{(k)})^\top$. 
 
Under this approximation, the evidence lower bound (ELBO) associated with the generalized posterior is given by
\[
  \mathrm{ELBO} = -\mathbb{E}_{q(f_y)}\!\left[\ell(f_y)\right] - \mathrm{KL}\!\left(q(U_y) \| p(U_y)\right),
\]
where $q(f_y) = \int p(f_y \mid U_y) q(U_y) d U_y$.
The expected loss is estimated via Monte Carlo with $S$ reparametrized samples,
\[
  \mathbb{E}_{q}[\ell(f_y)] \approx \frac{1}{S}\sum_{s=1}^{S} \ell\!\left(f_y^{(s)}\right), \quad f_y^{(s)} \sim q(f_y).
\]
The ELBO is maximized jointly over variational parameters $\{m^{(k)}, L^{(k)}\}_{k=1}^D$, kernel hyperparameters, and inducing inputs $U_x$, using Adam. Training uses full-batch gradient descent and $S = 500$ Monte Carlo samples per iteration.

The posterior mean embedding for regimen $x_{ij}$ at visit $j$ of patient $i$ is then
\[
  y_{ij} = \mathbb{E}_q[f_y(x_{ij})] \approx k(x_{ij}, U_x)\,K_{U_x,U_x}^{-1}\,m \in \mathbb{R}^D,
\]
where $k(x_{ij}, U_x)$ is the $1 \times M_u$ vector of kernel evaluation between $x_{ij}$ and the inducing inputs and 
$K_{U_x, U_x}$ is the $M_u \times M_u$ kernel matrix among inducing inputs, and $m$ concentrates $\{m^{(k)}\}_{k=1}^D$.


\subsection{Technical details for stage 2 of LOPEL} \label{sec:s-stage2}

This section provides technical details supplementing Section \ref{sec:method_stage2} of the main text, covering the multitask GP regression used to represent individual trajectories, the diagonal approximation to the Wasserstein distance that defines the trajectory kernel, and a summary of inference procedure for LOPEL stage 2.

\renewcommand{\theequation}{S.\arabic{equation}}

\paragraph{Multitask Gaussian Process regression for each individual}

Section \ref{sec:method_stage2} models each participant's longitudinal embedding trajectory $g_i(t)$ as a multitask Gaussian process and represents the resulting posterior as a Gaussian distribution $q_i$ on a common evaluation grid. Here we specify the full model.
For participant $i$, let $y_{ij} \in \mathbb{R}^D$ denote the visit-level embedding at time $t_{ij}$, and define the latent trajectory $g_i(t) \in \mathbb{R}^D$. We model $g_i(\cdot)$ using a multitask Gaussian process under the intrinsic coregionalization model (ICM):
\begin{equation} \label{eq:icm_model_prior}
g_i(\cdot) \sim \mathcal{GP}\!\left(0, K_g(\cdot,\cdot)\right), \quad
K_g(t,t') = B \, k(t,t'),
\end{equation}
where $k(t,t') = \exp\!\left(-\|t - t'\|^2 / (2\gamma^2)\right)$ is an Radial Basis Function (RBF) kernel and $B \in \mathbb{R}^{D \times D}$ is a positive semidefinite coregionalization matrix.

Observations follow
\begin{equation} \label{eq:icm_model_likelihood}
y_{ij} = g_i(t_{ij}) + \varepsilon_{ij}, \quad
\varepsilon_{ij} \sim \mathcal{N}(0, \Sigma_\varepsilon),
\end{equation}
where $\Sigma_\varepsilon$ is the task noise covariance. In implementation, $\Sigma_\varepsilon$ is learned via a multitask Gaussian likelihood, which may be diagonal or low-rank correlated across embedding dimensions.

For each participant, we fit the multitask GP by maximizing the exact marginal likelihood with respect to the kernel hyperparameters, the coregionalization matrix $B$, and the noise covariance $\Sigma_\varepsilon$. In implementation, this optimization is performed using the \texttt{gpytorch} library, where the exact marginal log-likelihood is optimized via stochastic gradient descent (Adam) with automatic differentiation.

Let $t_i = (t_{i1}, \dots, t_{iv_i})$ and define the stacked observation vector
\[
y_i = \mathrm{vec}\!\left([y_{i1}, \dots, y_{iv_i}]^\top\right) \in \mathbb{R}^{v_i D}.
\]
Let $t^\star = (t_1^\star, \dots, t_G^\star)$ denote a fixed evaluation grid shared across all participants. Define
\[
K_\star = \big(k(t_a^\star, t_{ib})\big)_{a=1,\dots,G;\,b=1,\dots,v_i}, 
\quad
K_{\star\star} = \big(k(t_a^\star, t_b^\star)\big)_{a,b=1}^G.
\]
The posterior distribution of the latent trajectory evaluated on $t^\star$ is
\begin{equation} \label{eq:qi_post}
    q_i = g_i(t^\star)\mid y_i \sim \mathcal{N}(m_i, \Sigma_i),
\end{equation}
with
\[
m_i = (K_\star \otimes B)\left(K_t \otimes B + I_{v_i} \otimes \Sigma_\varepsilon\right)^{-1} y_i,
\]
\[
\Sigma_i = (K_{\star\star} \otimes B)
- (K_\star \otimes B)\left(K_t \otimes B + I_{v_i} \otimes \Sigma_\varepsilon\right)^{-1}(K_\star^\top \otimes B),
\]
where $K_t \in \mathbb{R}^{v_i \times v_i}$ is the temporal kernel matrix with entries $k(t_{ia}, t_{ib})$.
Each $q_i$ is thus a Gaussian distribution on $\mathbb{R}^{p}$ with $p = G D$.

\paragraph{Diagonal approximation and Wasserstein distance based trajectory kernel.}
Section \ref{sec:method_stage2} introduces a diagonal approximation for computational scalability of the trajectory kernel.
Specifically, we approximate $\Sigma_i$ by its diagonal:
\[
\Sigma_i \approx \tilde{\Sigma}_i = \mathrm{diag}(\sigma_{i,1}^2, \dots, \sigma_{i,p}^2),
\]
where $\sigma_{i,\ell}^2 = [\Sigma_i]_{\ell\ell}$.
Under this approximation, the squared 2-Wasserstein distance between $q_i = \mathcal{N}(m_i,\tilde{\Sigma}_i)$ and $q_j = \mathcal{N}(m_j,\tilde{\Sigma}_j)$ simplifies to
\begin{equation} \label{eq:W2-diag}
    W_2^2(q_i,q_j) = \|m_i - m_j\|^2 + \sum_{\ell=1}^p (\sigma_{i,\ell} - \sigma_{j,\ell})^2.
\end{equation}

We define the trajectory kernel
\begin{equation} \label{eq:traj-kernel}
    K_{W}(q_i,q_j) = \exp\!\left(-\frac{W_2^2(q_i,q_j)}{2\tau^2}\right),
\end{equation}
where $\tau > 0$ is a learnable length-scale.
The resulting $n \times n$ kernel matrix $K_{W}$ is used as input to the second-stage GPSSL model, which learns subject-level embeddings $z_i \in \mathbb{R}^d$ using the same generalized Bayesian framework as in Stage 1.

\paragraph{Inference procedure for Stage 2 of LOPEL.}
For clarity, we summarize the complete inference procedure for Stage~2 of LOPEL:
 
\begin{enumerate}[leftmargin=2em]
  \item \textbf{Fit per-participant multitask GP.} For each participant~$i$ with visit-level embeddings $\{y_{ij}\}_{j=1}^{v_i}$ at times $\{t_{ij}\}_{j=1}^{v_i}$ from Stage~1, fit the ICM Gaussian process model in \eqref{eq:icm_model_prior}-\eqref{eq:icm_model_likelihood}.
 
  \item \textbf{Compute posterior on common grid.} On a common evaluation grid $t^\star = (t_1^\star, \ldots, t_G^\star)$, for each participant, compute the posterior mean $m_i$ and covariance $\Sigma_i$ via \eqref{eq:qi_post}.
 
  \item \textbf{Construct trajectory kernel matrix.} Approximate $\Sigma_i$ by its diagonal, yielding the marginal standard deviation vector $\bm{\sigma}_i \in \mathbb{R}^p$. For all participant pairs $(i, j)$, compute $W_2^2(q_i, q_j)$ via \eqref{eq:W2-diag}. Form the $n \times n$ kernel matrix with entries $K_{W}(q_i, q_j)$ using \eqref{eq:traj-kernel}.
 
  \item \textbf{Learn subject-level embeddings via GPSSL.} Apply the generalized Bayesian GPSSL framework (as in Stage~1, but with $K_{\mathrm{traj}}$ as the GP kernel) to learn $d$-dimensional subject-level embeddings $z_i \in \mathbb{R}^d$ for each participant. The GPSSL loss from Equation~(2) in the main text is used with the same variance and covariance regularization terms.
\end{enumerate}

%% file: table1.tex
{\renewcommand{\arraystretch}{0.75}
\begin{table}[htbp!]
\centering
\footnotesize

\resizebox{\textwidth}{!}{
\begin{tabular}{lcccccc}
\toprule
    & \textbf{PWH Overall} & \textbf{PWH-C1} & \textbf{PWH-C2} & \textbf{PWH-C3} & \textbf{PWH-C4} & \textbf{PWOH} \\
\textbf{Characteristic, N (\%)} & (N = 427) & (N = 106) & (N = 102) & (N = 130) & (N = 89) & (N = 155) \\
\midrule
\rowcolor{gray!20}
\multicolumn{7}{l}{\textbf{Demographic Factors}} \\
Age, years, median (IQR) & 54 (46–62) & 57 (51–63) & 55 (49–63) & 54 (44–62) & 49 (40–56) & 53 (44–59) \\
Education level $>=12$ yrs & 319 (75) & 86 (81) & 70 (69) & 99 (76) & 64 (72) & 150 (97) \\
\multicolumn{7}{l}{Sex} \\
\hspace{2em}Male & 244 (57) & 61 (58) & 57 (56) & 70 (54) & 56 (63) & 97 (63) \\
\hspace{2em}Female & 183 (43) & 45 (42) & 45 (44) & 60 (46) & 33 (37) & 58 (37) \\
\multicolumn{7}{l}{Race/ethnicity} \\
\hspace{2em}White & 234 (55) & 64 (60) & 61 (60) & 66 (51) & 43 (48) & 46 (30) \\
\hspace{2em}Black& 123 (29) & 29 (27) & 25 (25) & 39 (30) & 30 (34) & 17 (11) \\
\hspace{2em}Other & 70 (16) & 13 (12) & 16 (16) & 25 (19) & 16 (18) & 92 (59) \\
\rowcolor{gray!20}
\multicolumn{7}{l}{\textbf{Comorbidities}} \\
Hepatitis C Virus infection & 94 (22) & 28 (26) & 21 (21) & 30 (23) & 15 (17) & 7 (5) \\
Diabetes mellitus & 67 (16) & 36 (34) & 18 (18) & 12 (9) & 1 (1) & 15 (10) \\
Hyperlipidemia & 184 (43) & 73 (69) & 50 (49) & 43 (33) & 18 (20) & 40 (26) \\
Hypertension & 184 (43) & 72 (68) & 49 (48) & 46 (35) & 17 (19) & 35 (23) \\
Renal dysfunction (eGFR $<$ 60) & 18 (4) & 7 (7) & 7 (7) & 3 (2) & 1 (1) & 3 (2) \\
Depressive symptoms (BDI-II $\geq$ 16) & 119 (28) & 40 (38) & 31 (30) & 29 (22) & 19 (21) & 5 (3) \\
Obesity (body mass index $\geq$ 30 kg/m$^2$) & 87 (20) & 25 (24) & 24 (24) & 23 (18) & 15 (17) & 40 (26) \\
\rowcolor{gray!20}
\multicolumn{7}{l}{\textbf{HIV Disease-Related Characteristics}} \\
Current CD4 (cells/$\mu$L), median (IQR) & 543 (322–808) & 596 (405–883) & 500 (327–809) & 568 (295–777) & 551 (284–768) & 821 (662–1146) \\
Nadir CD4 (cells/$\mu$L), median (IQR) & 172 (42–347) & 192 (85–326) & 150 (23–300) & 133 (28–300) & 226 (29–421) & 826 (660–1130) \\
\rowcolor{gray!20}
\multicolumn{7}{l}{\textbf{Medications}} \\
\multicolumn{7}{l}{Degree of Polypharmacy} \\
\hspace{2em}Low (0--3 drugs) & 131 (31) & 2 (2) & 20 (20) & 45 (35) & 64 (72) & 128 (83) \\
\hspace{2em}Moderate (4--9 drugs) & 166 (39) & 28 (26) & 48 (47) & 66 (51) & 24 (27) & 27 (17) \\
\hspace{2em}High (10+ drugs) & 130 (30) & 76 (72) & 34 (33) & 19 (15) & 1 (1) & 0 (0) \\
\bottomrule
\end{tabular}
}
\begin{tablenotes}
\scriptsize
\item \textit{Note.} Values are N (\%) for binary variables and median (IQR) for continuous variables.
\item Abbreviations: PWH = people with HIV; PWOH = people without HIV;
eGFR = estimated glomerular filtration rate; BDI-II = Beck Depression Inventory, Second Edition.
\end{tablenotes}
\caption{Baseline characteristics across identified subgroups among PWH and PWOH.}
\label{tab:table1}
\end{table}
}

%% file: bibfile.bib
@article{crothers2006increased,
  title={{Increased COPD among HIV-positive compared to HIV-negative veterans}},
  author={Crothers, Kristina and Butt, Adeel A and Gibert, Cynthia L and Rodriguez-Barradas, Maria C and Crystal, Stephen and Justice, Amy C and Veterans Aging Cohort 5 Project Team and others},
  journal={Chest},
  volume={130},
  number={5},
  pages={1326--1333},
  year={2006},
  publisher={Elsevier}
}

@article{rousseeuw1987silhouettes,
  title={{Silhouettes: a graphical aid to the interpretation and validation of cluster analysis}},
  author={Rousseeuw, Peter J},
  journal={Journal of Computational and Applied Mathematics},
  volume={20},
  pages={53--65},
  year={1987},
  publisher={Elsevier}
}

@article{bissiri2016general,
  title={{A general framework for updating belief distributions}},
  author={Bissiri, Pier Giovanni and Holmes, Chris C and Walker, Stephen G},
  journal={Journal of the Royal Statistical Society Series B: Statistical Methodology},
  volume={78},
  number={5},
  pages={1103--1130},
  year={2016},
  publisher={Oxford University Press}
}

@inproceedings{kornblith2019similarity,
  title={{Similarity of neural network representations revisited}},
  author={Kornblith, Simon and Norouzi, Mohammad and Lee, Honglak and Hinton, Geoffrey},
  booktitle={International conference on machine learning},
  pages={3519--3529},
  year={2019},
  organization={PMlR}
}

@misc{rxclass,
  author = {{U.S. National Library of Medicine}},
  title = {RxClass},
  year = {2026},
  howpublished = {\url{https://rxnav.nlm.nih.gov/RxClass/}},
}

@article{prattValidityRxRiskComorbidity2018,
  title = {The Validity of the {{Rx-Risk Comorbidity Index}} Using Medicines Mapped to the {{Anatomical Therapeutic Chemical}} ({{ATC}}) {{Classification System}}},
  author = {Pratt, Nicole L. and Kerr, Mhairi and Barratt, John D. and Kemp-Casey, Anna and Ellett, Lisa M. Kalisch and Ramsay, Emmae and Roughead, Elizabeth Ellen},
  date = {2018-04-01},
  journal = {BMJ Open},
  volume = {8},
  number = {4},
  eprint = {29654048},
  eprinttype = {pmid},
  pages = {e021122},
  publisher = {British Medical Journal Publishing Group},
  issn = {2044-6055, 2044-6055},
  doi = {10.1136/bmjopen-2017-021122},
  urldate = {2026-03-20},
  langid = {english},
  year = {2018}
}

@article{widagdoValidityUpdatedRxRisk2025,
  title = {Validity of the {{Updated Rx-Risk Index}} as a {{Disease Identification}} and {{Risk-Adjustment Tool}} for {{Use}} in {{Observational Health Studies}}},
  author = {Widagdo, Imaina and Kerr, Mhairi and Kalisch Ellett, Lisa and Schlegel, Clement and Sadeqzadeh, Elham and Wang, Alvin and Clarke, Allison Louise and Pratt, Nicole},
  date = {2025},
  journal = {Clinical Interventions in Aging},
  shortjournal = {Clin Interv Aging},
  volume = {20},
  eprint = {40124172},
  eprinttype = {pmid},
  pages = {309--323},
  issn = {1178-1998},
  doi = {10.2147/CIA.S494145},
  langid = {english},
  pmcid = {PMC11930017},
  year = {2025}
}

@article{petersApproximateMatchingMethod2011a,
  title = {An {{Approximate Matching Method}} for {{Clinical Drug Names}}},
  author = {Peters, Lee and Kapusnik-Uner, Joan E. and Nguyen, Thang and Bodenreider, Olivier},
  year = {2011},
  journal = {AMIA Annual Symposium Proceedings},
  shortjournal = {AMIA Annu Symp Proc},
  volume = {2011},
  eprint = {22195172},
  eprinttype = {pmid},
  pages = {1117--1126},
  issn = {1942-597X},
  urldate = {2026-03-20},
  pmcid = {PMC3243188}
}

@article{heatonHIVassociatedNeurocognitiveDisorders2010a,
  title = {{{HIV-associated}} Neurocognitive Disorders Persist in the Era of Potent Antiretroviral Therapy: {{CHARTER Study}}},
  shorttitle = {{{HIV-associated}} Neurocognitive Disorders Persist in the Era of Potent Antiretroviral Therapy},
  author = {Heaton, R. K. and Clifford, D. B. and Franklin, D. R. and Woods, S. P. and Ake, C. and Vaida, F. and Ellis, R. J. and Letendre, S. L. and Marcotte, T. D. and Atkinson, J. H. and Rivera-Mindt, M. and Vigil, O. R. and Taylor, M. J. and Collier, A. C. and Marra, C. M. and Gelman, B. B. and McArthur, J. C. and Morgello, S. and Simpson, D. M. and McCutchan, J. A. and Abramson, I. and Gamst, A. and Fennema-Notestine, C. and Jernigan, T. L. and Wong, J. and Grant, I. and {CHARTER Group}},
  date = {2010-12-07},
  journal = {Neurology},
  shortjournal = {Neurology},
  volume = {75},
  number = {23},
  eprint = {21135382},
  eprinttype = {pmid},
  pages = {2087--2096},
  issn = {1526-632X},
  doi = {10.1212/WNL.0b013e318200d727},
  year = {2010},
  langid = {english},
  pmcid = {PMC2995535}
}

@misc{WelcomeHNRPa,
  title = {Welcome to the {{HNRP}}},
  author  = {{HIV Neurobehavioral Research Program}},
  urldate = {2026-03-20},
  year = {2026},
  howpublished = {\url{https://hnrp.hivresearch.ucsd.edu/index.php}}
}

@article{zhabokritsky2024non,
  title={{Non-AIDS-defining comorbidities impact health related quality of life among older adults living with HIV}},
  author={Zhabokritsky, Alice and Klein, Marina and Loutfy, Mona and Guaraldi, Giovanni and Andany, Nisha and Guillemi, Silvia and Falutz, Julian and Arbess, Gordon and Tan, Darrell HS and Walmsley, Sharon},
  journal={Frontiers in Medicine},
  volume={11},
  pages={1380731},
  year={2024},
  publisher={Frontiers Media SA}
}

@article{marcus2020comparison,
  title={{Comparison of overall and comorbidity-free life expectancy between insured adults with and without HIV infection, 2000-2016}},
  author={Marcus, Julia L and Leyden, Wendy A and Alexeeff, Stacey E and Anderson, Alexandra N and Hechter, Rulin C and Hu, Haihong and Lam, Jennifer O and Towner, William J and Yuan, Qing and Horberg, Michael A and others},
  journal={JAMA Network Open},
  volume={3},
  number={6},
  pages={e207954},
  year={2020}
}

@article{morgello2001national,
  title={{The National NeuroAIDS Tissue Consortium: a new paradigm in brain banking with an emphasis on infectious disease}},
  author={Morgello, S and Gelman, BB and Kozlowski, PB and Vinters, HV and Masliah, E and Cornford, M and Cavert, W and Marra, C and Grant, I and Singer, EJ},
  journal={Neuropathology and Applied Neurobiology},
  volume={27},
  number={4},
  pages={326--335},
  year={2001},
  publisher={Wiley Online Library}
}

@article{world2017atc,
  title={{ATC}},
  author={World Health Organization and others},
  journal={WHO Drug Information},
  volume={31},
  number={4},
  pages={629--634},
  year={2017},
  publisher={World Health Organization}
}

@incollection{lochmann2019selective,
  title={{Selective serotonin reuptake inhibitors}},
  author={Lochmann, Dee and Richardson, Tara},
  booktitle={Antidepressants: From biogenic amines to new mechanisms of action},
  pages={135--144},
  year={2019},
  publisher={Springer}
}

@article{do2014excess,
  title={{Excess burden of depression among HIV-infected persons receiving medical care in the United States: data from the medical monitoring project and the behavioral risk factor surveillance system}},
  author={Do, Ann N and Rosenberg, Eli S and Sullivan, Patrick S and Beer, Linda and Strine, Tara W and Schulden, Jeffrey D and Fagan, Jennifer L and Freedman, Mark S and Skarbinski, Jacek},
  journal={PloS One},
  volume={9},
  number={3},
  pages={e92842},
  year={2014},
  publisher={Public Library of Science San Francisco, USA}
}

@article{liu2005rxnorm,
  title={{RxNorm: prescription for electronic drug information exchange}},
  author={Liu, Simon and Ma, Wei and Moore, Robin and Ganesan, Vikraman and Nelson, Stuart},
  journal={IT Professional},
  volume={7},
  number={5},
  pages={17--23},
  year={2005}
}

@article{rubin2022degree,
  title={{Degree of polypharmacy and cognitive function in older women with HIV}},
  author={Rubin, Leah H and Neijna, Ava G and Shi, Qiuhu and Hoover, Donald R and Tamraz, Bani and Anastos, Kathryn and Edmonds, Andrew and Fischl, Margaret A and Gustafson, Deborah and Maki, Pauline M and others},
  journal={AIDS Research and Human Retroviruses},
  volume={38},
  number={7},
  pages={571--579},
  year={2022},
  publisher={Mary Ann Liebert, Inc., publishers 140 Huguenot Street, 3rd Floor New~…}
}

@article{greene2014polypharmacy,
  title={{Polypharmacy, drug--drug interactions, and potentially inappropriate medications in older adults with human immunodeficiency virus infection}},
  author={Greene, Meredith and Steinman, Michael A and McNicholl, Ian R and Valcour, Victor},
  journal={Journal of the American Geriatrics Society},
  volume={62},
  number={3},
  pages={447--453},
  year={2014},
  publisher={Wiley Online Library}
}

@article{duan2025self,
  title={{Self-Supervised Learning with Gaussian Processes}},
  author={Duan, Yunshan and Williamson, Sinead},
  journal={arXiv preprint arXiv:2512.09322},
  year={2025}
}

@article{zhao2021longitudinal,
  title={{Longitudinal self-supervised learning}},
  author={Zhao, Qingyu and Liu, Zixuan and Adeli, Ehsan and Pohl, Kilian M},
  journal={Medical Image Analysis},
  volume={71},
  pages={102051},
  year={2021},
  publisher={Elsevier}
}

@article{qiu2025deep,
  title={{Deep representation learning for clustering longitudinal survival data from electronic health records}},
  author={Qiu, Jiajun and Hu, Yao and Li, Li and Erzurumluoglu, Abdullah Mesut and Braenne, Ingrid and Whitehurst, Charles and Schmitz, Jochen and Arora, Jatin and Bartholdy, Boris Alexander and Gandhi, Shrey and others},
  journal={Nature Communications},
  volume={16},
  number={1},
  pages={2534},
  year={2025},
  publisher={Nature Publishing Group UK London}
}

@incollection{nahler2009anatomical,
  title={{Anatomical therapeutic chemical classification system (ATC)}},
  author={Nahler, Gerhard},
  booktitle={Dictionary of pharmaceutical medicine},
  pages={8--8},
  year={2009},
  publisher={Springer}
}

@article{yang2023model,
  title={{Model-based clustering of high-dimensional longitudinal data via regularization}},
  author={Yang, Luoying and Wu, Tong Tong},
  journal={Biometrics},
  volume={79},
  number={2},
  pages={761--774},
  year={2023},
  publisher={Oxford University Press}
}

@article{lu2024sparse,
  title={{A sparse factor model for clustering high-dimensional longitudinal data}},
  author={Lu, Zihang and Chandra, Noirrit Kiran},
  journal={Statistics in Medicine},
  volume={43},
  number={19},
  pages={3633--3648},
  year={2024},
  publisher={Wiley Online Library}
}

@article{nagin2018group,
  title={{Group-based multi-trajectory modeling}},
  author={Nagin, Daniel S and Jones, Bobby L and Passos, Valeria Lima and Tremblay, Richard E},
  journal={Statistical Methods in Medical Research},
  volume={27},
  number={7},
  pages={2015--2023},
  year={2018},
  publisher={SAGE Publications Sage UK: London, England}
}

@article{lin2000latent,
  title={{A latent class mixed model for analysing biomarker trajectories with irregularly scheduled observations}},
  author={Lin, Haiqun and McCulloch, Charles E and Turnbull, Bruce W and Slate, Elizabeth H and Clark, Larry C},
  journal={Statistics in Medicine},
  volume={19},
  number={10},
  pages={1303--1318},
  year={2000},
  publisher={Wiley Online Library}
}

@article{xia2019bayesian,
  title={{Bayesian analysis for mixture of latent variable hidden Markov models with multivariate longitudinal data}},
  author={Xia, Ye-Mao and Tang, Nian-Sheng},
  journal={Computational Statistics \& Data Analysis},
  volume={132},
  pages={190--211},
  year={2019},
  publisher={Elsevier}
}

@article{delara2022prevalence,
  title={{Prevalence and factors associated with polypharmacy: a systematic review and meta-analysis}},
  author={Delara, Mahin and Murray, Lauren and Jafari, Behnaz and Bahji, Anees and Goodarzi, Zahra and Kirkham, Julia and Chowdhury, Mohammad and Seitz, Dallas P},
  journal={BMC Geriatrics},
  volume={22},
  number={1},
  pages={601},
  year={2022},
  publisher={Springer}
}

@article{chang2020polypharmacy,
  title={{Polypharmacy, hospitalization, and mortality risk: a nationwide cohort study}},
  author={Chang, Tae Ik and Park, Haeyong and Kim, Dong Wook and Jeon, Eun Kyung and Rhee, Connie M and Kalantar-Zadeh, Kamyar and Kang, Ea Wha and Kang, Shin-Wook and Han, Seung Hyeok},
  journal={Scientific Reports},
  volume={10},
  number={1},
  pages={18964},
  year={2020},
  publisher={Nature Publishing Group UK London}
}

@article{sakoe2003dynamic,
  title={{Dynamic programming algorithm optimization for spoken word recognition}},
  author={Sakoe, Hiroaki and Chiba, Seibi},
  journal={IEEE Transactions on Acoustics, Speech, and Signal Processing},
  volume={26},
  number={1},
  pages={43--49},
  year={2003},
  publisher={IEEE}
}

@article{zhou2023clustermld,
  title={{ClusterMLD: An efficient hierarchical clustering method for multivariate longitudinal data}},
  author={Zhou, Junyi and Zhang, Ying and Tu, Wanzhu},
  journal={Journal of Computational and Graphical Statistics},
  volume={32},
  number={3},
  pages={1131--1144},
  year={2023},
  publisher={Taylor \& Francis}
}

@article{knoblauch2019generalized,
  title={Generalized variational inference: Three arguments for deriving new posteriors},
  author={Knoblauch, Jeremias and Jewson, Jack and Damoulas, Theodoros},
  journal={arXiv preprint arXiv:1904.02063},
  year={2019}
}

@inproceedings{hensman2015scalable,
  title={Scalable variational {G}aussian process classification},
  author={Hensman, James and Matthews, Alexander and Ghahramani, Zoubin},
  booktitle={Artificial intelligence and statistics},
  pages={351--360},
  year={2015},
  organization={PMLR}
}
